\documentclass[aps,pre,superscriptaddress,amsmath,amssymb]{revtex4}
\usepackage{epsfig}
\newcommand{\figsize}{5.5 in}
\newcommand{\bc}{\begin{center}}
\newcommand{\ec}{\end{center}}
\newcommand{\be}{\begin{eqnarray}}
\newcommand{\ee}{\end{eqnarray}}
\newcommand{\la}{\langle}
\newcommand{\ra}{\rangle}
\newcommand{\no}{\nonumber}
\newcommand{\bmath}{\begin{mathletters}} 
\newcommand{\emath}{\end{mathletters}}
\newcommand{\h}{\hat}

\begin{document}

\title{A Stochastic Theory of Single Molecule Spectroscopy\footnote{
{\it Adv. Chem. Phys.}, vol.~123, Chap.~4, pp.~119--266 (2002)}}

\author{YounJoon Jung}

\affiliation{Department of Chemistry, Massachusetts Institute of
  Technology, Cambridge, MA 02139} 

\affiliation{Department of Chemistry, University of California,
  Berkeley, CA, 94720-1460} 

\author{Eli Barkai}

\affiliation{Department of Chemistry, Massachusetts Institute of
  Technology, Cambridge, MA 02139} 

\affiliation{Department of Chemistry and Biochemistry, 
University of Notre Dame,  Notre Dame, IN, 46556}

\author{Robert J. Silbey}

\affiliation{Department of Chemistry, Massachusetts Institute of
  Technology, Cambridge, MA 02139} 


\begin{abstract}
A theory is formulated for time dependent fluctuations of 
the spectrum of a single molecule in a dynamic environment. 
In particular, we investigate 
the photon counting statistics
of a single molecule undergoing a spectral diffusion process. 
Based on the stochastic optical Bloch equation, 
fluctuations are characterized by Mandel's $Q$ parameter 
yielding the variance of number of emitted photons and the 
second order intensity correlation function, $g^{(2)}(t)$. 
Using a semi-classical approach and linear response theory, we show
that the $Q$ parameter 
can be described by a three-time dipole correlation function. This approach generalizes the 
Wiener-Khintchine formula that gives the average number of fluorescent 
photons in terms of a one-time dipole correlation function.
We classify the time ordering properties of 
the three-time dipole correlation function, 
and show that it can be
represented by three different pulse shape functions 
similar to those used in the context of nonlinear spectroscopy.
An exact solution is found for a single molecule
whose absorption frequency undergoes a two state random telegraph process 
(i.e., the Kubo-Anderson sudden jump process.) 
Simple expressions are obtained from the exact solution 
in the slow and fast modulation regimes based on appropriate 
approximations for each case. 
In the slow modulation regime $Q$ can be large even 
in the long time limit, while 
in the fast modulation regime it becomes small.
\end{abstract}

\maketitle

\section{Introduction}
In recent years, a new approach to condensed phase spectroscopy has emerged,
that focuses on the spectral properties of {\it a single molecule} (SM)
embedded in a condensed phase\cite{moerner-prl-89,moerner-sci-99,orrit-prl-90,tamarat-jpca-00}.
Thanks to experimental advances made in optics and microscopy\cite{smbook-96}, it is
now possible to perform single molecule spectroscopy (SMS) in many different
systems. Motivations for SMS arise from a fundamental point of view (e.g., the
investigation of the field-matter interaction at the level of a SM, the
verification of statistical assumptions made in ensemble spectroscopy, etc) and
from the possibility of applications (e.g., the use of SMS as a probe for large
biomolecules for which a SM is attached as a fluorescent marker).

In general, the spectral properties of each individual molecule vary from
molecule to molecule due to differences in the local environments with which
each SM is
interacting\cite{ambrose-nature-91,ambrose-jcp-91,zumbusch-prl-93,fleury-jlum-93,kozankiewicz-jcp-94,vacha-jcp-97,boiron-cp-99,naumov-prb-01,geva-jpcb-97,brown-jcp-98,barkai-prl-00}.
With its unique ability to detect dynamical phenomena occurring at the level of
an individual molecule surrounded by its local environment, SMS has uncovered
the statistical distributions of microscopic quantities of the environment that
are hidden in traditional ensemble averaged spectroscopy. In particular, a
single molecule spectrum measured for a finite time necessarily ``sees'' the
temporal fluctuations of the host environment that occur on timescales
comparable to the measurement timescale, and therefore lead, in many cases, to
a stochastically fluctuating single molecule spectrum. Time dependent
fluctuation phenomena in SMS occur in many ways, such as spectral diffusion\cite{ambrose-nature-91,ambrose-jcp-91,zumbusch-prl-93,fleury-jlum-93,boiron-cp-99,bach-prl-99}
and fluorescence intermittency\cite{bernard-jcp-93,basche-nature-95,brouwer-prl-98,kuno-jcp-00,neuhauser-prl-00,shimizu-prb-01}.
The physical mechanisms causing these fluctuation phenomena vary depending on
the dynamical processes a SM is undergoing, including: triplet state
dynamics\cite{bernard-jcp-93,basche-nature-95,brouwer-prl-98}, energy transfer
processes\cite{ha-pnas-96,vandenbout-sci-97,yip-jpca-98}, exciton transfer
processes\cite{bach-prl-99}, chemical reactions\cite{wang-prl-95,chernyak-jcp-99,berezhkovskii-jpc-00,xie-acr-96,jia-pnas-97,lu-sci-98},
conformational changes\cite{schenter-jpca-99,agmon-jpcb-00,cao-cpl-00,yang-jpcb-01}, rotational
dynamics\cite{ha-prl-96,ha-prl-98}, and diffusion
processes\cite{edman-jpca-00}. Thus SMS provides a unique microscopic tool to
investigate the dynamical processes that a SM and its environment undergo
during the measurement time.

One important process responsible for time--dependent fluctuations in SMS is
spectral diffusion, i.e., perturbations or excitations in the environment of
the SM produce random changes in the transition frequency of the
SM\cite{ambrose-nature-91,ambrose-jcp-91,zumbusch-prl-93,fleury-jlum-93,boiron-cp-99,bach-prl-99,reilly-prl-93,reilly1-jcp-94,zumofen-cpl-94,tanimura-jcp-98,osadko-jcp-00},
leading to a time--dependent spectrum. Spectral diffusion processes have been
observed in various systems including dye molecules in a molecular
crystal\cite{ambrose-nature-91,ambrose-jcp-91} and in a low temperature
glass\cite{zumbusch-prl-93,fleury-jlum-93}, quantum
dots\cite{empedocles-prl-96}, light harvesting
systems\cite{vanoijen-sci-99,bopp-pnas-97}, and
dendrimers\cite{hofkens-jacs-00}. Since the spectral diffusion process directly
reflects both (i) the interaction between the SM and its environment and (ii)
the local dynamics of the latter, careful analysis of the time--dependent
fluctuations of SMS illuminates the interplay between various dynamical
processes in the condensed phase. In this work we formulate a stochastic theory
of SMS undergoing a spectral diffusion process. In particular, we address the
issue of the counting statistics of emitted photons produced by a SM undergoing
a spectral diffusion process. Our studies show how the fluctuations in SMS can
be used to probe the dynamics of SM and its interaction with the excitations of
the environment.

Previously, the photon counting statistics of an ensemble of molecules, studied
by various methods, for example, the fluorescence correlation spectroscopy\cite{elson-biopolym-74,ehrenberg-cp-74,koppel-pra-74,qian-bpchem-90,chen-bpj-99}
has proved useful for investigating dynamical processes of various systems. The
photon statistics of a SM is clearly different from that of the ensemble of
molecules due both to the absence of inhomogeneous broadening and to the
correlation between fluorescence photons that exists only on the SM level. In
some SMS experiments, the measurement time is limited due to photobleaching,
where the emission of a SM is quenched suddenly because of various reasons, for
example, reaction with oxygen. Thus it is not an easy task in general to
collect a sufficient number of photon counts to have good statistics. However,
many SM--host systems have been found to remain stable for long enough time to
measure photon statistics\cite{brunel-prl-99,fleury-prl-00,lounis-nature-00,nonn-prl-00,edman-jpca-00,molski-cpl-00b,novikov-jcp-01}.
In view of these recent experimental activities, a theoretical investigation of
the counting statistics of photons produced by single molecules, in particular
{\it when there is a spectral diffusion process} is timely and important. 

We will analyze the SM spectra and their fluctuations semiclassically using the
stochastic Bloch equation in the limit of a weak laser field.
The Kubo-Anderson sudden jump approach\cite{anderson-jpsj-54,kubo-jpsj-54,kubo-fluct-62}
is used to describe the spectral diffusion process. 
For several decades this model has been a useful tool for 
understanding line
shape phenomena, namely, of the average number of
counts $\langle n \rangle$ per measurement time $T$,
and has found many applications mostly in 
ensemble measurements, e.g., NMR,\cite{kubo-statphys2-91} and nonlinear
spectroscopy.\cite{mukamel-nonlinear-95}
More recently, it was applied to model SMS
in low temperature glass systems 
in order to describe the static properties of line shapes
\cite{geva-jpcb-97,brown-jcp-98,barkai-prl-00,barkai-jcp-00}  
and also to model the time dependent fluctuations of SMS. 
\cite{plakhotnik-prl-98,plakhotnik-jlum-99,plakhotnik-prb-99}

Mandel's $Q$ parameter quantitatively describes the deviation of the photon
statistics from the Poissonian case\cite{saleh-photoelec-78,mandel-optical-95},
\begin{equation}
Q= {\langle n^2 \rangle - \langle n \rangle^2 \over \langle n \rangle} - 1,
\label{eq:1}
\end{equation}
where $n$ is the random number of photon counts, and the average is taken over
stochastic processes involved. In the case of Poisson counting statistics $Q=0$
while our semiclassical results show super-Poissonian behavior ($Q>0$) for a SM
undergoing a spectral diffusion process. For short enough times, fluorescent
photons emitted by a SM show anti-bunching phenomena ($-1<Q<0$), a
sub-Poissonian nonclassical effect\cite{fleury-prl-00,lounis-nature-00,basche-prl-92,michler-nature-00,short-prl-83,schubert-prl-92}.
Our semiclassical approach is valid when the number of photon counts is large.
Further discussion of the validity of our approach is given in Section \ref{SecDisc}.

One of the other useful quantities to characterize dynamical processes in SMS
is the fluorescence intensity correlation function, also called the
second--order correlation function, $g^{(2)}(t)$, defined by\cite{cohentann-atom-93,loudon-qlight-83} \be g^{(2)}(\tau)={\la
I(t+\tau)I(t)\ra \over \la I\ra^2}. \label{g2intro} \ee This correlation
function has been used to analyze dynamical processes involved in many SMS
experiments\cite{yip-jpca-98,zumbusch-prl-93,weston-jcp-98,osadko-jetp-98,osadko-jetp-99}.
Here $I(t)$ is the random fluorescence intensity observed at time $t$. It is
well known that for a stationary process there is a simple relation between
$g^{(2)}(t)$ and $Q(T)$\cite{fleury-prl-00,short-prl-83,kim-pra-87}
\be
Q(T)={2\la I\ra \over T} \int_0^{T}{\rm d}t_1 \int_{0}^{t_1}{\rm
d}t_2g^{(2)}(t_2)-\la I\ra T, \label{Qg2intro} \ee
where $T$ is the measurement time.

The essential quantity in the present formulation is a three--time correlation
function, $C_3\left( \tau_1 , \tau_2, \tau_3\right)$, which is similar to the
nonlinear response function investigated in the context of four wave mixing
processes\cite{mukamel-nonlinear-95}. The three--time correlation function
contains all the microscopic information relevant for the calculation of the
lineshape fluctuations described by $Q$. It has appeared as well in a recent
paper of Plakhotnik\cite{plakhotnik-prb-99} in the context of
intensity--time--frequency--correlation technique. In the present work,
important time ordering properties of this function are fully investigated, and
an analytical expression for $Q$ is found. The relation between $C_3\left(
\tau_1 , \tau_2, \tau_3\right)$  and lineshape fluctuations described by $Q$
generalizes the Wiener--Khintchine theorem, that gives the relation between the
one--time correlation function and the averaged lineshape.

The timescale of the bath fluctuations is an important issue in SMS. Bath
fluctuations are typically characterized as being in either fast or slow
modulation regimes (to be defined later)\cite{kubo-fluct-62}. If the bath is
very slow a simple adiabatic approximation is made based on the steady state
solution of time--independent Bloch equation. Several studies have considered
this simple limit in the context of
SMS\cite{zumbusch-prl-93,fleury-jlum-93,reilly-prl-93,reilly1-jcp-94}. From a
theoretical and also experimental point of view it is interesting to go beyond
the slow modulation case. In the fast modulation case it is shown that a
factorization approximation for the three--time correlation function yields a
simple limiting solution. In this limit the lineshape exhibits the well known
behavior of motional narrowing (as timescale of the bath becomes short, the
line is narrowed). By considering a simple spectral diffusion process, we show
that $Q$ exhibits a more complicated behavior than the lineshape does. When the
timescale of the bath dynamics goes to zero, we find Poissonian photon
statistics. Our exact results can be evaluated for an arbitrary timescale of
the bath and are shown to interpolate between the fast and slow modulation
regimes.

This paper is organized as follows. 
In Sec.~\ref{SecOBE}, the stochastic Bloch equation is
presented and a brief discussion of its physical interpretation
is given, and in Sec.~\ref{SecNoise} 
the prescription for the relation between the solution 
of the optical Bloch equation and 
the discrete photon counts is described.
We briefly review several results on counting statistics, 
which will later clarify the meaning of some of our results. 
Section~\ref{SecSim} presents simple simulation 
results of SM spectra in the presence of the spectral diffusion 
to demonstrate a generic physical situation 
to which the present theory is applicable.    
In Sec.~\ref{SecTT}, an important relationship between $Q$ and 
the three-time correlation function is found, and the general properties of 
the latter are investigated. 
An exact solution for a simple spectral
diffusion process is found in Sec.~\ref{SecTSJM}. 
In Sec.~\ref{SecQ} we analyze the exact solution in various limiting cases 
so that the physical meaning of our results becomes
clear. Connection of the present theory to experiments 
is made in Sec.~\ref{SecExp}. In Sec.~\ref{SecDisc} 
we further discuss the validity of the present model 
in connection with other approaches. 
We conclude in Sec.~\ref{SecCon}. Many of the mathematical 
derivations are relegated to the Appendices.

\section{Theory}
\label{SecTheory}
Our theory presented in this section consists of two parts;
first, we model the time evolution of a
SM in a dynamic environment by the stochastic optical Bloch equation, and
second, we introduce the photon counting statistics of a SM by
considering the shot noise process due to the discreteness of photons.

\subsection{Stochastic Optical Bloch Equation}
\label{SecOBE}
We assume a simple nondegenerate
two level SM in an external classical laser field.
The electronic excited state $|e\ra$ is located at energy $\omega_0$ above the
ground state $| g\ra$. We consider the time--dependent SM Hamiltonian
\begin{equation}
H = { 1 \over 2} \hbar\omega_0 \sigma_z + \sum_{j=1}^J {1 \over 2}\hbar \Delta
\omega_j (t) \sigma_z -{\bf d}\cdot {\bf E}_0 \cos\left(\omega_L t \right),
\label{eq:H}
\end{equation}
where $\sigma_z$ is the Pauli matrix. The second term reflects the effect of
the fluctuation of the environment on the absorption frequency of the SM
coupled to $J$ perturbers. The stochastic frequency shifts $\Delta
\omega_{j}(t)$ (i.~e.~the spectral diffusion) are random functions whose
properties will be specified later. The last term in Eq.~(\ref{eq:H}) describes
the interaction between the SM and the laser field (frequency $\omega_L$),
where $ {\bf d} \equiv {\bf d}_{e g} \sigma_x$ is the dipole operator with the
real matrix element ${\bf d}_{eg} = \la e | {\bf d} | g \ra$. We assume that
the molecule does not have any permanent dipole moment either in the ground or
in the excited state, $\la g | {\bf d} | g \ra=\la e|{\bf d}|e\ra =0$.

In the limit of a weak external field the model Hamiltonian describes the well
known Kubo--Anderson random frequency modulation process whose properties are
specified by statistics of $\Delta \omega_j(t)$\cite{anderson-jpsj-54,kubo-jpsj-54,kubo-fluct-62}. When the fluctuating part
of the optical frequency $\Delta \omega _j$ is a two state random telegraph
process, the Hamiltonian describes a SM (or spin of type $A$) coupled to $J$
bath molecules(or spins of  type $B$), these being two level systems. Under
certain conditions this Hamiltonian describes a SM interacting with many two
level systems in low temperature glasses that has been used to analyze SM
lineshapes\cite{geva-jpcb-97,brown-jcp-98,barkai-prl-00,barkai-jcp-00,plakhotnik-jlum-99,plakhotnik-prb-99}.

The molecule is described by $2\times 2$ density matrix $\rho$ whose elements
are $\rho_{gg},\rho_{ee},\rho_{ge},$ and $\rho_{eg}$. Let us define
\be
u&\equiv&
{1\over 2}{( \rho_{ge} e^{ - i \omega_L t} + \rho_{eg}e^{ i \omega_L t}) }, \label{eq:u} \\
v&\equiv&
{1\over 2i}{( \rho_{ge}e^{ - i \omega_L t} - \rho_{eg}e^{i \omega_L t})}, \label{eq:v}  \\
w&\equiv&{1\over 2}{  (\rho_{ee} - \rho_{gg})}, \label{eq:w} \ee
and note that from the normalization condition $\rho_{ee} + \rho_{gg} =1$, we
have  $\rho_{ee} = w +1/2$. By using Eq.~(\ref{eq:H}) the stochastic Bloch
equations in the rotating wave approximation are given by\cite{shore-josab-84,colmenares-theochem-97}
\be
\dot{u} &=& \delta_L(t) v -  {\Gamma u\over 2}, \label{eq:udot} \\
\dot{v} &=& -\delta_L(t) u -  { \Gamma v \over 2} -
\Omega w, \label{eq:vdot} \\
\dot{w} &=& \Omega v - \Gamma w - {\Gamma \over 2}. \label{eq:wdot} \ee
$1/\Gamma$ is the radiative lifetime of the molecule added phenomenologically
to describe spontaneous emission, $\Omega=- {\bf d}_{eg}\cdot {\bf E}_0/\hbar$
is the Rabi frequency, and the detuning frequency is defined by
\be
\delta_L(t)&=& \omega_L - \omega_0 - \Delta \omega(t), \label{eq:dlt} \\
\Delta\omega(t)&=&\sum_{j=1}^{J}\Delta\omega_{j}(t).
\label{eq03a}
\ee
Besides the natural relaxation process described by
$\Gamma$, other $T_1$ and $T_2$ processes
can easily be included in the present theory.
$w$ represents half the difference between the populations of the
state $|e\ra$ and $|g\ra$, while $u$ and $v$ give the mean value of
the dipole moment ${\bf d}$,
\begin{equation}
\mbox{Tr}\left( \rho {\bf d} \right)=
2 {\bf d}_{ge}\left[ u \cos\left( \omega_L t \right) - v \sin\left(\omega_L t\right) \right].
\label{eqTr}
\end{equation}
In recent studies\cite{lounis-prl-97,brunel-prl-98}
it has been demonstrated that the {\em deterministic}
two level optical Bloch equation approach
captures the essential features of SMS in condensed phases,
which further justifies our assumptions.

The physical interpretation of the optical Bloch equation in the absence of
time--dependent fluctuations is well
known\cite{mukamel-nonlinear-95,cohentann-atom-93}. Now that the stochastic
fluctuations are included in our theory we briefly discuss the additional
assumptions needed for standard interpretation to hold. The time--dependent
power absorbed by the SM due to work of the driving field is,
\begin{equation}
{{\rm d} {\cal {W}} \over {\rm d} t} = \cos\left( \omega_L t\right)
{\bf E}_0\cdot {{\rm d} \over {\rm d} t} \mbox{Tr}\left(\rho {\bf d} \right).
\label{eq04}
\end{equation}
As usual, additional averaging (denoted with overbar) of Eq.~(\ref{eq04}) over
the optical period of the laser  is made. This averaging process is clearly
justified for an ensemble of molecules each being out of phase. For a SM, such
an additional averaging is meaningful when the laser timescale, $1/\omega_L$,
is much shorter than any other timescale in the problem (besides $1/\omega_0$,
of course). By using Eq.~(\ref{eqTr}) this means, \be \overline{ v
\cos^2(\omega_L t )}\approx v \overline{\cos^2(\omega_L t)}, \ee under the
conditions $|\dot{u}(t)|\ll\omega_L |v(t)|$ etc, and hence we have
\begin{equation}
{{\rm d} \overline{\cal{W}} \over {\rm d} t} \simeq \hbar \Omega \omega_L v(t).
\label{eq05}
\end{equation}
The absorption photon current (unit 1/[time]) is\cite{cohentann-atom-93}
\begin{equation}
  I(t)  = { 1  \over \hbar \omega_L} {{\rm d}
\overline{\cal{W}} \over {\rm d} t} = \Omega v(t).
\label{eqInt1}
\end{equation}
Neglecting photon shot noise (soon to be considered), $\int_0^T I(t) {\rm d} t$
has the meaning of  the  number of absorbed photons in the time interval
$(0,T)$ (i.e., since $\int_0^T ({{\rm d} \overline{\cal{W}} \over {\rm d}t})
{\rm d} t$ is the total work and each photon carries energy $\hbar \omega_L$).
By using Eqs.~(\ref{eq:udot})--(\ref{eq:wdot}), we have
\begin{equation}
 \dot{\rho}_{ee} = \Omega v - \Gamma \rho_{ee}.
\label{eq000}
\end{equation}
In the steady state, $\dot{\rho}_{ee}=0$,
we have $\Omega v = \Gamma \rho_{ee}$,
and since $\Omega v$
has a meaning of absorbed photon current, $\Gamma \rho_{ee}$ has
the meaning of photon emission current.
For the  stochastic
Bloch equation, a steady photon flux is never reached;
however, integrating Eq.~(\ref{eq000})
over the counting time interval $T$,
\begin{equation}
\left[ \rho_{ee}(T) - \rho_{ee}(0) \right]
+ \Gamma \int_0^T \rho_{ee} (t') {\rm d} t' = \Omega \int_0^T v(t') {\rm d} t',
\label{eq234}
\end{equation}
and using $|\rho_{ee}(T) - \rho_{ee}(0)| \le 1$ we find for large $T$ that the
absorption and emission photon counts are approximately equal,
\begin{equation}
\underbrace{\Gamma \int_0^T \rho_{ee} (t) {\rm d} t}_{\mbox{emitted photons}}
\simeq
\underbrace{ \int_0^T I(t) {\rm d} t\mbox{,} }_{\mbox{absorbed photons}}
\label{eqid1}
\end{equation}
provided that $\int_0^T I(t) {\rm d} t \gg 1$. Eq.~(\ref{eqid1}) is a necessary
condition for the present theory to hold, and it means that the large number of
absorbed photons is approximately equal to the large number of emitted photons
(i.~e.~we have neglected any non--radiative decay channels). When there are
non--radiative decay channels involved, one may modify Eq.~(\ref{eqid1})
approximately by taking into account the fluorescence quantum yield, $\phi$,
the ratio of the number of emitted photons to the number of absorbed photons,
\be \Gamma\int_{0}^{T}\rho_{ee}(t) {\rm d}t \simeq \phi\int_{0}^{T}I(t) {\rm
d}t. \label{eqid2} \ee

\subsection{Classical Shot Noise}
\label{SecNoise}

 Time dependent fluctuations are produced not only
by the fluctuating environment in SMS.
In addition, an important source
of fluctuations is the discreteness of the photon, i.e.,
shot noise.  Assuming a {\em classical}  photon emission
process,
the probability of having a single photon emission event
in time interval $(t,t+ {\rm d} t)$ is\cite{mandel-optical-95}
\be
\mbox{Prob}(t, t+ {\rm d} t) = \Gamma \rho_{ee}(t){\rm d} t.
\label{eq1111}
\ee
While this equation is certainly valid for ensemble of molecules all subjected
to a hypothetical identical time--dependent environment, the validity of this
equation for a SM is far from being obvious. In fact, as we discuss below, only
under certain conditions we can expect this equation to be valid. By using Eq.~(\ref{eq1111})
the probability of recording $n$ photons in time interval
$(0,T)$ is given by the classical counting formula\cite{mandel-optical-95}
\begin{equation}
p(n,T) = { W^n \over n!} \exp\left( - W \right),
\label{eqMandel}
\end{equation}
with
\begin{equation}
W = \eta \int_0^T I(t) {\rm d} t,
\label{eqMandel2}
\end{equation}
where $\eta$ is a suitable constant depending on the detection efficiency. For
simplicity we set $\eta=1$ here, but will re-introduce it later in our final
expressions. Here $W=\eta \overline{\cal{W}}/ \hbar \omega_L$ is $\eta$ times
the work done by the driving laser field whose frequency is $\omega_L$ divided
by the energy of one photon $\hbar \omega_L$ [see Eq.~(\ref{eqInt1})]. It is a
dimensionless time--dependent random variable, described by a probability
density function $P(W,T)$, which at least in principle can be evaluated based
on the statistical properties of the spectral diffusion process and the
stochastic Bloch equations. From Eq.~(\ref{eqMandel}) and for a specific
realization of the stochastic process $\Delta \omega(t)$, the averaged number
of photons counted in time interval $(0,T)$ is given by $W$,
\begin{equation}
\langle n \rangle_{s} = \sum_{n=0}^{\infty} n p(n,T) = W,
\label{eqbaba}
\end{equation}
where the shot noise average is
$\langle ... \rangle_s =\sum_{n=0}^{\infty}...p(n,T)$.
Since $W$ is random, additional
averaging over the stochastic process $\Delta \omega(t) $ is necessary and
statistical properties of the photon count
are determined by $\langle p(n,T) \rangle$,
where $\langle \cdots \rangle$ denotes averaging with respect
to the spectral diffusion (i.e., not including the shot noise),
\be
\la p(n,T) \ra = \left\la {W^n\over n!} \exp(-W) \right \ra.
\label{eqavrp}
\ee
Generally the calculation of $\langle p(n,T) \rangle$ is nontrivial; however,
in some cases simple behavior can be found. Assuming temporal fluctuations of
$W$ occur on the timescale $\tau_c$, then we have

{\bf (a) }  for counting intervals $T \gg \tau_c$ and
for ergodic systems,
\begin{equation}
\lim_{T\to \infty} {\la W \ra\over T}
=\lim_{T \to \infty} {1 \over T} \int_0^T  I(t) {\rm d} t \equiv \langle I\left( \omega_L\right) \rangle,
\label{eqErgo}
\end{equation}
which is the fluorescence lineshape of the molecule, that is, the averaged
number of photon counts per unit time when the excitation laser frequency is
$\omega_L$ [later we suppress $\omega_L$ in $ \langle I\left( \omega_L\right)
\rangle$]. Several authors\cite{chen-bpj-99,mandel-optical-95,schenzle-pra-86}
have argued quite generally (though not in the SM context) that in the long
measurement time limit we may use the approximation $W \simeq \langle I \rangle
T$ (i.e., neglect the fluctuations), and hence photon statistics becomes
Poissonian\cite{mandel-optical-95},
\begin{equation}
\langle p(n,T)\rangle
\simeq {(\langle I \rangle T)^n \over n!}\exp( -  \langle I \rangle T).
\label{eqSB}
\end{equation}
At least in principle $\langle I \rangle$ can be calculated  based  on standard
lineshape theories, (e.g., in Appendix A we calculate $\langle I
\rangle$ for our working example considered in Section \ref{SecTSJM}).
Eq.~(\ref{eqSB}) implies that a single measurement of the lineshape (i.e.,
averaged number of emitted photons as a function of laser frequency) determines
the statistics of the photon count in the limit of long measurement time. In
fact, it tells us that in this case, counting statistics beyond the average
will not reveal any new information on the SM interacting with its dynamical
environment. Mathematically this means that the distribution $P(W,T)$
satisfying
\begin{equation}
\langle p(n,T) \rangle = \int_0^{\infty} {\rm d} W P(W,T)
{ W^n \over n!}  \exp\left( - W \right)
\label{eqrrp}
\end{equation}
converges to
\begin{equation}
P(W,T) \rightarrow  \delta\left( W -  \langle I \rangle T \right),
\label{eqrrp1}
\end{equation}
when $T \to \infty$. We argue below, however, using the central limit theorem,
for cases relevant for SMS, $P(W,T)$ is better described by a Gaussian
distribution. The transformation Eq.~(\ref{eqrrp}) is called the Poisson
transform of $P(W,T)$\cite{saleh-photoelec-78}.

{\bf (b) } in the opposite limit, $T \ll \tau_c$ we may use the
approximation\cite{walls-quantopt}
\begin{equation}
W \simeq I_0 T,
\label{eqShort}
\end{equation}
where $I_0=I(T=0)$.
In a steady state $\langle p(n,T) \rangle$ can be calculated if the
distribution of intensity (i.e., photon current)
is known
\begin{equation}
\langle p(n,T) \rangle \simeq \int_0^\infty {\rm d} I_0 P(I_0) e^{ -  I_0 T}
{ \left(  I_0  T\right)^n \over n!}.
\end{equation}
 For example, assume
that $I(t)$ is a two state process, i.~e.~the case when a SM is coupled to a
single slow two level system in a glass, then
\begin{equation}
P(I) = p_1 \delta\left( I - I_1\right) + p_2 \delta\left( I - I_2\right),
\label{eqTwo}
\end{equation}
and
\begin{equation}
 \langle p(n,T) \rangle = \sum_{i=1,2}
 p_i {\left(  I_i T\right)^n\over n!} \exp\left( - I_i T \right),
\label{eqTwo1}
\end{equation}
and if, for example, $I_2\sim 0$, the SM is either ``on'' or ``off'', a case
encountered in several
experiments\cite{lu-sci-98,empedocles-prl-96,xie-acr-96}.

{\bf (c) } a more challenging case is  when $T \simeq \tau_c$; later, we
address this case in some detail.

We would like to emphasize that the photon statistics we consider is classical,
while the Bloch equation describing dynamics of the SM has quantum mechanical
elements in it (i.e., the coherence). In the weak laser intensity case the
Bloch equation approach allows a classical interpretation based on the Lorentz
oscillator model as presented in Appendix B.

\section{Simulation}
\label{SecSim} To illustrate combined effects of the spectral diffusion and the
shot noise on the fluorescence spectra of a SM, we present simulation results
of spectral trails of a SM, where the fluorescence intensity of a SM is
measured as a function of the laser frequency as the spectral diffusion
proceeds\cite{boiron-cp-99,ambrose-jcp-91}.

First, we present a simple algorithm for generating random fluorescence based
on the theory presented in Section \ref{SecTheory}, by using the stochastic
Bloch equation, Eqs.~(\ref{eq:udot})-(\ref{eq:wdot}), and the classical photon
counting distribution, Eq.~(\ref{eqMandel}). A measurement of the spectral
trail is performed from $t=0$ to $t=t_{\rm end}$. As in the experimental
situation, we divide $t_{\rm end}$ into $N$ time bins each of which has a
length of time $T$. For each bin time $T$, a random number of photon counts is
recorded. Simulations are performed following the steps described below:

{\bf Step (1)} Generate a spectral diffusion process $\Delta\omega(t)$
from $t=0$ to $t=t_{{\rm end}}$.

{\bf Step (2)} Solve the stochastic Bloch equation,
Eqs.~(\ref{eq:udot})-(\ref{eq:wdot}), for a random realization of the spectral
diffusion process generated in Step ($1$) for a given value of $\omega_L$
during the time period $0<t<t_{\rm {end}}$.

{\bf Step (3)} Determine
$W(t_{k},\omega_L)$ during the $k$th time bin ($k=1,2,3,\cdots N$),
$(k-1)T<t<kT$, with a measurement time
$T$ according to Eq.~(\ref{eqMandel2}),
\be
W(t_{k},\omega_L)=\int_{(k-1)T}^{kT} {\rm d}t I(t).
\ee

{\bf Step (4)} Generate a random number $0< x < 1$
using a uniform random number generator, and then the random count
$n$ is found using the criterion,
\begin{equation}
\sum_{j=0}^{n-1} p\left(j,W\right) < x \le \sum_{j=0}^n p\left(j,W\right).
\label{eqAL01}
\end{equation}
According to Eq.~(\ref{eqMandel}) we find
\begin{equation}
{\Gamma\left( n , W \right) \over \left( n - 1\right)!} < x \le
{\Gamma\left( n+1 , W \right) \over n !},
\label{eqAL02}
\end{equation}
where $\Gamma(a,z)$ is the incomplete gamma function.
Steps ($1-4$) must be repeated many times to get good statistics.

For an illustration purpose we choose a simple model of the spectral diffusion,
which is called a two state jump process or a dichotomic process. We assume
that the frequency modulation can be either $\Delta\omega(t)=\nu$ or
$\Delta\omega(t)=-\nu$, and the flipping rate between these two frequency
modulations is given by $R$. This model will be used as a working example for
which an analytical solution is obtained later in Section \ref{SecTSJM}.

\begin{figure}
\bc \epsfig{file=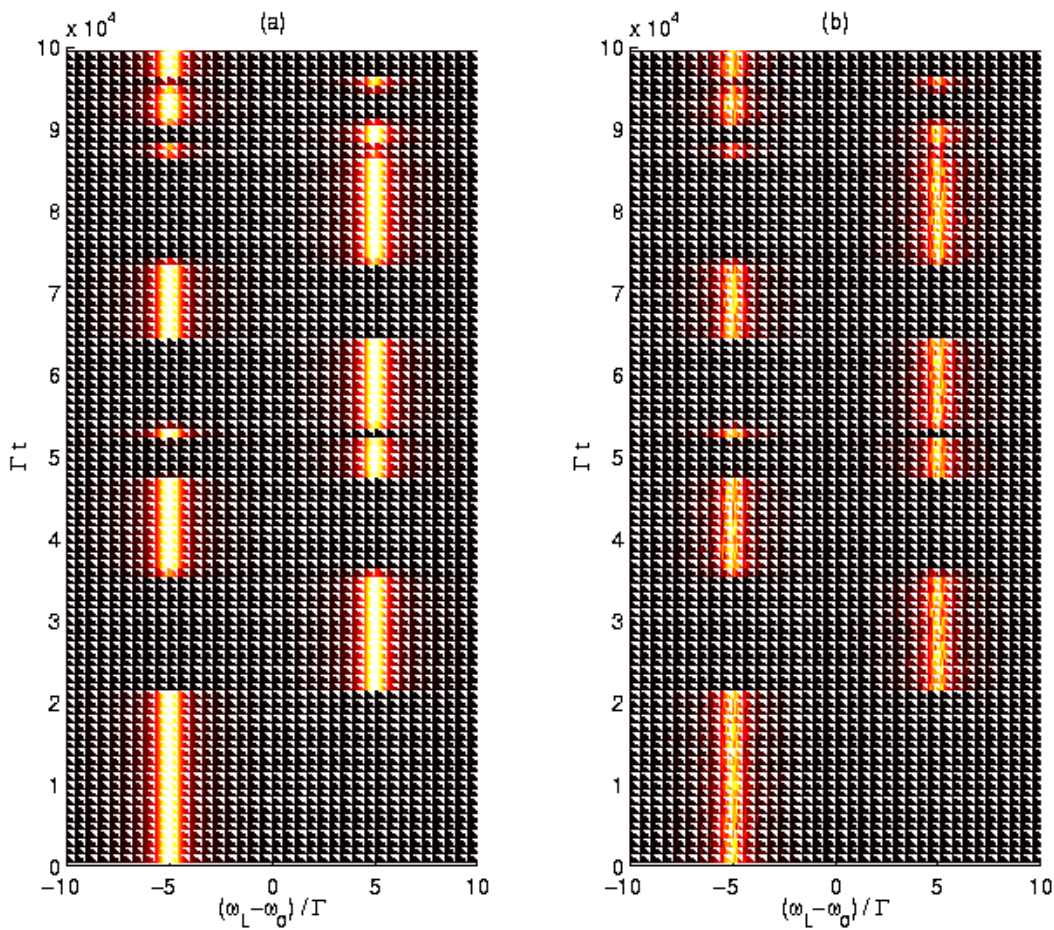,width=\figsize} \ec
\caption[Spectral trails of a SM undergoing a slow spectral diffusion process]
{Spectral trails of a SM undergoing a very slow spectral diffusion process
described by the two state jump model are shown for (a) $W$ (without the shot
noise) and (b) $n$ (with the shot noise). Parameters are chosen as
$\Omega=0.2\Gamma$, $R=10^{-4}\Gamma$, $\nu=5\Gamma$, $T=10^{3}\Gamma^{-1}$,
$t_{\rm end}=10^5 \Gamma^{-1}=100T$, and $\Gamma=1$. } \label{sdslow}
\end{figure}

In Fig.~\ref{sdslow} we present a simulation result of one realization of a
spectral diffusion process when the fluctuation rate $R$ is much smaller than
$\Gamma$ and $\nu$ (slow modulation regime to be defined later). Parameters are
given in the figure caption. Fig.~\ref{sdslow}(a) shows $W(t_{k},\omega_L)$,
demonstrating the effects of the spectral diffusion process on the fluorescence
spectra. Note that $W(t_{k},\omega_L)$ has been defined {\it without the shot
noise}. One can clearly see that the resonance frequency of a SM is jumping
between two values as time goes on, and $W$ shows its maximum values either at
$\omega_L-\omega_0=\nu$ or at  $\omega_L-\omega_0=-\nu$. Since shot noise is
not considered in Fig.~\ref{sdslow}(a), $W$ appears smooth and regular between
the flipping events. In Fig.~\ref{sdslow}(b) we have taken into account the
effects of shot noise as described in Step ($4$) and plotted the random counts
$n$ as a function of $\omega_L$ and $t$. Compared to Fig.~\ref{sdslow}(a), the
spectral trail shown in Fig.~\ref{sdslow}(b) appears more fuzzy and noisy due
to the shot noise effect. It looks similar to the experimentally observed
spectral trails (see, for example, Ref.~\cite{boiron-cp-99}).

\begin{figure}
\bc \epsfig{file=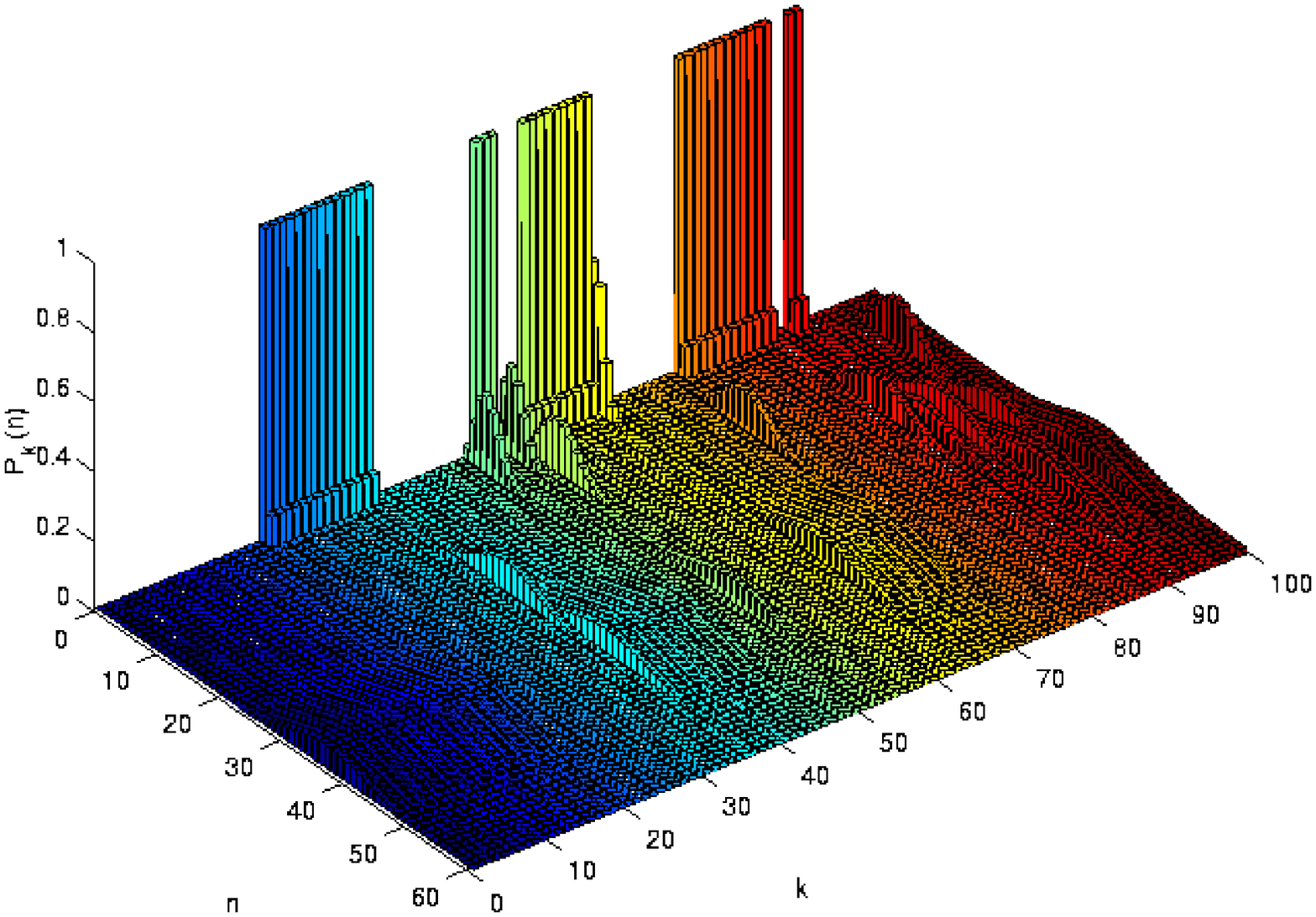,width=\figsize} \ec
\caption[Photon counting distributions $P_{k}(n)$ for a slow
spectral diffusion process]{Time evolution of photon counting distributions
$P_{k}(n)$ for the slow modulation case corresponding to
$\omega_L-\omega_0=-\nu$ in Fig.~\ref{sdslow}. Other parameters are given in
Fig.~\ref{sdslow}.} \label{pnslow}
\end{figure}

In Fig.~\ref{pnslow} we show the evolution of the photon counting distribution
$P_{k}(n)$, Eq.~(\ref{eqMandel}), at a fixed laser frequency, chosen here as
$\omega_L-\omega_0=-\nu=-5\Gamma$. The spectral diffusion process is identical
to that shown in Fig.~\ref{sdslow}. Here $k$ denotes the measurement performed
during the $k$th time bin as described in Step ($3$). Notice that two distinct
forms of the photon counting distributions appear. During the dark period at
the chosen frequency $\omega_L-\omega_0=-\nu$ in Fig.~\ref{sdslow} (e.g.,
$2.1\times 10^4 < \Gamma t< 3.5\times 10^4$ corresponding to $21 < k < 35$),
$P_{k}(n)$ reaches its maximum at $n=0$ with $P_k(n=0)\simeq 1$ and
$P_{k}(n>1)\ll 1$, meaning that the probability for a SM not to emit any photon
during each time bin in the dark period is almost one. However, during the
bright period (for example, $3.5\times10^4 <\Gamma t<4.6\times 10^4$
corresponding to $35 <  k < 46$), $P_{k}(n)$ shows a wide distribution with
$\la n\ra_s \simeq 35$, meaning that on the average $\la n\ra_s\simeq 35$
number of photons are emitted per bin during this period. As the spectral
diffusion proceeds, one can see the corresponding changes in the photon
counting distribution, $P_{k}(n)$, typically among these two characteristic
forms.

\begin{figure}
\bc \epsfig{file=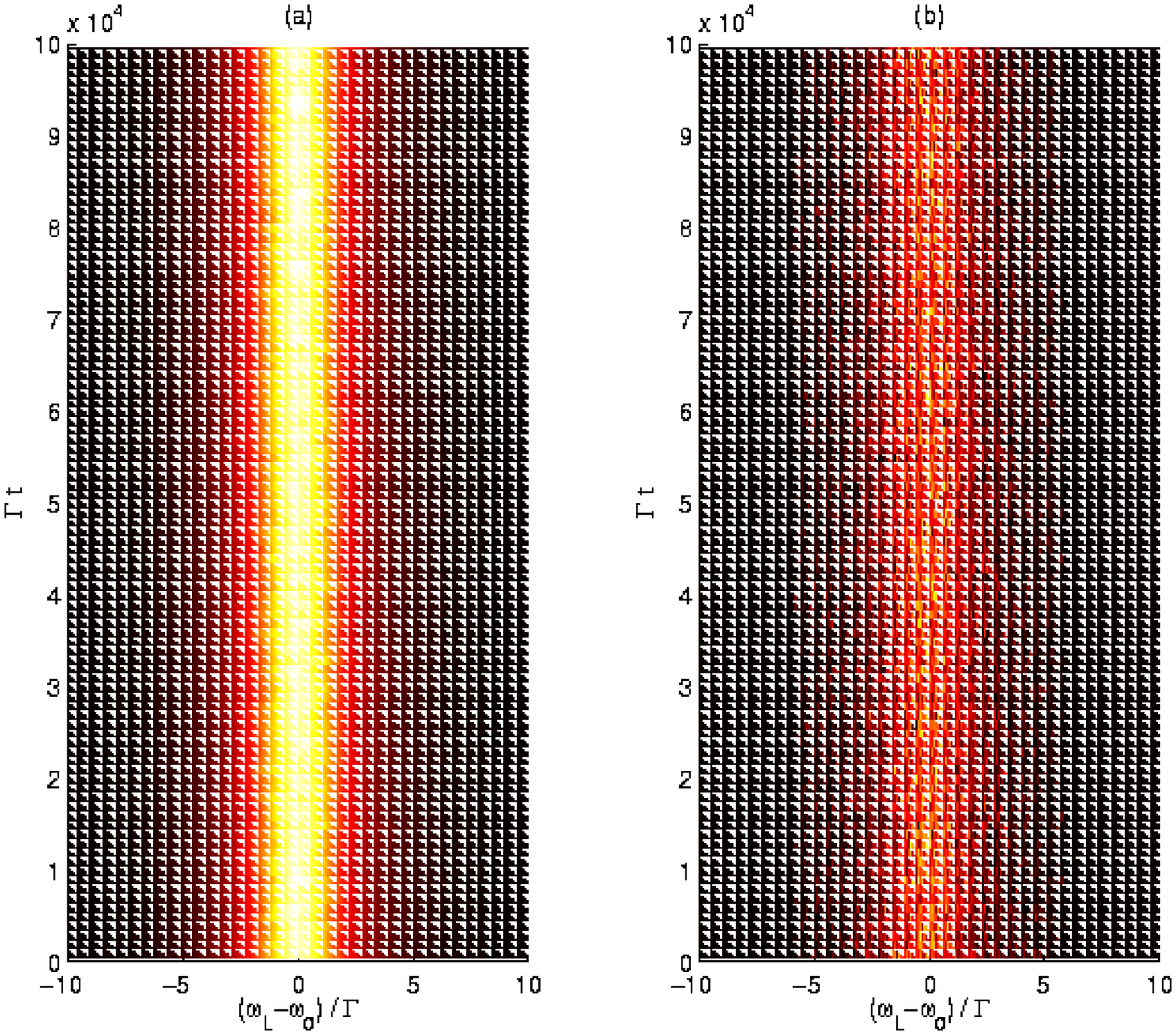,width=\figsize}\ec
\caption[Spectral trails of a SM undergoing a fast spectral diffusion process]
{Spectral trails of a SM undergoing a fast spectral diffusion process described
by the two state jump model are shown for (a) $W$ (without the shot noise) and
(b) $n$ (with the shot noise). Parameters are chosen as $\Omega=0.2\Gamma$,
$R=10\Gamma$, $\nu=5\Gamma$, $T=10^{3}\Gamma^{-1}$, $t_{\rm
end}=10^5\Gamma^{-1}=100T$, and $\Gamma=1$. } \label{sdfast}
\end{figure}

In Fig.~{\ref{sdfast}} we present a simulation result of a spectral trail for
the case when the resonance frequency of a SM fluctuates very quickly compared
with $\Gamma$ and $\nu$ (i.~e.~ $R\gg \nu \gg \Gamma$). In this case, since the
frequency modulation is so fast compared with the spontaneous emission rate, a
large number of frequency modulations are realized during the time $1/\Gamma$,
and the frequency of the SM where the maximum photon counts are observed is
dynamically averaged between $\omega_0-\nu$ and $\omega_0+\nu$ (i.~e.~a
motional narrowing phenomenon)\cite{kubo-fluct-62,talon-josab-92}. The width of
the spectral trail is approximately $\Gamma$, and no splitting is observed even
though the frequency modulation $\nu$ is larger than the spontaneous decay rate
$\Gamma$ ($\nu=5\Gamma$ in this case). This behavior is very different from the
slow modulation case shown in Fig.~\ref{sdslow}, where two separate trails
appear at $\omega_L=\omega_0\pm\nu$.

\begin{figure}
\bc \epsfig{file=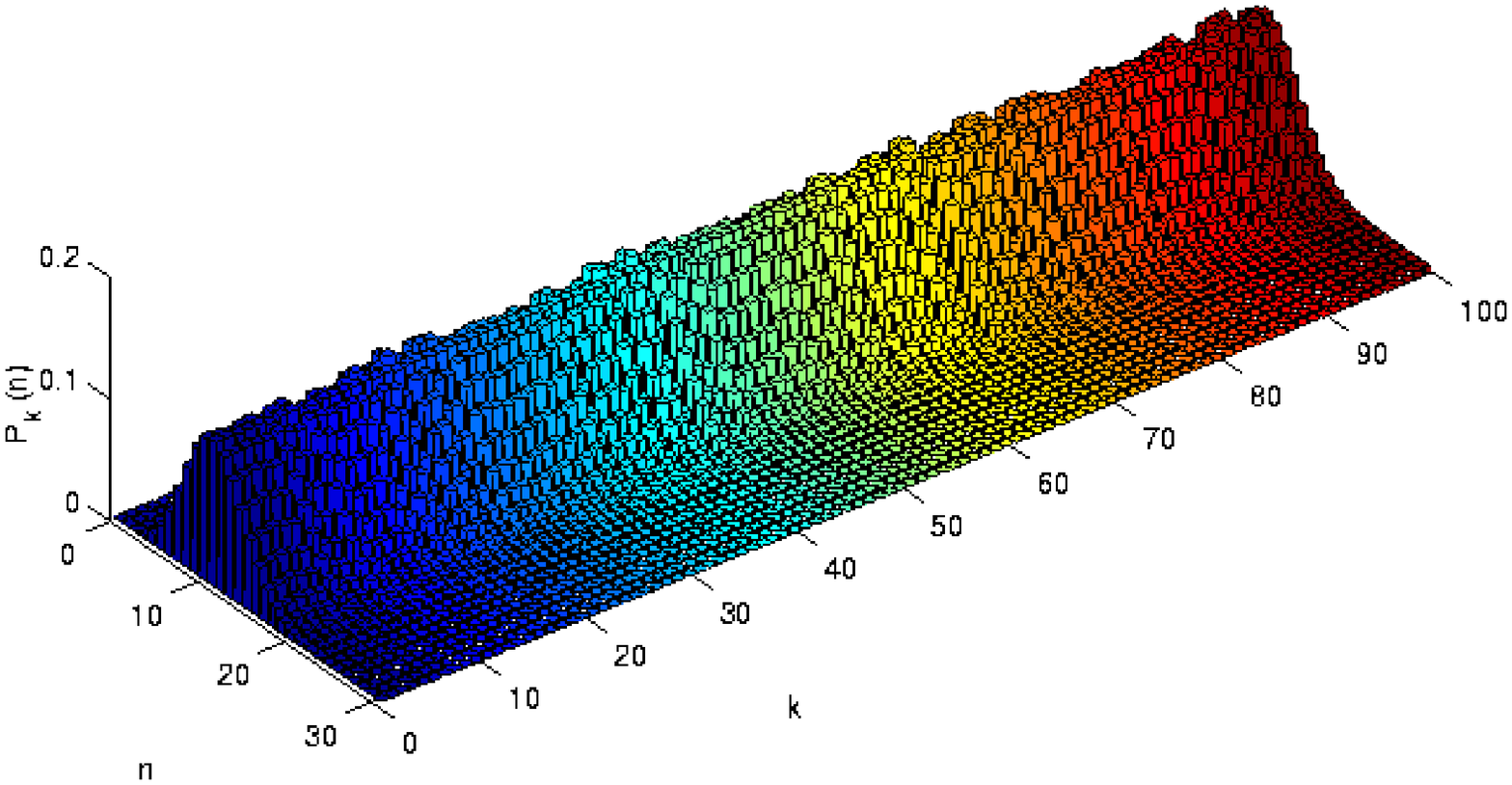,width=\figsize}\ec
\caption[Photon counting distributions $P_{k}(n)$ for a fast
spectral diffusion process] {Time evolution of photon counting distributions
$P_{k}(n)$ for the fast modulation case corresponding to $\omega_L=\omega_0$ in
Fig.~\ref{sdfast}. Other parameters are given in Fig.~\ref{sdfast}.}
\label{pnfast}
\end{figure}

In Fig.~\ref{pnfast} we also show the evolution of the photon counting
distribution $P_{k}(n)$ during a spectral diffusion process at a fixed
frequency $\omega_L=\omega_0$ where the lineshape reaches its maximum in the
fast fluctuation case shown in Fig.~\ref{sdfast}. Unlike the slow modulation
case in Fig.~\ref{pnslow}, where one can see large fluctuations of the photon
counting distributions, the fluctuations of the photon counting distributions
are much smaller in the fast modulation case, and $P_{k}(n)$ shows a broad
Gaussian-like behavior, centered at $\la n\ra_s\simeq 12$.

\section{$Q$ and Three--Time Correlation Function}
\label{SecTT} Having observed the interplay between spectral diffusion and shot
noise on the fluorescence spectra of a SM in simple simulation results of the
previous section, it is natural to ask how one can analyze theoretically the
photon counting statistics of a SM in the presence of a spectral diffusion
process. The probability density of the number of photon counts $\la
p(n,T)\ra$, or equivalently $P(W,T)$ in Eq.~(\ref{eqrrp}), would give complete
information of the dynamical processes of a SM undergoing a spectral diffusion
process, but is difficult to calculate in general. In order to obtain dynamical
information, we will consider the mean $\la W\ra$ and the second moment $\la
W^2\ra$ of the random photon counts.

It is easy to show that the average number of
photons counted in time interval $(0,T)$ is given from Eq.~(\ref{eqavrp}),
\begin{equation}
\langle \langle n \rangle_{s} \rangle = \sum_{n=0}^{\infty} n
\langle  p(n,T) \rangle = \langle W \rangle,
\label{eqAVV03}
\end{equation}
and the second factorial moment of the photon counts in time interval $(0,T)$
is given by \be \la \la n(n-1) \ra_{s} \ra =\sum_{n=0}^{\infty} n(n-1) \la
p(n,T) \ra =\la W^2 \ra. \label{eqW2} \ee The Mandel $Q$ parameter is now
introduced to characterize the fluctuations\cite{mandel-optical-95},
\begin{equation}
{ \langle \langle n^2 \rangle_{s} \rangle - \langle \langle n \rangle_{s} \rangle^2 \over \langle \langle n \rangle_{s} \rangle }  = 1 + Q,
\label{eqDefQ}
\end{equation}
and it is straightforward to show that\cite{mandel-optical-95}
\begin{equation}
Q = { \langle W^2  \rangle - \langle W \rangle^2 \over \langle W  \rangle }.
\label{eqQsol}
\end{equation}
This equation is important relating $Q$ to the variance of the stochastic
variable $W$. We see that $Q\ge0$, indicating that photon statistics is
super-Poissonian.
For our classical case we anticipate:\\
{\bf (a)} for an ergodic system, when $T/\tau_c \to \infty$, and {\it if
Eq.~(\ref{eqSB}) is strictly valid}, $Q\to 0$ (i.e., Poissonian statistics).
However, below we find an analytical expression for $Q$ which is non--zero and
in some cases large even in the limit of $T \to \infty$. We will discuss this
subtle issue later;
\\
{\bf (b)} in the opposite limit, $T \ll \tau_c$,
\begin{equation}
Q= { \langle I_0^2 \rangle - \langle I_0 \rangle^2 \over \langle  I_0 \rangle} T;
\label{eqQQq9}
\end{equation}
{\bf (c) } if $I(t)= I$, independent of time, $Q=0$, as expected;
\\
{\bf (d)} it is easy to see that $Q \propto \eta$, hence when $\eta\to 0$,
counts recorded in the measuring device tend to follow the
Poissonian counting statistics.

We now consider the important limit of weak laser intensity. In this limit the
Wiener--Khintchine theorem relating the lineshape to the one--time correlation
function holds. As we shall show now, a three--time correlation function is the
central ingredient of the theory of fluctuations of SMS in this limit. In
Appendix B we perform a straightforward perturbation expansion with
respect to the Rabi frequency $\Omega$ in the Bloch equation,
Eqs.~(\ref{eq:udot})-(\ref{eq:wdot}), to find
\be
v  \left( t \right)=
{ \Omega \over 2}
\mbox{Re}\left\{  \int_0^t {\rm d} t_1
\exp\left[ - i \int_{t_1}^t {\rm d} t' \delta_L (t')  -{ \Gamma \over 2}(t - t_1)
 \right]
\right\}.
\label{eqA03bmmm}
\ee
According to the discussion in Section \ref{SecTheory} the random number of
photons absorbed in time interval $(0,T)$ is determined by
 $W=\Omega \int_0^T v(t) {\rm d} t$ (see Eqs.~(\ref{eqInt1}) and (\ref{eqMandel2})),
and from Eq.~(\ref{eqA03bmmm}) we find
\be
W={\Omega^2 \over 2} \mbox{Re} \left[ \int_0^T {\rm d} t_2 \int_0^{t_2} {\rm d}
t_1 e^{ - i \omega_L (t_2 - t_1) - \Gamma(t_2 - t_1)/2 + i \int_{t_1}^{t_2}
{\rm d} t' \Delta \omega(t') } \right], \label{eqApzz1} \ee where we have
neglected terms of higher order than $\Omega^2$. In standard lineshape theories
Eq.~(\ref{eqApzz1}) is averaged over the stationary stochastic process and the
long time limit is taken, leading to the well known result for the
(unnormalized) lineshape
%
\begin{equation}
 \langle  I(\omega_L) \rangle  = \lim_{T \to \infty} {\langle  W
 \rangle  \over T}=
{\Omega^2 \over 2} \mbox{Re} \left[ \int_0^{\infty} {\rm d} \tau e^{ - i \omega_L
\tau - \Gamma \tau/2} C_1^{-1} \left( \tau\right) \right],
\label{eqApzz3mmm}
\end{equation}
where we have set $\omega_0=0$. The one--time correlation function
$C_1^{l}(\tau)$ is defined by
\be
C_1^{l}(\tau) =\langle e^{ -i l\int_0^{\tau} \Delta \omega (t') {\rm d} t ' } \rangle,
\label{eqC1}
\ee
where $l=\pm1$ and $\langle \cdots \rangle$ is an average over the stochastic
trajectory $\Delta \omega$. Eq.~(\ref{eqApzz3mmm}) is the celebrated
Wiener--Khintchine formula relating the one--time correlation function to the
average number of photon counts, i.~e.~the averaged lineshape of a SM. We now
investigate lineshape fluctuation by considering the statistical properties of
$W$.

Using Eq.~(\ref{eqA03bmmm}) we show in Appendix B that
\be &&\langle W^2 \rangle  = {\Omega^4 \over 16 }
\int_0^T\int_0^T\int_0^T\int_0^T {\rm d} t_1 {\rm d} t_2 {\rm d} t_3 {\rm d}
t_4 e^{ - i \omega_L ( t_2 - t_1 + t_3 - t_4) - \Gamma(|t_1 -t_2| + |t_3 -
t_4|)/
2} \no \\
&&\mbox{\makebox[0.5in] {  }}\times\left\langle \exp \left[ i \int_{t_1}^{t_2}
\Delta \omega (t') {\rm d} t'
  - i \int_{t_3}^{t_4} \Delta \omega (t') {\rm d} t'\right] \right\rangle.
\label{eqApzz4}
\ee
As can be seen from Eq.~(\ref{eqApzz4}) the key quantity of the theory of
lineshape fluctuation is the three--time correlation function,
\be
C_3 \left(\tau_1 , \tau_2 , \tau_3\right)=
\left \langle \exp\left[ i \int_{t_1}^{t_2} \Delta \omega (t') {\rm d} t' -
 i \int_{t_3}^{t_4} \Delta \omega (t') {\rm d} t' \right] \right \ra,
\label{eqC3}
\ee
%
which depends on the time ordering of $t_1,t_2,t_3,t_4$. In Eq.~(\ref{eqC3}) we
have defined the time ordered set of $\{t_1, t_2, t_3, t_4\}$ as $\{t_{I},
t_{II}, t_{III}, t_{I V}\}$ such that $t_I < t_{II} < t_{III} < t_{I V}$, and
$\tau_1=t_{II}-t_{I}$, $\tau_2=t_{III}-t_{II}$ and $\tau_3=t_{IV}-t_{III}$. Due
to the stationarity of the process $C_3\left(\tau_1 , \tau_2 , \tau_3\right)$
does not depend on the time elapsing between start of observation $t=0$ and
$t_I$. It has a similar mathematical structure to that of the nonlinear
response function used to describe four wave mixing spectroscopies such as
photon echo or hole burning\cite{mukamel-nonlinear-95}.

\begin{table}
\begin{center}
\setlength{\unitlength}{0.35cm}
\begin{center}
\begin{tabular}{|c|c|c|c|}
\hline
 $\ m\ $ & $S_m(t)$  &  $C_3^{m}(\tau_1,\tau_2,\tau_3)$ &  time ordering \\
\hline $\ 1\ $ &\ \setlength{\unitlength}{0.35cm}
\begin{picture}(8.5,3)
\put(0.0,-0.3){-1} \put(0.2,0.7){0} \put(0.2,1.7){1} \put(1,0){\line(1,0){2}}
\put(3,0){\line(0,1){1.}} \put(3,1){\line(1,0){2}} \put(5,1){\line(0,1){1.}}
\put(5,2){\line(1,0){2}} \put(.5,-1.5){$t_I$} \put(2.5,-1.5){$t_{II}$}
\put(4.5,-1.5){$t_{III}$} \put(6.5,-1.5){$t_{IV}$}
\end{picture} &\
$\ {1\over 2}\sum P_{ij}^{-1}(\tau_1)P_{jk}^0(\tau_2) P_{kl}^1(\tau_3)\ $ &
$\ t_1<t_2<t_3<t_4\ $ \\
$ $ &$ $ & $ $ & $ $ \\
\hline $\ 2\ $ &\ \setlength{\unitlength}{0.35cm}
\begin{picture}(8.5,3)
\put(0.2,0.7){1} \put(0.2,-0.3){0} \put(0.0,-1.3){-1}
\put(1,1){\line(1,0){2}} \put(3,1){\line(0,-1){1}} \put(3,0){\line(1,0){2}}
\put(5,0){\line(0,-1){1}} \put(5,-1){\line(1,0){2}}
\end{picture} &\
$\ {1\over 2}\sum P_{ij}^{1}(\tau_1)P_{jk}^0(\tau_2) P_{kl}^{-1}(\tau_3)\ $ &
$\ t_2<t_1<t_4<t_3\ $ \\
$ $ &$ $ & $ $ & $ $ \\
\hline $\ 3\ $ &\ \setlength{\unitlength}{0.35cm}
\begin{picture}(8.5,3)
\put(0.2,.7){1} \put(0.2,-.3){0}
\put(1,1){\line(1,0){2}} \put(3,1){\line(0,-1){1}} \put(3,0){\line(1,0){2}}
\put(5,0){\line(0,1){1}} \put(5,1){\line(1,0){2}}
\end{picture}\  &\
$\ {1\over 2}\sum P_{ij}^{1}(\tau_1)P_{jk}^0(\tau_2) P_{kl}^{1}(\tau_3)\ $ &
$\ t_2<t_1<t_3<t_4\ $ \\
$ $ &$ $ & $ $ & $ $ \\
\hline $\ 4\ $ &\ \setlength{\unitlength}{0.35cm}
\begin{picture}(8.5,3)
\put(0.0,-.3){-1} \put(0.2,.7){0}
\put(1,0){\line(1,0){2}} \put(3,0){\line(0,1){1}} \put(3,1){\line(1,0){2}}
\put(5,1){\line(0,-1){1}} \put(5,0){\line(1,0){2}}
\end{picture} &\
$\ {1\over 2}\sum P_{ij}^{-1}(\tau_1)P_{jk}^0(\tau_2) P_{kl}^{-1}(\tau_3)\ $ &
$\ t_1<t_2<t_4<t_3\ $ \\
$ $ &$ $ & $ $ & $ $ \\
\hline $\ 5\ $ &\ \setlength{\unitlength}{0.35cm}
\begin{picture}(8.5,3)
\put(0.2,-0.3){1} \put(0.2,0.7){2}
\put(1,0){\line(1,0){2}} \put(3,0){\line(0,1){1}} \put(3,1){\line(1,0){2}}
\put(5,1){\line(0,-1){1}} \put(5,0){\line(1,0){2}}
\end{picture} &\
$\ {1\over 2}\sum P_{ij}^{1}(\tau_1)P_{jk}^2(\tau_2) P_{kl}^{1}(\tau_3)\ $ &
$\ t_2<t_3<t_1<t_4\ $ \\
$ $ &$ $ & $ $ & $ $ \\
\hline $\ 6\ $ &\ \setlength{\unitlength}{0.35cm}
\begin{picture}(8.5,3)
\put(0.0,0.7){-1} \put(0.0,-0.3){-2}
\put(1,1){\line(1,0){2}} \put(3,1){\line(0,-1){1}} \put(3,0){\line(1,0){2}}
\put(5,0){\line(0,1){1}} \put(5,1){\line(1,0){2}}
\end{picture} &\
$\ {1\over 2}\sum P_{ij}^{-1}(\tau_1)P_{jk}^{-2}(\tau_2) P_{kl}^{-1}(\tau_3)\ $
&
$\ t_1<t_4<t_2<t_3\ $ \\
$ $ &$ $ & $ $ & $ $ \\
\hline
\end{tabular}
\end{center}

\caption{
Three--time correlation functions $C_3^{m}$ represented by six
different pulse shape functions $S_m(t)$ for $24$ time orderings schemes.
Values of $S_{m}(t)$ for each time interval are shown to the left of the pulse
shape. Only one representative time ordering scheme for each class is shown in
the third column. Three other time ordering schemes belonging to the same class
are obtained by exchanging $t_1\leftrightarrow t_4$ and/or $t_2\leftrightarrow
t_3$. For example, three other time orderings belonging to $m=1$ are
$t_4<t_2<t_3<t_1$, $t_1<t_3<t_2<t_4$, and $t_4<t_3<t_2<t_1$. In the second
column expressions of the three--time correlation functions $C_3^{m}$ are shown
in terms of the weight functions considered in Section \ref{SecTSJM}. Note
that $C_3^{2n-1}(\tau_1,\tau_2,\tau_3)$ and $C_3^{2n}(\tau_1,\tau_2,\tau_3)$
$(n=1,2,3)$ are complex conjugates of each other. \label{table2-1}}
\end{center}
\end{table}

In Eq.~(\ref{eqC3}) there are $4!=24$ options for the time ordering of
$(t_1,t_2,t_3,t_4)$; however, as we show below, only three of them (plus their
complex conjugates) are needed. It is convenient to rewrite the three--time
correlation function as a characteristic functional,
\begin{equation}
C_3^{m} \left(\tau_1 , \tau_2 , \tau_3\right)=
\left \langle\exp\left[ - i \int_{t_I}^{t_{I V}} S_{m}(t') \Delta\omega(t') {\rm d} t' \right] \right \rangle,
\label{eqC3m}
\end{equation}
where $S_m(t)$ ($m=1,2,\cdots,6$) is defined in Table \ref{table2-1} as the
pulse shape function corresponding to the $m$th time ordering. Let us consider
as an example the case $m=1$, $t_1<t_2<t_3<t_4$ (for which
$t_1=t_I,t_2=t_{II},\cdots$, and $\tau_1=t_2 - t_1$, $\tau_2=t_3 -t_2$, and
$\tau_3=t_4 - t_3$). Then the pulse shape function is given by
\begin{equation}
S_1(t) =\left\{
\begin{array}{c c}
-1 \ & t_{I} < t < t_{II} \\
0 \ & t_{II}< t < t_{III} \\
1 \ & t_{III}< t <t_{IV},
\end{array}
\right.
\label{S1}
\end{equation}
and the shape of this pulse is shown in the first line of Table \ref{table2-1}.
Similarly, other pulse shapes describe the other time orderings.

The four dimensional integration in Eq.~(\ref{eqApzz4}) is over  $24$ time
orderings. We note, however, that
\be
&&
e^{ - i \omega_L ( t_2 - t_1 + t_3 - t_4) - \Gamma(|t_1 -t_2| + |t_3 - t_4|)/2}
\left \langle \exp \left[ i \int_{t_1}^{t_2}{\rm d} t' \Delta \omega (t')
 - i \int_{t_3}^{t_4}{\rm d} t' \Delta \omega (t') \right] \right\rangle
\no
\ee
in Eq.~(\ref{eqApzz4}) has two important properties: {\bf (a)} the expression
is invariant when we replace $t_1$ with $t_4$ and $t_2$ with $t_3$, and {\bf
(b)} the replacement of $t_1$ with $t_2$ (or $t_3$) and of $t_3$ with $t_4$(or
$t_1$) has a meaning of taking the complex conjugate. Hence it is easy to see
that only three types of time orderings (plus their complex conjugates) must be
considered. Each time ordering corresponds to different pulse shape function,
$S_{m}(t)$. In Table \ref{table2-1}, for all six time ordering schemes, the
corresponding pulse shape functions are presented. We also give expressions of
$C^{m}_3(t)$ for the working example to be considered soon in Section \ref{SecTT}.

We note that if $\tau_1=\tau_3$ the pulse in Eq.~(\ref{S1}) is identical to
that in the three--time photon echo experiments. The important relation between
lineshape fluctuations and nonlinear spectroscopy has been pointed out by
Plakhotnik in Ref.~\cite{plakhotnik-prb-99} in the context of
intensity--time--frequency--correlation measurement technique.

\section{Two State Jump Model: Exact Solution}
\label{SecTSJM}
\subsection{Model} 
\label{SecTSJM-a}
\begin{figure}
\bc
\epsfig{file=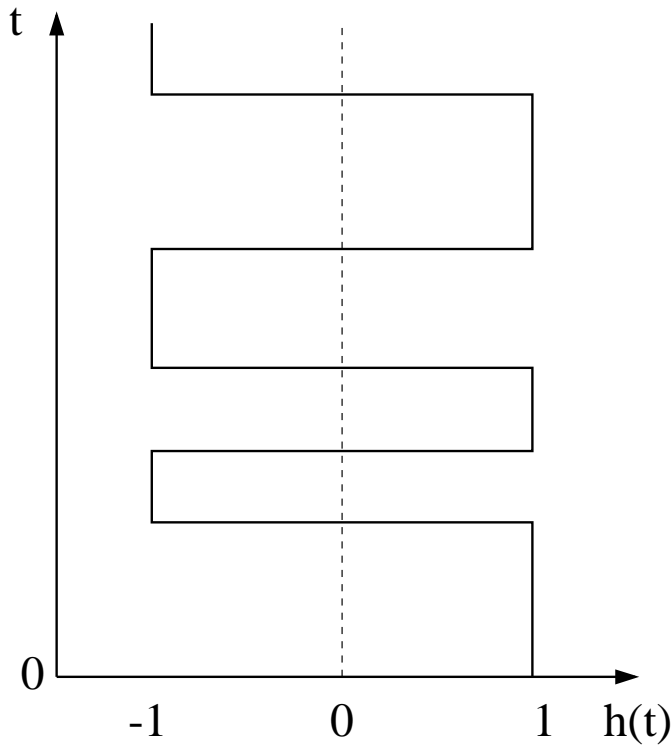,width=3.5in}
\ec \caption[Schematic representation of the spectral diffusion process] { A
schematic representation of the spectral diffusion process modeled by two state
jump process(random telegraph noise). The fluctuating transition frequency is
given by $\Delta\omega(t)=\nu h(t)$. } \label{specdiff}
\end{figure}

In order to investigate basic properties of lineshape
fluctuations we consider a simple situation. We assume that there is only one
bath molecule that is coupled to a chromophore by setting $J=1$ and $\Delta
\omega_1(t)=\nu h(t)$ in Eq.~(\ref{eq:H}), where $\nu$ is the magnitude of the
frequency shift and measures the interaction of the chromophore with the bath,
and $h(t)$ describes a random telegraph process $h(t)=1$ or $h(t)=-1$ depending
on the bath state, up or down, respectively. For simplicity, the transition
rate from up to down and vice versa are assumed to be $R$. The generalization
to the case of different up and down transition rates is important, but not
considered here. A schematic representation of the spectral diffusion process
is given in Fig.~\ref{specdiff}.

This model introduced by Kubo and Anderson in the context of stochastic
lineshape
theory\cite{anderson-jpsj-54,kubo-jpsj-54} is
called the sudden jump model, and it describes a stochastic process that
describes fluctuation phenomena arising from Markovian transitions between
discrete
states\cite{anderson-jpsj-54,kubo-jpsj-54}. For
several decades the Kubo--Anderson sudden jump model has been a useful tool for
understanding lineshape phenomena, namely, of the average number of counts
$\langle n \rangle$ per measurement time $T$, and has found many applications
mostly in ensemble measurements, e.g., NMR\cite{kubo-fluct-62} and nonlinear
spectroscopy\cite{mukamel-nonlinear-95}. More recently, it was applied to model
SMS in low temperature glass systems in order to describe the static properties
of lineshapes\cite{geva-jpcb-97,brown-jcp-98,barkai-prl-00} and
also to model the time--dependent fluctuations of
SMS\cite{plakhotnik-prl-98,plakhotnik-jlum-99,plakhotnik-prb-99}. In this
paper, we will consider this model as a working example to study properties
of lineshape fluctuations.

The above model can describe a single molecule coupled to a single two level
system in low temperature glasses as explained in Section \ref{SecExp}. In this
case $\nu$ depends on the distance between the SM and the two level
system\cite{esquinazi-tunn-98}. Another physical example of this model is the
following: consider a chromophore that is attached to a macromolecule, and
assume that conformational fluctuations exist between two conformations of the
macromolecule. Depending on the conformation of the macromolecule, the
transition frequency of the chromophore is either $\omega_0-\nu$ or
$\omega_0+\nu$\cite{jia-pnas-97}.

\subsection{Solution}
\label{SecTSJM-b} By using a method of Su\'arez and Silbey\cite{suarez-cpl-94},
developed in the context of photon echo experiments,  we now analyze the
properties of the three--time correlation function. We first define the weight
functions,
\begin{equation}
P^{a}_{if}(t) = \left \langle \exp \left[ - i a \int_0^t \Delta \omega (t') {\rm d} t'
\right] \right \rangle_{if},
\label{eq09}
\end{equation}
where the initial
(final) state of the stochastic process $\Delta \omega$ is $i$
 ($f$)
and $a=0$ or $a=\pm 1$ or $a=\pm 2$. For example, $P^{ - 1}_{++} (t)$ is the
value of $\langle e^{i \int_0^t \Delta \omega (t') {\rm d} t' } \rangle$ for a
path restricted to have $\Delta \omega (0) =\nu$ and $\Delta \omega (t) =\nu$.
The one--time correlation function $C_1^{l}(\tau)$ defined in Eq.~(\ref{eqC1})
can be written as sum of these weights, \be C_1^{l}\left(\tau\right)={1\over 2
}\sum_{i,j}P_{ij}^{l}(\tau), \label{C1} \ee where a prefactor $1\over 2$ is due
to the symmetric initial condition and the summation is over all the possible
paths during time $\tau$(i.~e.~$i=\pm,j=\pm$). Also by using the Markovian
property of the process, we can express all the $C_3^{m}$ functions in terms of
the weights. For example, for the pulse shape in Eq.~(\ref{S1})
%
\be
C_3^{1}\left(\tau_1 , \tau_2 , \tau_3\right)=
{1\over 2} \sum_{i,j,k,l} P^{-1}_{ij}\left( \tau_1 \right)
P^0_{jk}\left( \tau_2 \right) P_{kl}^{1}\left( \tau_3\right),
\label{eq10}
\ee
where the summations are over all possible values of $i=\pm$, $j=\pm$, $k=\pm$,
and $l=\pm$. The other $C_3^{m}$ functions are expressed in terms of weights in
Table \ref{table2-1}. Explicit expressions of the weights for the working model
are given by
\be
P_{if}^0(t)&=&{1\over 2}[1+(-1)^{\delta_{if}+1}\exp(-2 R t)], \\
P_{+-}^1(t)&=&P_{-+}^1(t)= R \exp(- R t)\sin(Y_1 t)/Y_1, \\
P_{\pm \pm}^1(t)&=&\exp(- R t) \left[ \cos\left(Y_1 t \right)
 \mp i \nu {\sin\left( Y_1 t \right) \over Y_1} \right],
\label{eq11}
\ee
$P_{if}^{-1}(t)= \mbox{C.C.}\left[ P_{if}^1\left( t \right) \right]$
where $\mbox{C.C.}$ denotes
complex conjugate,
and $Y_1=\sqrt{\nu^2 - R^2}$.
$P^{\pm 2}_{if} ( t) $ is given by the same
expressions as the corresponding $P^{\pm 1}_{if}(t)$ with $\nu$ replaced
by $2 \nu$.

Now we can evaluate $\la W\ra$ and $\la W^2 \ra$ explicitly. First we consider
$\la W^2\ra$ and, in particular, focus on the case $m=1$,
$0<t_1<t_2<t_3<t_4<T$. By using Table \ref{table2-1}, the contribution of
$\langle W^2 \rangle$ to Eq.~(\ref{eqApzz4}) is
%
\be &&\langle W^2 \rangle_{1234}={\Omega^4 \over 16 } \int_0^T {\rm d}
t_4\int_0^{t_4}{\rm d} t_3\int_0^{t_3} {\rm d} t_2 \int_0^{t_2}{\rm d} t_1
e^{ - i \omega_L ( t_2 - t_1 + t_3 - t_4) - \Gamma(|t_1 -t_2| + |t_3 - t_4|)/2} \no \\
&&\mbox{\makebox[0.8in]{ }}\times {1 \over 2} \sum_{i,j,k,l} P_{ij}^{-1}(t_2 -
t_1) P_{jk}^0(t_3-t_2) P_{kl}^1(t_4 - t_3).
\label{eqApzz4b}
\ee
We use the convolution theorem of Laplace transform four times
and find
%
\be
\langle W^2 \rangle_{1234}
=
{\Omega^4 \over 32} {\cal L}^{-1}\left\{
{1\over s^2} \sum_{i,j,k,l}
\hat{P}_{ij}^{-1}\left( s + \Gamma/2 + i \omega_L\right)
\hat{P}_{jk}^0\left( s \right)
\hat{P}_{kl}^1 \left( s + \Gamma/2 - i \omega_L \right)\right\},  \label{eq2233}
\ee
where ${\cal L}^{-1}$ denotes the inverse Laplace transform,
where the Laplace $T \to s$ transform is defined by
\begin{equation}
\hat{f}(s)  = \int_0^\infty f(T) e^{ - s T} {\rm d} T,
\label{eqbvcb}
\end{equation}
and the Laplace transforms of
the functions $\hat{P}_{ij}^{a}(s)$
are listed in Eqs.~(\ref{eqLTW1})-(\ref{eqLTW}).
Considering the other 23 time orderings we find
\be &&\langle W^2 \rangle = { \Omega^4 \over 16}  {{\cal L}}^{-1} \left\{
{1\over s^2}
\sum_{i,j,k,l}
\left[ \hat{P}_{ij}^{-1} \left( s + \Gamma/2 + i \omega_L\right)
\hat{P}_{jk}^0\left( s \right) \hat{P}_{kl}^{+1} \left( s + \Gamma/2 - i \omega_L\right)
\right.  \right.
\nonumber  \\
&&\mbox{\makebox[0.3in]{ }} \left. \left.  + \hat{P}_{ij}^{-1} \left( s +
\Gamma/2 + i \omega_L\right) \hat{P}_{jk}^0 \left( s + \Gamma \right)
\hat{P}_{kl}^{+1} \left( s + \Gamma/2 - i \omega_L\right) \right. \right.
\nonumber  \\
&& \mbox{\makebox[0.3in]{ }}
 \left. \left.
 + \hat{P}_{ij}^{+1} \left( s + \Gamma/2 - i \omega_L\right) \hat{P}_{jk}^0\left( s \right) \hat{P}_{kl}^{+1} \left( s + \Gamma/2 - i \omega_L\right)
\right. \right.
\nonumber \\
&& \mbox{\makebox[0.3in]{ }}
 \left. \left. +\hat{P}_{ij}^{+1} \left( s +
\Gamma/2 - i \omega_L\right) \hat{P}_{jk}^0\left( s + \Gamma \right)
\hat{P}_{kl}^{+1} \left(  s + \Gamma/2 - i \omega_L\right) \right. \right.
\nonumber \\
&& \mbox{\makebox[0.3in]{ }}
 \left. \left. + 2 \hat{P}_{ij}^{+1} \left( s +
\Gamma/2 - i \omega_L\right) \hat{P}_{jk}^{+2} \left( s + \Gamma - 2 i \omega_L
\right)
\hat{P}_{kl}^{+1} \left( s + \Gamma/2 - i \omega_L\right)
 + \rm{C.C.} \right]
\right\}. \no \\
\label{eqmain} \ee
%
Eq.~(\ref{eqmain}), which can be used to describe the lineshape fluctuations,
is our main result so far. In Appendix A we invert this equation
from the Laplace $s$ domain to the time $T$ domain using straightforward
complex analysis. Our goal is to investigate Mandel's $Q$ parameter, Eq.
(\ref{eqQsol}); it is calculated using Eqs.~(\ref{eqmain}) and
\begin{equation}
\langle  W \rangle = {\Omega^2 \over 8} {\cal L}^{-1}
\left\{
{1 \over s^2} \sum_{i,j} \hat{P}_{ij}^{-}\left( s + \Gamma/2 + i \omega_L \right) + \rm{C.C.}
\right\},
\label{eqmain1}
\end{equation}
which is also evaluated in Appendix A. As mentioned,
Eq.~(\ref{eqmain1}) is the celebrated Wiener--Khintchine formula for the
lineshape (in the limit of $T \to \infty$) while Eq.~(\ref{eqmain}) describes
the fluctuations of the lineshape within linear response theory. Note that
$\langle W^2 \rangle$ and $\langle W \rangle$ in Eqs.~(\ref{eqmain}) and
(\ref{eqmain1}) are time--dependent, and these time--dependences are of
interest only when the dynamics of the environment is slow (see more details
below). Exact time--dependent results of $\langle W^2 \rangle$ and $\langle W
\rangle$, and thus $Q$, are obtained in Appendix A. The limit $T\to
\infty$ has, of course, special interest since it is used in standard lineshape
theories, and does not depend on an assumption of whether the frequency
modulations are slow. The exact expression for $Q$ in the limit of $T \to
\infty$ is given in Appendix B, Eqs.~(\ref{QltDen}) and
(\ref{QltNum}). These equations are one of the main results of this paper. It
turns out that $Q$ is not a simple function of the model parameters; however,
as we show below, in certain limits, simple behaviors are found.

\section
{Analysis of Exact Solution} \label{SecQ}
In this section, we investigate the behavior of $Q(\omega_L,T)$
for several physically important cases. In the two state model considered in
Section \ref{SecTSJM}, in addition to two control parameters $\omega_L$ and
$T$ we have three model parameters that depend on the chromophore and the bath
: $\Gamma$, $R$, and $\nu$. Depending on their relative magnitudes we consider
six different limiting cases:
\begin{center}
1.  $R \ll \Gamma \ll \nu$  \\
2.  $R \ll \nu \ll \Gamma$ \\
3. $\Gamma \ll \nu \ll  R$ \\
4. $\nu \ll \Gamma \ll R$ \\
5. $\Gamma \ll  R \ll \nu$ \\
6. $\nu \ll R \ll \Gamma$
\end{center}
We discuss all these limits in this section.

\subsection{Slow Modulation Regime : $R\ll \nu, \Gamma$}   
\label{SecSlow} We first consider the slow modulation regime, $R\ll \nu,
\Gamma$, where the bath fluctuation process is very slow compared with the
radiative decay rate and the frequency fluctuation amplitude. In this case the
foregoing Eqs.~(\ref{eqmain}) and (\ref{eqmain1}) can be simplified. This case
is similar to situations in several SM experiments in condensed
phases\cite{ambrose-jcp-91,ambrose-nature-91,fleury-jlum-93,zumbusch-prl-93,bach-prl-99,boiron-cp-99}.

\begin{figure}
\bc
\epsfig{file=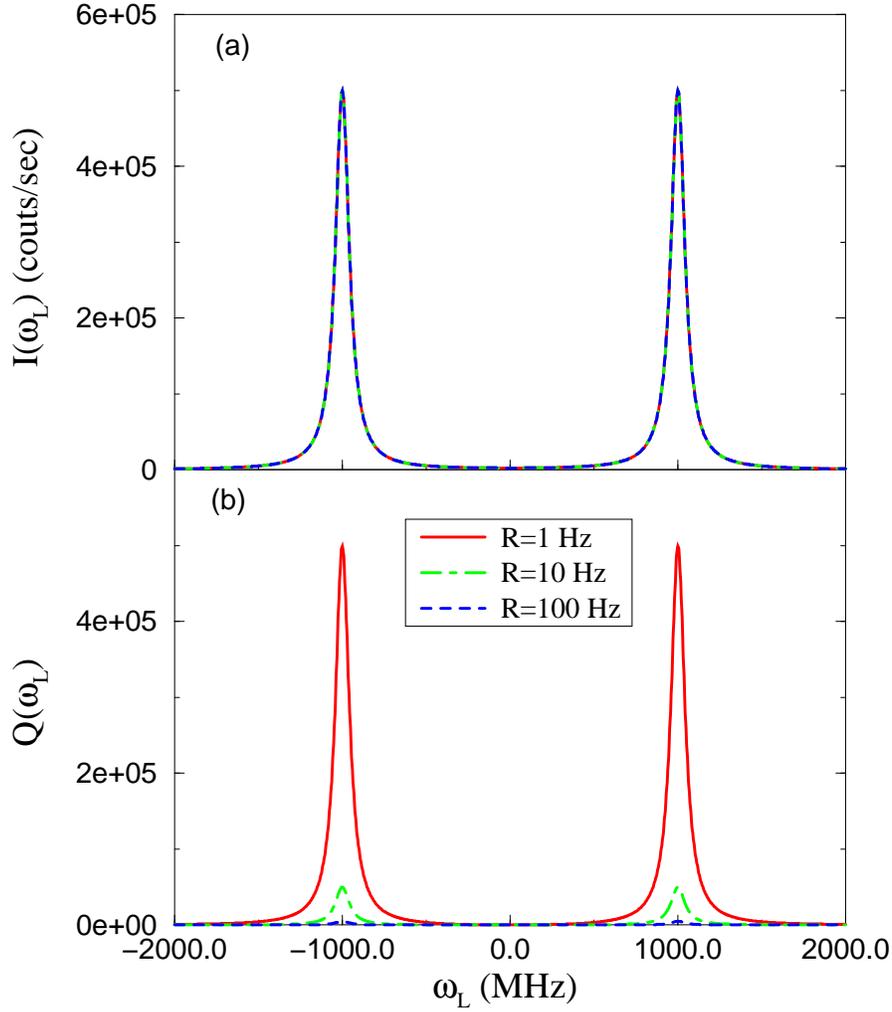,width=\figsize}
\ec \caption[Lineshape and $Q$ in case 1 ($R\ll \Gamma \ll \nu$) in the
steady-state limit]{ Case 1 ($R\ll \Gamma \ll \nu$) in the steady-state limit.
Exact results for lineshape, Eq.~(\ref{Iex}), and for $Q$, Eqs.~(\ref{QltDen}),
(\ref{QltNum}), are shown for the strong, slow modulation case as functions of
$\omega_L$. Parameters are chosen as $\nu=1\mbox{GHz}$, $\Gamma=100\mbox{MHz}$,
$\Omega=\Gamma/10$, $R=1\mbox{Hz}-100\rm{Hz}$, and $T\to\infty$. They mimic a
SM coupled to a single slow two level system in a glass. Note that lineshape
does not change with $R$ while $Q$ does.} \label{strong_slow}
\end{figure}

Within the slow modulation regime, we can have two distinct behaviors of the
lineshape depending on the magnitude of the frequency modulation, $\nu$. When
the frequency modulation is slow but strong such that $R\ll \Gamma\ll \nu$
[case 1], the lineshape exhibits a splitting with the two peaks centered at
$\omega_L=\pm \nu$ (see Fig.~\ref{strong_slow}(a)). On the other hand, when the
frequency modulation is slow and weak, $R\ll \nu \ll \Gamma$ [case 2] a single
peak centered at $\omega_L=0$ appears in the lineshape (see
Fig.~\ref{weak_slow}(a)). From now on we will term the case $\Gamma \ll \nu$ as
the strong modulation limit and the other case $\Gamma \gg \nu$ as the weak
modulation limit. The same distinction can also be applied to the fast
modulation regime considered later.

In the slow modulation regime, we can find $Q(T)$ using random walk theory and
compare it with the exact result obtained in Appendix A. In this
regime, the molecule can be found either in the up $+$ or in the down $-$
state, if the transition times (i.e., typically ${\cal{O}}(1/R)$) between these
two states are long; the rate of photon emission in these two states is
determined by the {\em stationary} solution of time--independent Bloch equation
in the limit of weak laser intensity\cite{cohentann-atom-93} [see also
Eqs.~(\ref{eqA4}) and (\ref{eqA5})]
\begin{equation}
I_{\pm}(\omega_L) =
{\Omega^2 \Gamma \over
4[(\omega_L\mp\nu)^2+{\Gamma^2\over 4}]}.
\label{eqSML04}
\end{equation}
Now the stochastic variable $W=\int_0^T I(t) {\rm d} t$ must be considered,
where $I(t)$ follows a two state process, $I(t)=I_{+}$ or $I(t)=I_{-}$ with
transitions $+ \to -$ and $- \to +$ described by the rate $R$. One can map this
problem onto a simple two state random walk problem\cite{weiss-random-94},
where a particle moves with a ``velocity'' either $I_+$ or $I_-$ and the
``coordinate'' of the particle is $W$. Then from the random walk theory it is
easy to see that for long times ($T\gg 1/R$),
\begin{equation}
\langle W \rangle \simeq \left({I_{+} + I_{-}\over 2}\right)T=\la I\ra T,
\label{eqSML05}
\end{equation}
meaning that the line is composed of two Lorentzians centered
at $\pm \nu$ with a width determined by the lifetime
of the molecule,
\be
\la I\ra = {1\over 2} (I_++I_-),
\label{eqSML00}
\ee
and the ``mean square displacement'' is given by
\begin{equation}
\langle W^2 \rangle - \langle W \rangle^2 \simeq{\left( I_{+} - I_{-} \right)^2\over 4R}{T}.
\label{eqSML06}
\end{equation}
After straightforward algebra Eqs.~(\ref{eqSML05}), (\ref{eqSML06}), and
(\ref{eqQsol}) yield \be Q= {\Omega^2 \Gamma \nu^2 \omega_L^2 \over R
(\omega_L^2+\nu^2+{\Gamma^2\over 4}) ((\omega_L-\nu)^2+{\Gamma^2\over 4})
((\omega_L+\nu)^2+{\Gamma^2\over 4})}, \label{eqSML01} \ee in the limit $T\gg
1/R$. The full time--dependent behaviors of $\la W\ra$ and $\la W^2 \ra$ are
calculated in Appendix C using the two state random walk
model\cite{weiss-random-94}. From Eq.~(\ref{AppE5}), we have $Q$ as a function
of the measurement time in the slow modulation regime \be Q(T) \simeq {\la
(\Delta I)^2 \ra \over \la I\ra R} \left(1+{e^{-2RT}-1\over 2RT} \right),
\label{eqSML09} \ee where the ``variance'' of the lineshape is defined by \be
\la (\Delta I)^2 \ra ={1\over2} [(I_{+}-\la I\ra)^2+(I_{-}-\la I\ra)^2]. \ee
Eq.~(\ref{eqSML09}) shows that $Q$ is factorized as a product of frequency
dependent part and the time--dependent part in the slow modulation regime. It
is important to note that Eq.~(\ref{eqSML09}) could also have been derived from
the exact result presented in Appendix A by considering the slow
modulation conditions. Briefly, from the exact expressions of $\la W\ra$ and
$\la W^2\ra$ in the Laplace domain given in Appendix A, by keeping
only the pole $s$'s that satisfies $|{\rm {Re}}(s)|\sim \cal{O}(R)$ and
neglecting other poles such that $|{\rm {Re}}(s)|\gg R$,
we can recover the result given in Eq.~(\ref{eqSML09}), thus confirming the
validity of the random walk model in the slow modulation regime. In the limits
of short and long times, we have \be Q\simeq\left\{
\begin{array}{cc}
{{\la (\Delta I)^2 \ra \over  \la I\ra }R^{-1}}  & \mbox{\makebox[1in]{$T\gg 1/R$}}
\vspace{.2cm} \\
{{\la (\Delta I)^2 \ra \over  \la I\ra } T}  & \mbox{\makebox[1in]{$T\ll 1/R$}}.       \label{Qs}
\end{array}\right.
\ee
We recover Eq.~(\ref{eqSML01}) from Eq.~(\ref{Qs})
in the limit of $T\to \infty$.
Eq.~(\ref{Qs}) for $T\ll 1/R$ is a special case of Eq.~(\ref{eqQQq9}).
Note that the results in this subsection can be easily generalized to
the case of strong external fields using Eqs.~(\ref{eqA4}) and (\ref{eqA5}).

\subsubsection{strong modulation : case 1}
When the $\Gamma \ll \nu$, case 1, the two intensities $I_{\pm}$ are well
separated by the amplitude of the frequency modulation, $2\nu$, therefore we
can approximate $\la I\ra\simeq I_{+}/2$ when $I_{-}\simeq 0$ and vice versa,
which leads to \be \la (\Delta I)^2 \ra \simeq \la I\ra^2.
\mbox{\makebox[1.in][r]{$\Gamma \ll \nu$}} \ee In this case $Q$ is given by \be
Q\simeq\left\{
\begin{array}{cc}
{\la I\ra R^{-1}}  & \mbox{\makebox[1in]{$T\gg 1/R$}}
\vspace{.2cm} \\
{\la I\ra T}  & \mbox{\makebox[1in]{$T\ll 1/R$}}.       \label{Qss}
\end{array}\right.
\ee

We show the lineshapes and $Q$ as functions of the laser frequency $\omega_L$
at the steady-state limit ($T\to \infty$) for case 1 in Fig.~\ref{strong_slow}.
In all the calculations shown in figures of the present work, we have assumed
an ideal measurement, $\eta=1$. Values of parameters are given in the figure
caption. This is relevant for the case that a chromophore is strongly coupled
to a single two level system in a low temperature glass\cite{zumbusch-prl-93,fleury-jlum-93}. Since $Q\simeq\la I\ra/R$ in this case,
both the lineshape (Fig.~\ref{strong_slow}(a)) and $Q$
(Fig.~\ref{strong_slow}(b)) are similar to each other; two Lorentzian peaks
located at $\omega_L=\pm\nu$ with widths $\Gamma$. Note that in this limit the
value of $Q$ is very large compared with that in the fast modulation regime
considered later. While the lineshape is independent of $R$ in this limit, the
magnitude of $Q$ decreases with $R$, hence it is $Q$ not $\la I\ra$ which
yields information on the dynamics of the environment.

\subsubsection{weak modulation : case 2}
On the other hand, in case 2, where the fluctuation is very weak ($\nu\ll
\Gamma$), we notice that $I_{+} \simeq I_{-}$ from Eq.~(\ref{eqSML04}) (note
that when there is no spectral diffusion, $\nu=0$, $I_{+}=I_{-}$, thus $Q=0$).
The lineshape is given by a single Lorentzian centered at $\omega_L=0$,
\be \la I\ra \simeq {\Omega^2 \Gamma / 4 \over \omega_L^2+\Gamma^2/4}.
\label{Ilor}
\ee
Since $\nu \ll \Gamma$ in this case we expand $I_{\pm}$ in
terms of $\nu$ to find
\be \la (\Delta I)^2\ra \simeq {\nu^2}\left({d\la I\ra\over d\omega_L}\right)^2
={\nu^2\Omega^4\Gamma^2\omega_L^2\over 4 (\omega_L^2+\Gamma^2/4)^4} .
\mbox{\makebox[1in]{$\nu \ll \Gamma$}} \label{Iweak} \ee
Therefore, in case 2, $Q$ is given by \be Q\simeq\left\{
\begin{array}{cc}
{{\nu^2\over \la I\ra} {\left (d\la I\ra\over d\omega_L\right)}^2{R^{-1}}} & \mbox{\makebox[1in]{$T\gg 1/R$}}
\vspace{.2cm} \\
{{\nu^2\over \la I\ra}  {\left (d\la I\ra\over d\omega_L\right)}^2 T} &
\mbox{\makebox[1in]{$T\ll 1/R$}}.
\end{array}\right. \label{Qsw}
\ee
Note that
$Q\propto {1\over \la I\ra}({d\la I\ra\over d\omega_L})^2$
in the weak, slow modulation case while
$Q\propto \la I\ra$ in the strong, slow modulation case.

\begin{figure}
\bc
\epsfig{file=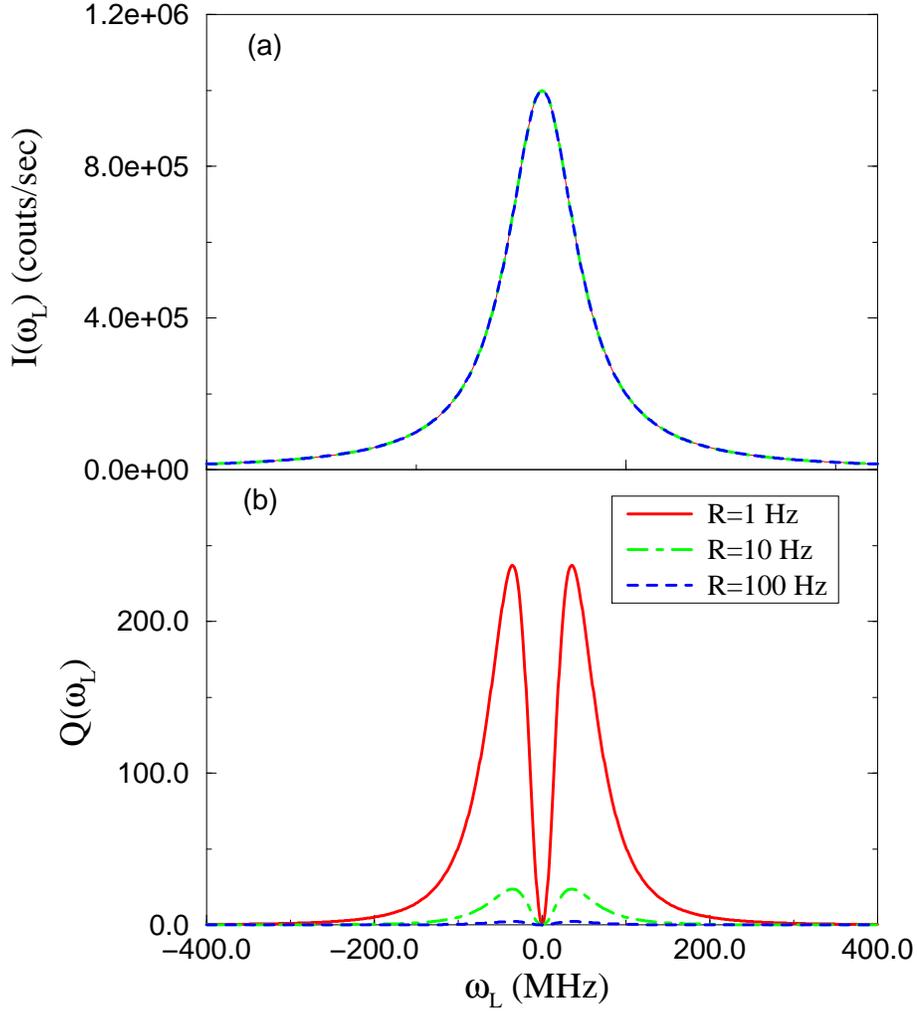,width=\figsize}
\ec \caption[Lineshape and $Q$ in case 2 ($R\ll \Gamma \ll \nu$) in the
steady-state limit]{ Case 2 ($R\ll \nu \ll \Gamma$) in the steady-state limit.
Exact results for lineshape, Eq.~(\ref{Iex}), and for $Q$, Eqs.~(\ref{QltDen}),
(\ref{QltNum}), are shown for the weak, slow modulation case as functions of
$\omega_L$. Parameters are chosen as $\nu=1\mbox{MHz}$, $\Gamma=100\mbox{MHz}$,
$\Omega=\Gamma/10$, $R=1\rm{Hz}-100\rm{Hz}$, and $T\to\infty$. }
\label{weak_slow}
\end{figure}

In Fig.~\ref{weak_slow} we show the lineshape and $Q$ for the weak, slow
modulation limit, case 2. The lineshape (Fig.~\ref{weak_slow} (a)) is a
Lorentzian with a width $\Gamma$, to a good approximation, thus the features of
the lineshape do not depend on the properties of the coupling of the SM to a
environment such as $\nu$ and $R$. On the other hand, $Q$ (Fig.~\ref{weak_slow}
(b)), shows a richer behavior. Recalling Eq.~(\ref{Qsw}) for $T\gg 1/R$ in
Eq.~(\ref{Iweak}), it exhibits doublet peaks separated by $\sim
{\cal{O}}(\Gamma)$ with the dip located at $\omega_L=0$, and its magnitude is
proportional to $1/R$. We will later show that this kind of a doublet and a dip
in $Q$ is a generic feature of the weak coupling case, found not only in the
slow but also in the fast modulation case considered in the next section. Note
that when the SM is not coupled to the environment ($\nu=0$), $Q=0$ as
expected.

Both in the strong (Fig.~\ref{strong_slow})
and the weak (Fig.~\ref{weak_slow}) cases,
we find $Q(\omega_L=0)\simeq 0$.
This is expected in the slow modulation case considered here.
Physically,
when the laser detuning frequency is exactly in the
middle of two frequency shifts, $\pm \nu$,
the rate of photon emissions is identical
whether the molecule is in the up state ($+\nu$)
or in the down state ($-\nu$).
Therefore, the effect of bath fluctuation on
the photon counting statistics is negligible,
which leads to Poissonian counting statistics
at $\omega_L=0$.

\subsubsection{time--dependence}

\begin{figure}
\bc
\epsfig{file=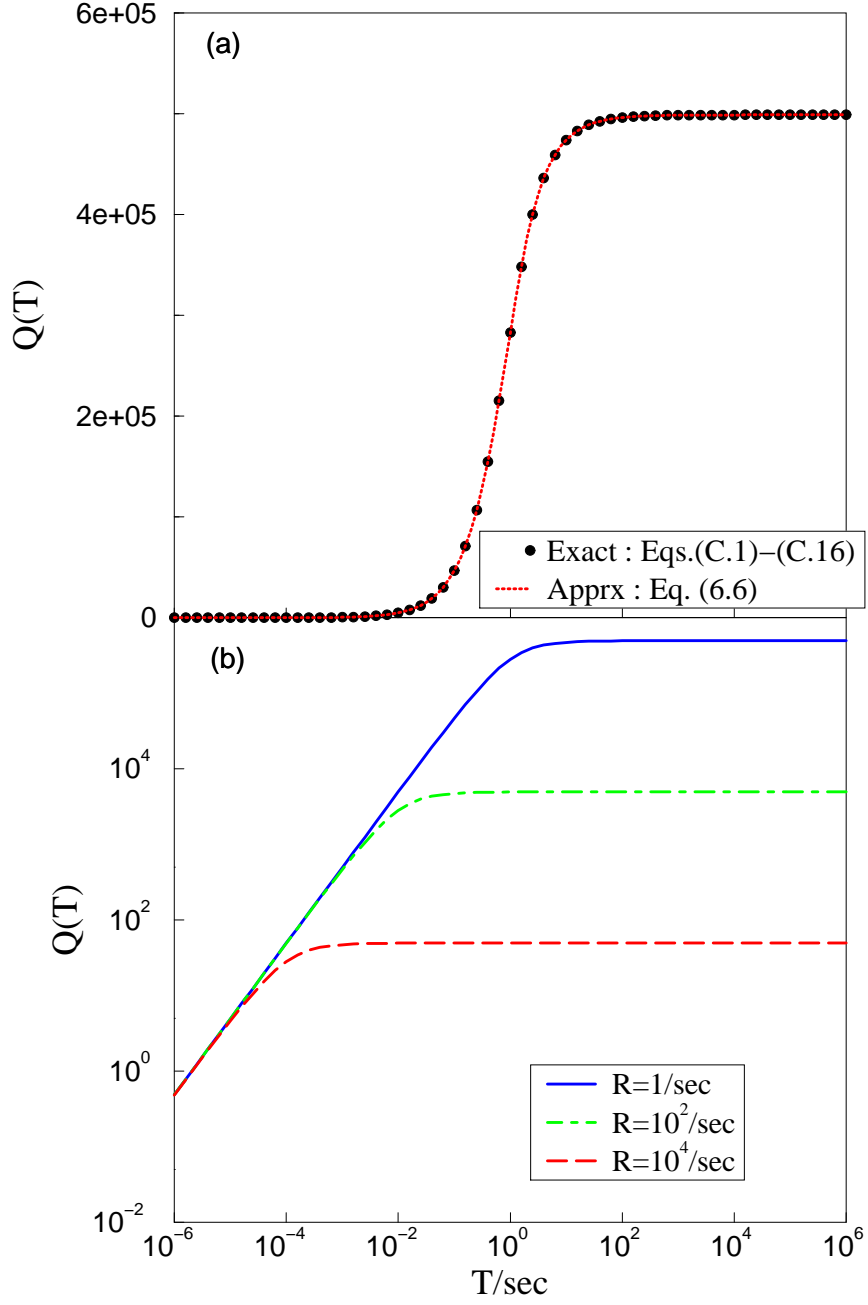,width=\figsize}
\ec \caption[$Q$ in case 1 ($R \ll \Gamma \ll \nu$) in the time--dependent
regime]{ Case 1 ($R \ll \Gamma \ll \nu$) in the time--dependent regime. (a)
$Q(T)$ is shown both for the exact result and for the approximate result given
in the linear-log scale. This figure demonstrates that our exact results given
in Appendix A are well reproduced by the approximation,
Eq.~(\ref{eqSML09}). The same parameters are used as in Fig.~\ref{strong_slow}
except $\omega_L=\nu=1\mbox{GHz}$, $R=1\mbox{Hz}$, and $T$ (varied). (b) $Q(T)$
for different values of $R$ in the slow modulation regime in the log-log scale.
All the parameters except for $R$ are chosen the same as those in (a). }
\label{slowT}
\end{figure}

Additional information on the environmental fluctuations can be gained by
measuring the time dependence of $Q$ in the slow modulation regime. In
Fig.~\ref{slowT} we show $Q$ versus the measurement time $T$ for the strong,
slow modulation limit, case 1, both for the exact result calculated in Appendix
A and for the approximate result, Eq.~(\ref{eqSML09}). We choose the
resonance condition $\omega_L=\nu$, and used parameters relevant to SMS in
glass systems. The approximate result based on the two state random walk model,
Eq.~(\ref{eqSML09}), shows a perfect agreement with the exact result in
Fig.~\ref{slowT}(a) as expected. When $T\ll 1/R$, $Q$ increases linearly with
$T$ as predicted from Eq.~(\ref{Qs}) for $RT \ll 1$, and it reaches the
steady-state value given by Eq.~(\ref{Qs}) when $T$ becomes $RT\gg 1$. We also
notice that even in the long measurement time limit the value of $Q$ is large:
$Q=5\times10^5\eta$ in the example given in Fig.~\ref{slowT}(a) (including the
detection efficiency). Therefore, even if we consider the imperfect detection
of the photon counting device (for example, $\eta=15\rm{\%}$ has been reported
recently\cite{ambrose-cpl-97}), deviation of the photon statistics from
Poissonian is observable in the slow modulation regime of SMS. This is
seemingly contrary to propositions made in the literature
\cite{mandel-optical-95,chen-bpj-99,schenzle-pra-86}, in which the claim is
made that the Poissonian distribution is achieved in the long time limit. We
defer discussion of this issue to the end of this section. We note that it has
been shown that $Q(T)$ can be very large ($Q\sim 10^4$) at long times in the
atomic three-level system with a metastable state but without a spectral
diffusion process\cite{kim-pra-87}. Fig.~\ref{slowT}(b) illustrates that the
steady-state value of $Q(T)$ is reached when $RT\gg 1$, and the magnitude of
$Q$ in the steady-state decreases as $1/R$ as predicted from Eq.~(\ref{Qs}) for
$RT\gg1$. This therefore  illustrates that valuable information on the bath
fluctuation timescales can be obtained by measuring the fluctuation of the
lineshape as a function of measurement time.

\begin{figure}
\bc
\epsfig{file=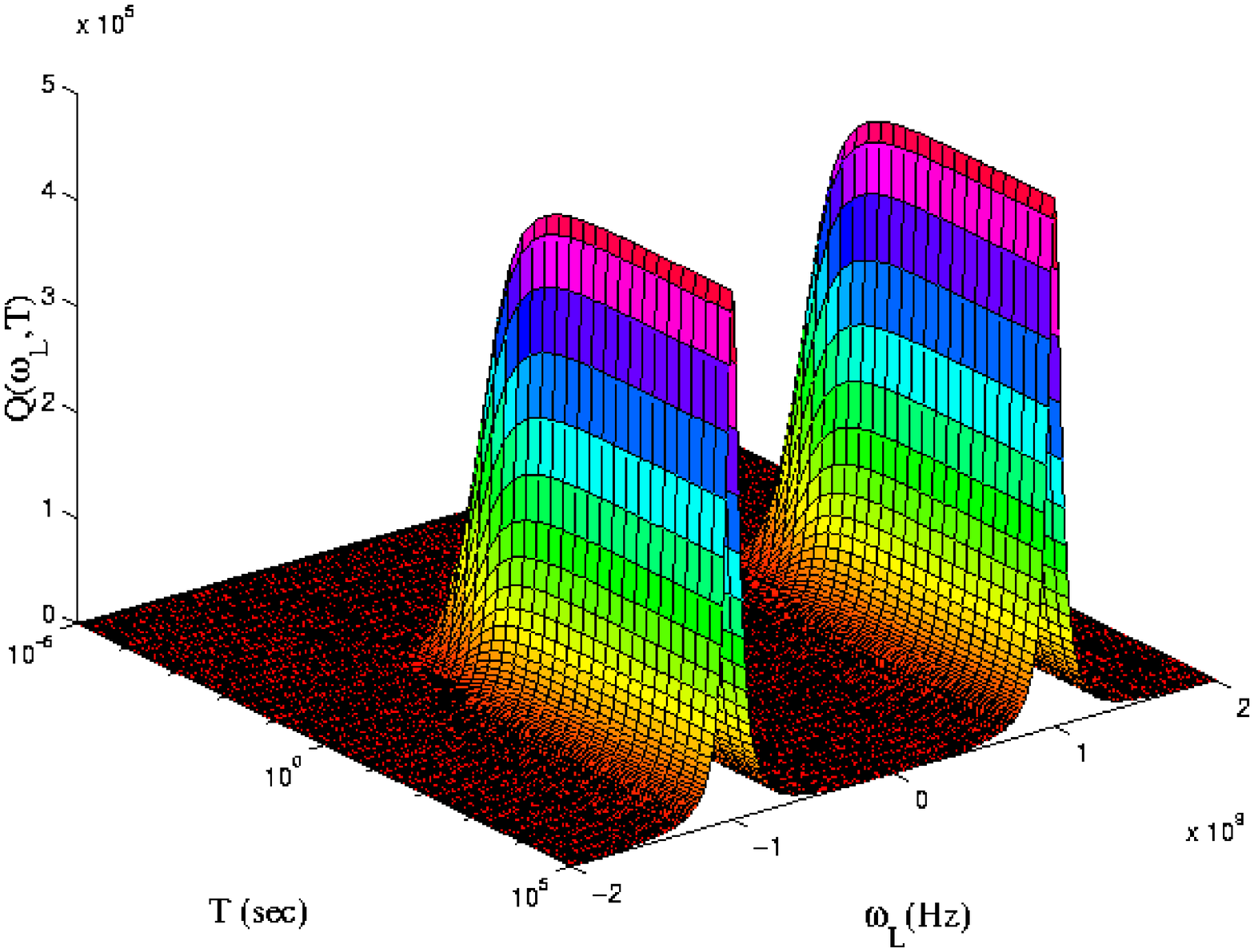,width=\figsize}
\ec \caption[Two dimensional plot of $Q(\omega_L,T)$ in the time dependent
case]{ Case 1 ($R \ll \Gamma \ll \nu$) in the time dependent case. Two
dimensional plot of $Q(\omega_L,T)$ for the slow modulation as a function of
$\omega_L$ and $T$. The same parameters are used as in Fig.~\ref{slowT}(a)
except $\omega_L$ (varied).
}
\label{slow2d}
\end{figure}

Fig.~\ref{slow2d} shows a two--dimensional plot of $Q$ as a function of the
laser frequency and the measurement time for the parameters chosen in
Fig.~\ref{slowT}(a). We observe that two Lorentzian peaks located at resonance
frequencies become noticeable when $T\simeq 1/R$. Similar time dependent
behavior can be also found for the weak, slow modulation limit, case 2,
however,  then $Q$ along the $\omega_L$ axis shows doublet peaks separated by
$\sim {\cal{O}}(\Gamma)$ as shown in Fig.~\ref{weak_slow}.

We can extend our result and describe
photon statistics beyond the second moment. Based on the
central limit theorem, the probability density function
of the two state random walk variable $W$ in the long time limit is described by
\begin{equation}
P(W,T) \stackrel{RT\gg 1} \longrightarrow G(W,T) \equiv{1 \over \sqrt{ 4 \pi D T } }
\exp\left[ -{\left(W - V T\right)^2 \over 4 D T }\right]
\label{eqSML07}
\end{equation}
with $D=\la (\Delta I)^2 \ra/(2 R)$ and $V=\la I\ra$ when $\eta =1$. We also
note that $Q\simeq 2D/V$ in the long time limit. By using Eq.~(\ref{eqrrp})
\begin{equation}
\langle p(n,T) \rangle \simeq \int_{0}^{\infty} {\rm d} W  G(W,T){W^n \over n!}
\exp(-W).
\label{eqSML08}
\end{equation}
Eq.~(\ref{eqSML08}) shows that in the long time limit the photon statistics is
the Poisson transform\cite{saleh-photoelec-78} of a Gaussian. This
transformation can be found explicitly (one can slightly improve this
approximation by considering a normalized truncated Gaussian with $G(W,T) =0$
for $W<0$; we expect that our approximation will work well when $V^2 T \gg D$).
In contrast to Eq.~(\ref{eqSML07}), the proposed Eq.~(\ref{eqrrp1}) suggests
that $G(W,T)$ be replaced with a delta function, for which a single parameter
$V$ controls the photon statistics, while according to our approach both $V$
and $D$, or equivalently, $\la W\ra$ and $Q$ are important. Mathematically,
when $T\to \infty$ the Gaussian distribution, Eq.~(\ref{eqSML07}) may be said
to ``converge'' to the delta function distribution, Eq.~(\ref{eqrrp1}), in a
sense that the mean is $\la W\ra\sim {\cal{O}}(T)$ while the standard deviation
is $\Delta W =(\la W^2\ra-\la W\ra^2)^{1/2}\sim {\cal{O}}(T^{1/2})$. This
argument corresponds to the proposition made in the literature,
Eq.~(\ref{eqrrp1}), and then the photon statistics is only determined by the
mean $\la W\ra\simeq VT$. However, physically it is more informative and
meaningful to consider not only the mean $\la W\ra$ but also the variance
$(\Delta W)^2\simeq 2DT \simeq Q\la W\ra$ since it contains information on the
bath fluctuation. In a sense the delta function approximation might be
misleading since it implies that $Q=0$ in the long time limit, which is clearly
{\it not} true in general.

\subsection{Fast Modulation Regime : $\nu, \Gamma \ll R$} 
\label{SecFast}

In this section we consider the fast modulation regime, $\nu, \Gamma \ll R$.
Usually in this fast modulation regime, the dynamics of the bath (here modeled
with $R$) is so fast that only the long time limit of our solution should be
considered [i.e. Eqs.~(\ref{QltDen}) and (\ref{QltNum})]. Hence all our results
below are derived in the limit of $T \to \infty$, since the time dependence of
$Q$ is irrelevant. The fast modulation regime considered in this section
includes case 3 ($\Gamma \ll \nu \ll R$) as the strong, fast modulation case
and case 4 ($\nu \ll \Gamma \ll R$) as the weak, fast modulation case.

It is well known that the lineshape is Lorentzian\cite{mukamel-nonlinear-95} in
the fast modulation regime (soon to be defined precisely),
\begin{equation}
\langle I\left( \omega_L \right) \rangle=
{{\Omega}^2 (\Gamma + \Gamma_{f}) \over
4\omega_L^2 + {(\Gamma +   \Gamma_{f})^2}},
\label{eqFML00}
\end{equation}
where $\Gamma_{f}$ is the line width due to the dephasing induced by the bath
{\it fluctuations} and given by \be \Gamma_{f}= {\nu^2\over R}. \ee The
lineshape given in Eq.~(\ref{eqFML00}) exhibits motional narrowing, namely as
$R$ is increased the line becomes narrower and in the limit $R \to \infty$ the
width of the line is simply $\Gamma$.

Now we define the fast modulation regime considered in this work. When $R \to
\infty$ (and other parameters fixed) $Q\to 0$, so fluctuations become
Poissonian. This is expected and in a sense trivial because the molecule cannot
respond to the very fast bath, so the two limits $R \to \infty$ and $\nu = 0$
(i.e., no interaction with the bath) are equivalent. It is physically more
interesting to consider the case that $R\to \infty$ with $\Gamma_f/\Gamma$
remaining finite, which is the standard definition of the fast modulation
regime\cite{kubo-fluct-62}. In this fast modulation regime the well known
lineshape is Lorentzian as given in Eq.~(\ref{eqFML00}) with a width
$\Gamma+\Gamma_f$.

Here we present a simple calculation of $Q$ in the fast modulation regime based
on physically motivated approximations. We justify our approximations by
comparing the resulting expression with our exact result, Eqs.~(\ref{QltDen})
and (\ref{QltNum}). When the dynamics of the bath is very fast, the correlation
between the state of the molecule during one pulse interval with that of the
following pulse interval is not significant in the stochastic averaging
Eq.~(\ref{eqC3m}). Therefore in the fast modulation regime, we can
approximately factorize the three--time correlation function
$C_3^{m}(\tau_1,\tau_2,\tau_3)$ in Eq.~(\ref{eqC3m}) in terms of one--time
correlation functions $C_1^{l}(\tau)$ as was done by Mukamel and Loring in the
context of four wave mixing spectroscopy\cite{mukamel-josab-86}, \be
&&C_3^{m}(\tau_1,\tau_2,\tau_3) \no \\
&=&\left\la e^{
       -i a\int_{t_I}^{t_{II}}\Delta\omega(t'){\rm d}t'
      -i b\int_{t_{II}}^{t_{III}}\Delta\omega(t'){\rm d}t'
      -i c\int_{t_{III}}^{t_{IV}}\Delta\omega(t'){\rm d}t'}
\right\ra \no \\
&\simeq&
  \left\la e^{-i a \int_{t_I}^{t_{II}}\Delta\omega(t'){\rm d}t'}\right\ra
  \left\la e^{-i b \int_{t_{II}}^{t_{III}}\Delta\omega(t'){\rm d}t'}\right\ra
\left\la e^{-i c \int_{t_{III}}^{t_{IV}}\Delta\omega(t'){\rm d}t'}\right\ra \no \\
&=&C_1^{a}(\tau_1)C_1^{b}(\tau_2)C_1^{c}(\tau_3), \label{factor} \ee where $a$,
$b$, and $c$ are values of $S_m(t)$ in time intervals, $t_{I}<t<t_{II}$,
$t_{II}<t<t_{III}$, and $t_{III}<t<t_{IV}$, respectively.  For example, $a=-1$,
$b=0$, and $c=1$ for the case $m=1$ (see Table \ref{table2-1}). The one--time
correlation function is evaluated using the second order cumulant
approximation, which is also valid for the fast modulation
regime\cite{mukamel-josab-86},
\be
C_1^{l}(t)\simeq\exp\left(-l^2 \int_0^t {\rm d}t'\int_0^{t'}{\rm d}t''
              \la \Delta \omega(t')\Delta \omega(t'')\ra\right). \label{C1fast}
\ee
Note that these two approximations, Eqs.~(\ref{factor}) and (\ref{C1fast}), are
not limited to the two state jump model but are generally valid in the fast
modulation regime. The frequency correlation function is given by \be \la
\Delta \omega(t')\Delta \omega(t'')\ra= \nu^2 \exp(-2R |t'-t''|)
\label{freqcorr} \ee for the two state jump model. Substituting
Eq.~(\ref{freqcorr}) into Eq.~(\ref{C1fast}) we have \be
C_1^{l}(t)\simeq\exp\left[-l^2\nu^2\left({t\over 2R}+{e^{-2Rt}-1\over
(2R)^2}\right)\right]. \ee In the limit of $R\to \infty$ we can approximate
$C_1^{l}(t)$ further \be C_1^{l}(t)\simeq \exp(-l^2 \Gamma_f t/2),
\label{cumul} \ee and in turn evaluate all the three--time correlation
functions corresponding to different time orderings within these approximations
using Eqs.~(\ref{factor}) and (\ref{cumul}). We note that the lineshape shows a
Lorentzian behavior with the width $\Gamma+\Gamma_f$ in Eq.~(\ref{eqFML00})
because of the exponential decay of the one--time correlation function in
Eq.~(\ref{cumul}). Since we have simple analytical expressions of the
three--time correlation functions, it is straightforward to calculate $Q$ in
the fast modulation regime. After straightforward algebra described in Appendix
D, $Q$ in the fast modulation regime is
\be Q= {{\Omega^2 \Gamma_{f}}\left[ 8 \Gamma_{f} \omega_L^4 + 2 (\Gamma +
\Gamma_{f})(8\Gamma^2 + 15\Gamma \Gamma_f + 5 \Gamma_f^2) \omega_L^2 + \Gamma_f
(\Gamma + 2\Gamma_f)(\Gamma + \Gamma_f)^3\right] \over {16 \Gamma (\Gamma +
\Gamma_f) \left(\omega_L^2 + {(\Gamma + \Gamma_f)^2\over 4}\right)^2
\left(\omega_L^2 + {(\Gamma + 2\Gamma_f)^2\over 4}\right)}}. \no \\
\label{eqFML01}
\ee
Note that the exact expression for $Q$ obtained in Appendix
B yields the same expression as Eq.~(\ref{eqFML01}) when $R \to
\infty$ but $\Gamma_{f}$ remains finite, which justifies the approximations
introduced.

Let us now estimate the magnitude of these fluctuation. We consider
$\omega_L=0$ since the photon current is strongest for this case (i.e., the
lineshape has a maximum at $\omega_L=0$), then
\begin{equation}
 Q=\frac{ 4\, \eta \,\Omega^2 \,\Gamma_{f}^2}{\Gamma\,{\left( \Gamma + \Gamma_{f} \right) }^2\,\left( \Gamma + 2\,\Gamma_{f} \right) },
\label{eqFML02}
\end{equation}
where the detection efficiency $\eta$ has been restored. The maximum of $Q$ is
found when $\Gamma_{f}= \Gamma(1 + \sqrt{5})/2$ and then $Q\simeq 0.361
\eta\Omega^2/\Gamma^2$. Even if we take $|\Omega|/\Gamma=1/10$ and
$\eta=5\times 10^{-2}$ as reasonable estimates for a weak laser field and
detection efficiency we find $Q \simeq 2\times 10^{-4}\ll 1$ which shows the
difficulties of measuring deviations from Poissonian statistics in this limit.
We note, however, that values of $Q$ as small as $|Q|\le 10^{-4}$ have been
measured recently (although in the short time regime, not in the steady-state
case)\cite{fleury-prl-00}. Therefore it might also be possible to observe the
deviation from Poissonian photon statistics in the fast modulation limit under
appropriate experimental situations.

\begin{figure}
\bc
\epsfig{file=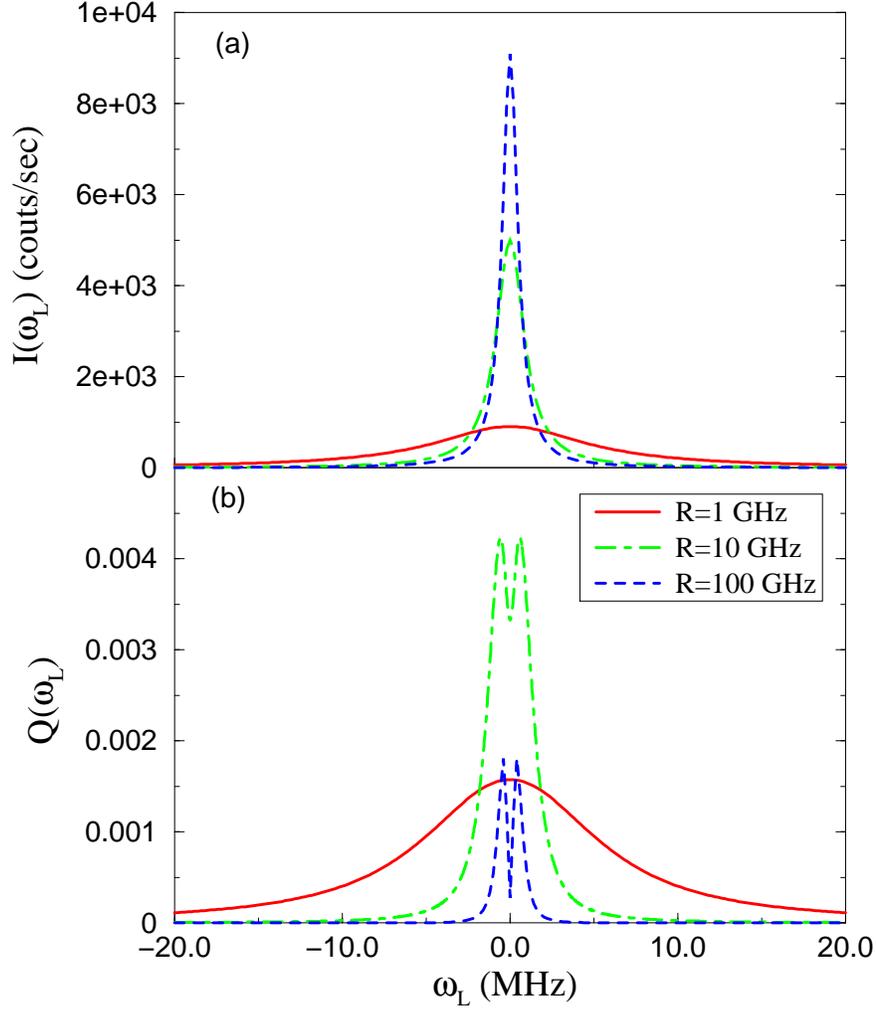,width=\figsize}
\ec \caption[Lineshape and $Q$ in case 3 ($\Gamma \ll \nu \ll R$) in the
steady-state limit] { Case 3 ($\Gamma \ll \nu \ll R$) in the steady-state
limit. Lineshape and $Q$ in the fast modulation regime are shown as functions
of $\omega_L$. Parameters are chosen as $\nu=100\rm{MHz}$, $\Gamma=1\rm{MHz}$,
$\Omega=\Gamma/10$, $R=1\rm{GHz}-100\rm{GHz}$, and $T\to\infty$. } \label{fast}
\end{figure}

In Fig.~\ref{fast} we show the results of exact steady-state calculations for
the lineshape and $Q$ [Eq.~(\ref{QltDen}) and (\ref{QltNum})] for different
values of the fluctuation rate $R$ in the fast modulation regime. We have
chosen the parameters as $\nu=100\rm{MHz}$, $\Gamma=1\rm{MHz}$, and
$\Omega=\Gamma/10$, and $R$ is varied from $1\rm{GHz}$ to $100\rm{GHz}$,
corresponding to case 3. For this parameter set we have checked that the fast
modulation approximation for $Q$ given in Eq.~(\ref{eqFML01}) agrees well with
the exact calculation, Eqs.~(\ref{QltDen}), (\ref{QltNum}).

The lineshape shown in Fig.~\ref{fast}(a) shows the well known motional
narrowing behavior. When $R=1\rm{GHz}$, the lineshape is a broad Lorentzian
with the width $\Gamma+\Gamma_f\simeq\Gamma_f$ (note that $\Gamma_f\gg \Gamma$
in this case). As $R$ is increased further the line becomes narrowed and
finally its width is given by $\Gamma$ as $R\to\infty$.

Compared with the lineshape, $Q$ in Fig.~\ref{fast}(b) shows richer behavior.
The most obvious feature is that when $R\to \infty$, $Q\to 0$. This is expected
since when the bath is very fast the molecule cannot respond to it, hence
fluctuations are Poissonian and $Q\to 0$. It is noticeable that, unlike
$I(\omega_L)$, $Q$ shows a type of narrowing behavior {\it with splitting} as
$R$ is increased. The lineshape remains Lorentzian regardless of $\Gamma_f$ in
the fast modulation case, while $Q$ changes from a broad Lorentzian line with
the width $\Gamma_f$ when $\Gamma_f\gg \Gamma$ to doublet peaks separated
approximately by $\Gamma$ when $\Gamma_f\ll \Gamma$, that will be analyzed in
the following [see, for example, Eq.~(\ref{Qfs})]. Therefore, although the
value of $Q$ is small in the fast modulation regime, it yields additional
information on the relative contributions of $\Gamma$ and $\Gamma_f$ which are
not differentiated in the lineshape measurement.

\begin{figure}
\bc
\epsfig{file=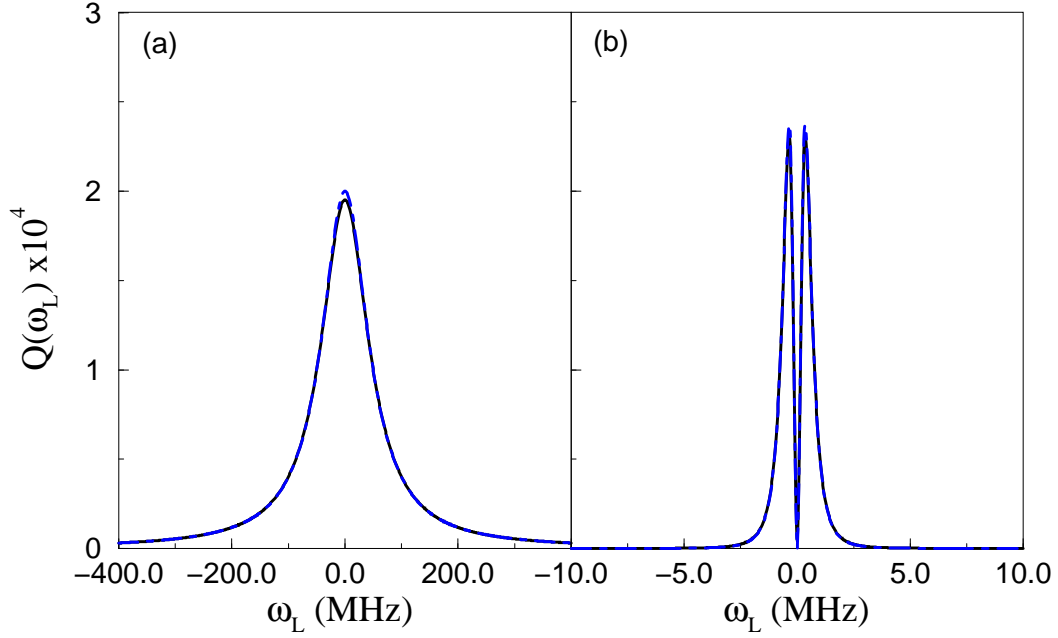,width=\figsize}
\ec \caption [Lorentzian and splitting behaviors of $Q$ in case 3 ($\Gamma \ll
\nu \ll R$) in the steady-state limit]{Case 3 ($\Gamma \ll \nu \ll R$) in the
steady-state limit. Exact calculations (solid line) of $Q$, Eqs.~(\ref{QltDen})
and (\ref{QltNum}) are compared with the approximate expressions,
Eq.~(\ref{Qfs}), (dashed line) for (a) the Lorentzian ($\Gamma_f\gg\Gamma$),
and (b) the splitting ($\Gamma_f\ll\Gamma$) cases in the fast modulation case.
Parameters are chosen as $\nu=1\rm{GHz}$, $\Gamma=1\rm{MHz}$,
$\Omega=\Gamma/10$, $T\to\infty$, and $R=10\rm{GHz}$ in (a) while
$R=10^5\rm{GHz}$ in (b). } \label{strong_fast}
\end{figure}

\subsubsection{strong modulation : case 3}
We further analyze the case that the bath fluctuation is both strong and fast,
case 3, $\Gamma \ll \nu \ll R$. The results shown in Fig.~\ref{fast} correspond
to this case. We find that Eq.~(\ref{eqFML01}) is further simplified in two
different limits, $\Gamma_f \gg \Gamma$ and $\Gamma_f \ll \Gamma$, \be
Q\simeq\left\{
\begin{array}{cc}
\frac{{\Omega}^2 \, \Gamma_{f} }
{ 2\Gamma \left(\omega_L^2 + \Gamma_{f}^2/4 \right) }
& \mbox{\makebox[1.in]{$\Gamma_f \gg \Gamma$}}
\vspace{.2cm} \\
\frac{{\Omega}^2\,\Gamma_{f} \,\Gamma \,\omega_L^2 }{\,\left( \omega_L^2+ \Gamma^2/4\right)^3 }
& \mbox{\makebox[1.in]{$\Gamma_f \ll \Gamma$}}
\end{array}\right.
\label{Qfs} \ee When $\Gamma_f \gg \Gamma$, both $Q$ and the lineshape are a
Lorentzian located at $\omega_L=0$ with a width $\Gamma_{f}$, which yields the
relation $Q\simeq 2\la I(\omega_L)\ra/\Gamma$, and both exhibit motional
narrowing behaviors. In the other limit, $\Gamma_f \ll \Gamma$, we have
neglected ${\cal{O}}((\Gamma_f/\Gamma)^2)$ terms with additional conditions for
$\omega_L^2$, ${\Gamma _f\over \Gamma} \ll
{\left({\omega_L\over\Gamma}\right)}^2 \ll {\Gamma\over \Gamma_f}$. In this
case $Q$ shows a splitting behavior at $|\omega_L|\sim  {\cal{O}}(\Gamma)$. We
note that Eq.~(\ref{Qfs}) for $\Gamma_f\ll\Gamma$ is the same as
Eq.~(\ref{Qsw})
This is the case because the very fast frequency modulation corresponds to the
weak modulation case, if we recall that the dephasing rate due to the bath
fluctuation is given by $\Gamma_f=\nu^2/R$ in the fast modulation regime.

In Fig.~\ref{strong_fast} we have checked the validity of the limiting
expressions of $Q$ for the Lorentzian and the splitting cases
(Eq.~(\ref{Qfs}) by comparing to the
exact results, Eqs.~(\ref{QltDen}), (\ref{QltNum}).
In Fig.~\ref{strong_fast}(a), the parameters are chosen
such that $\Gamma_f=100\Gamma$ while in Fig.~\ref{strong_fast}(b)
$\Gamma_f=\Gamma/100$.
Approximate expressions (dashed line)
show a good agreement with
exact expressions (solid line) in each case.

\subsubsection{weak modulation : case 4}
Now we consider the weak, fast modulation case, case 4 ($\nu \ll \Gamma \ll
R$). In this limit, the lineshape is simply a single Lorentzian peak given by
Eq.~(\ref{eqFML00}) with $\Gamma+\Gamma_f\simeq \Gamma$. We obtain the
following limiting expression for $Q$ from Eq.~(\ref{eqFML01}), noting
$\Gamma_f=\nu^2/R$, \be Q\simeq \frac{\nu^2 \, {\Omega}^2\,\Gamma \,\omega_L^2
}{R \left( \omega_L^2+ \Gamma^2/4\right)^3 }.  \label{Qfw} \ee Also here $Q$
shows splitting behavior. It is given by the same expression as that in the
strong, fast modulation case with $\Gamma_f \ll \Gamma$ (see Eq.~(\ref{Qfs}))
and also as that in the weak, slow  modulation case with $T\gg 1/R$ (see
Eq.~(\ref{Qsw})) (see also Table \ref{table2-2} in Section \ref{SecPhase}).
When $Q$ is plotted as a function of $\omega_L$, it would look similar to
Fig.~\ref{strong_fast}(b).

\subsection{Intermediate Modulation Regime}
So far we have considered four limiting cases: (i) strong, slow, (ii) weak,
slow, (iii) strong, fast, and (iv) weak, fast case. We now consider case 5
($\Gamma \ll R \ll \nu$) and case 6 ($\nu \ll R \ll \Gamma$). They are neither
in the slow nor in the fast modulation regime according to our definition.

\begin{figure}
\bc
\epsfig{file=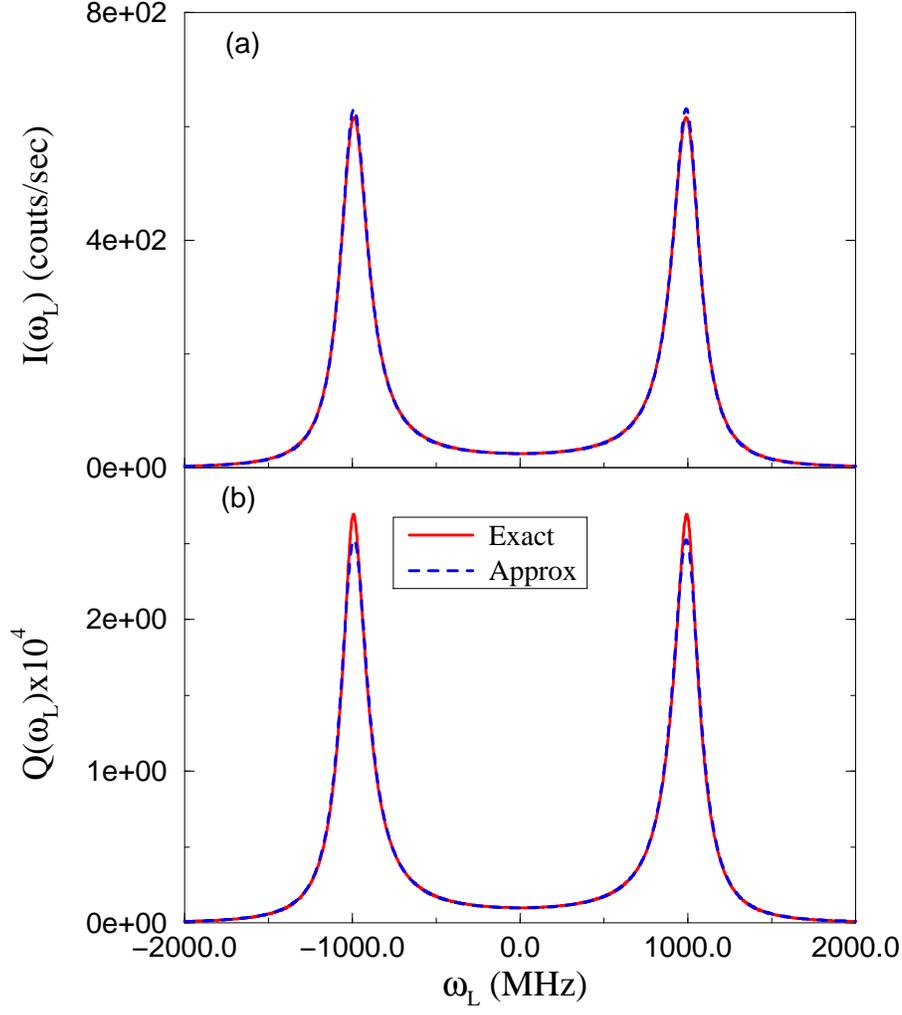,width=\figsize}
\ec \caption[Lineshape and $Q$ in case 5 ($\Gamma \ll R \ll \nu$) in the
steady-state limit] { Case 5 ($\Gamma \ll R \ll \nu$) in the steady-state
limit. Exact results for lineshape,  Eq.~(\ref{Iex}), and for $Q$,
Eqs.~(\ref{QltDen}), (\ref{QltNum}), are compared with approximations
corresponding to $\Gamma\rightarrow 0$ limit, Eqs.~(\ref{IG0}) and (\ref{QG0}),
respectively. Parameters are chosen as $\nu=1\mbox{GHz}$, $\Gamma=5\mbox{MHz}$,
$\Omega=\Gamma/10$, $R=100\mbox{MHz}$, and $T\to\infty$. } \label{case5}
\end{figure}

\subsubsection{case 5}
In case 5($\Gamma \ll R \ll \nu$), the bath fluctuation is fast compared with
the radiative decay rate but not compared with the fluctuation amplitude.
Because $\Gamma \ll R,\nu$ in this case we can approximate the exact results
for $\la I\ra$ and $Q$ by their limiting expressions corresponding to
$\Gamma\rightarrow 0$, yielding Eqs.~(\ref{IG0}) and (\ref{QG0}), and an
important relation holds in this limit, \be \lim_{\Gamma\rightarrow 0} Q
={2\over\Gamma}\lim_{\Gamma\rightarrow 0}\la I\ra. \label{limG0} \ee Note that
the same relation between $Q$ and $\la I\ra$ was also found to be valid in one
of the fast modulation regimes, Eq.~(\ref{Qfs}) with $\Gamma_f\gg \Gamma$.
Fig.~\ref{case5} shows that in this case the limiting expressions approximate
well the exact results.

\begin{figure}
\bc \epsfig{file=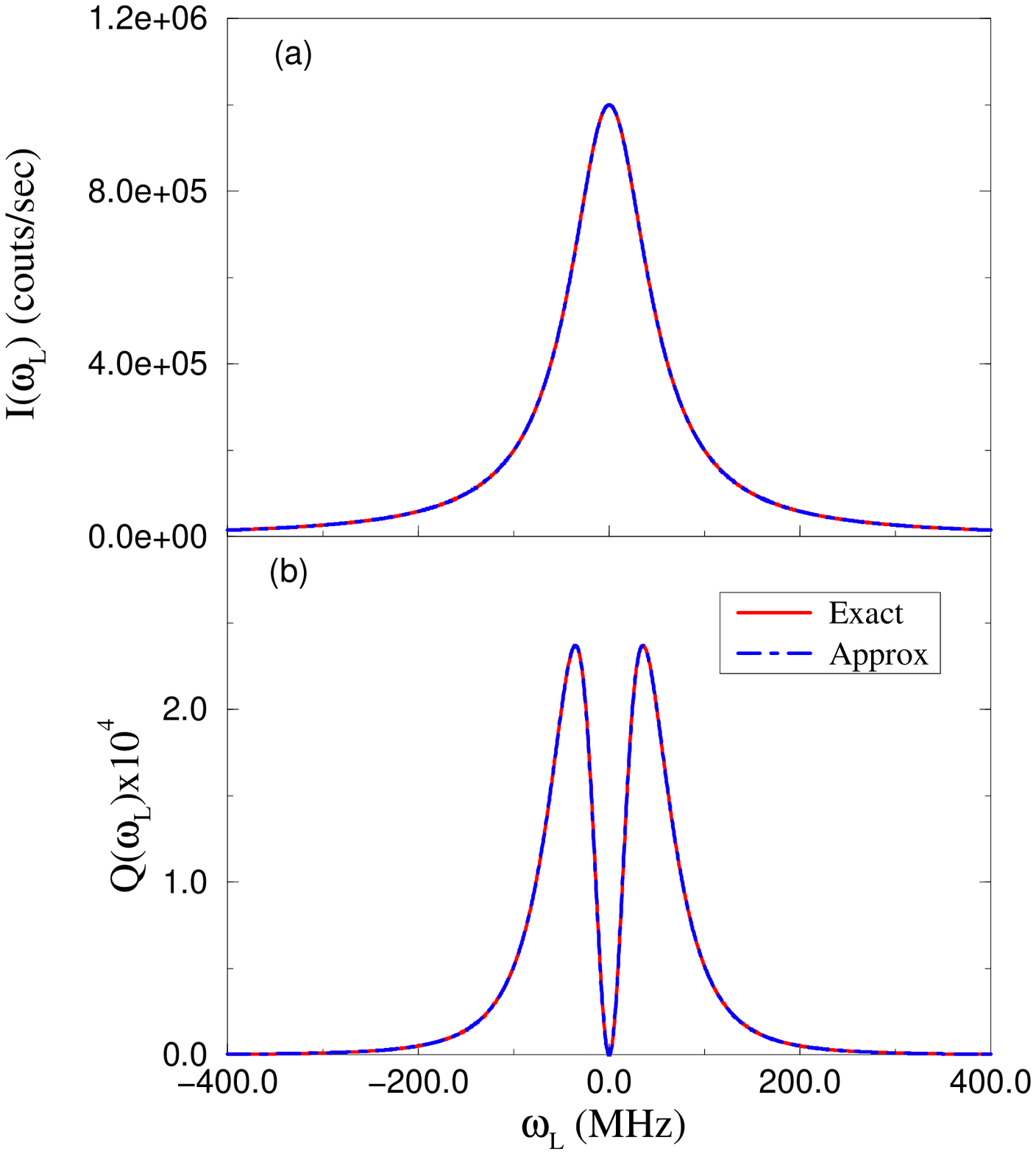,width=\figsize} \ec \caption[Lineshape
and $Q$ in case 6 ($\nu \ll R \ll \Gamma$) in the steady-state limit] { Case 6
($\nu \ll R \ll \Gamma$) in the steady-state limit. Exact results for
lineshape, Eq.~(\ref{Iex}), and for $Q$, Eqs.~(\ref{QltDen}), (\ref{QltNum}),
are compared with approximations corresponding to $\nu\rightarrow 0$ limit,
Eqs.~(\ref{Ilor}) and (\ref{Qnu0}), respectively, Parameters are chosen as
$\nu=1\mbox{MHz}$, $\Gamma=100\mbox{MHz}$, $\Omega=\Gamma/10$, $R=10
\mbox{MHz}$, and $T\to \infty$. } \label{case6}
\end{figure}

\subsubsection{case 6}
In case 6, since $\nu\ll R\ll \Gamma$ we can approximate the exact results by
considering small $\nu$ limit in Eqs.~(\ref{QltDen}) and (\ref{QltNum}) for
$Q$, and Eq.~(\ref{Iex}) for the lineshape. By taking this limit, we find that
the lineshape is well described by a single Lorentzian given by
Eq.~(\ref{Ilor}), and $Q$ by Eq.~(\ref{Qsw}). We note that for all weak
modulation cases, cases 2, 4, and 6, the lineshape and $Q$ behave in a unique
way described by Eq.~(\ref{Ilor}) and Eq.~(\ref{Qsw}) for $RT\gg 1$,
respectively, and both the slow and fast modulation approximate results are
valid in this case (see Table \ref{table2-2}). Also a simple relation between
$\la I\ra$ and $Q$ holds in the limit, $\nu\to 0$, \be \lim_{\nu\to 0}Q={\nu^2
\over R}\lim_{\nu\to 0} {1\over \la I\ra} \left(d\la I\ra\over
d\omega_L\right)^2. \label{Qnu0} \ee The exact results of lineshape and $Q$ in
Fig.~\ref{case6} show good agreement with the approximate results,
Eqs.~(\ref{Ilor}) and (\ref{Qnu0}).

\begin{figure}
\bc
\epsfig{file=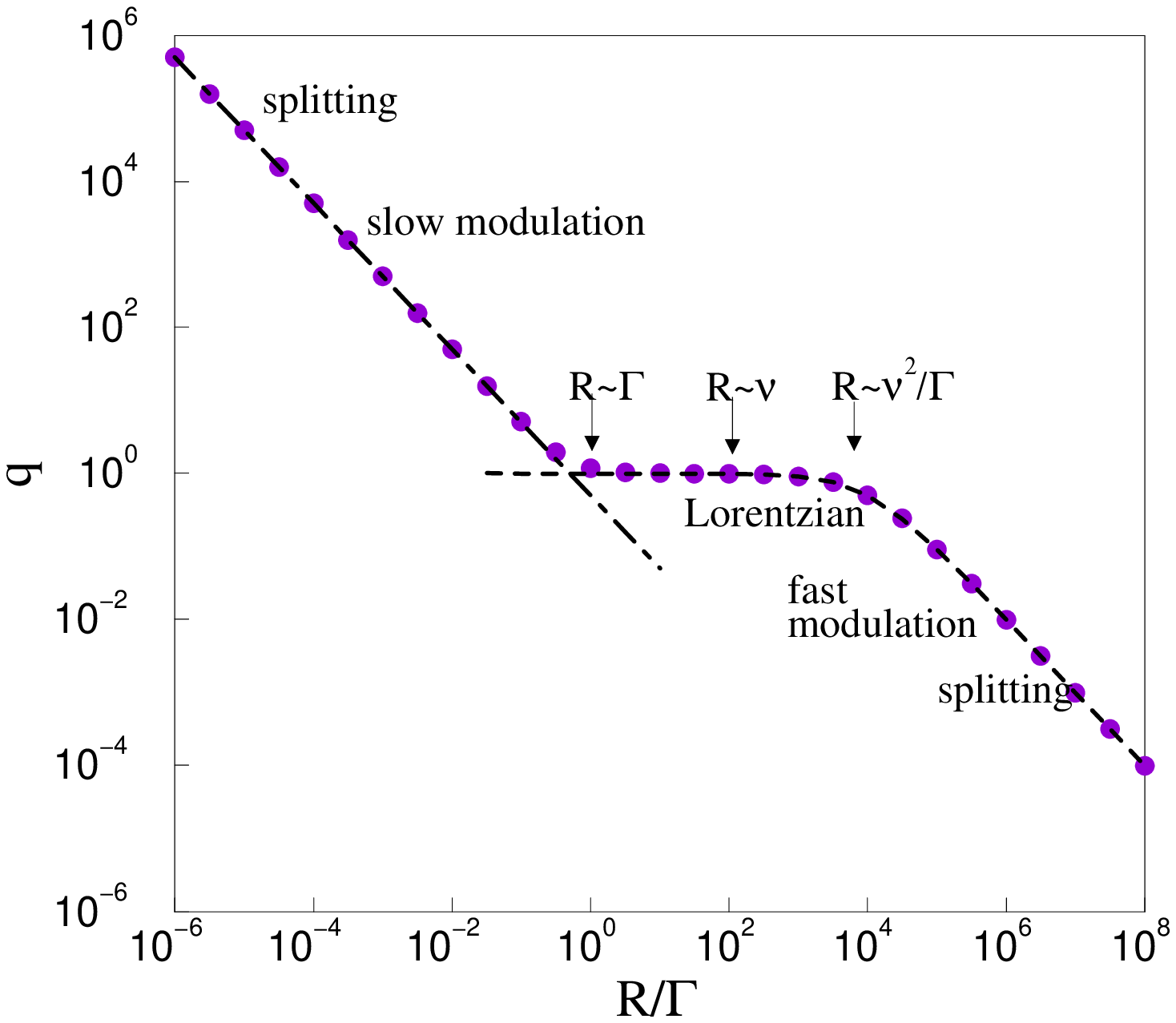,width=\figsize}
\ec \caption[$q$ versus $R/\Gamma$ in the strong modulation case]{ $q$ versus
$R/\Gamma$ in the strong modulation case. Filled circles are the result of
exact calculation based on Eqs.~(\ref{QltDen}) and (\ref{QltNum}) when
$\nu/\Gamma=100$. The dot-dashed and dashed curves are calculations based on
approximate expressions of $Q$, Eq.~(\ref{Qss}) (slow modulation case) and
Eq.~(\ref{eqFML01}) (fast modulation case), respectively. } \label{pd0}
\end{figure}

\subsection{Phase Diagram}
\label{SecPhase} 
\begin{figure}
\bc
\epsfig{file=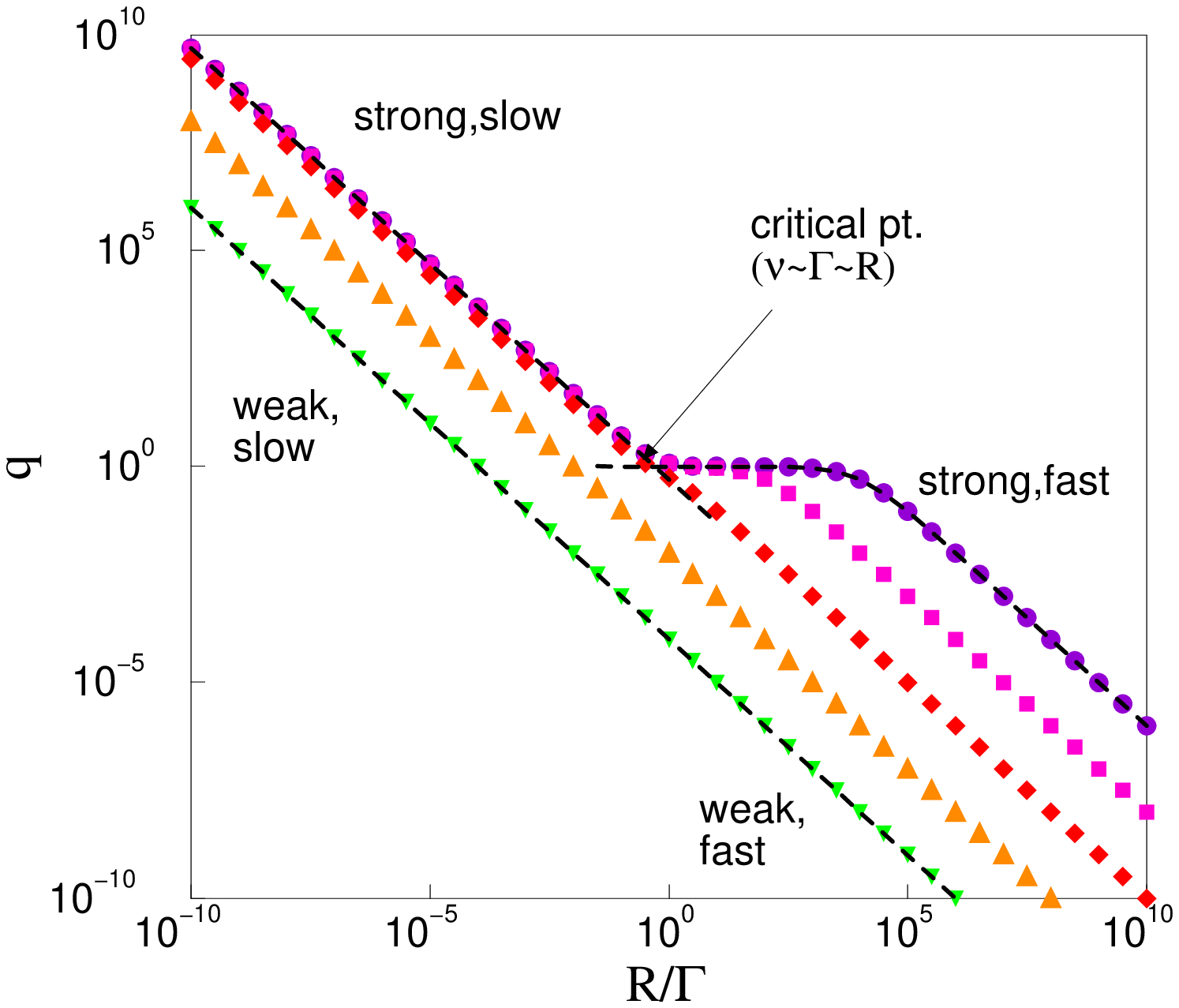,width=\figsize}
\ec \caption[Phase diagram of $q$ versus $R/\Gamma$]{ Phase diagram of $q$
versus $R/\Gamma$. Symbols are the results of exact calculations based on
Eqs.~(\ref{QltDen}) and (\ref{QltNum}) as $\nu/\Gamma$ is varied from the
lower-left to the upper-right cases: $0.01$ (downward triangle), $0.1$ (upward
triangle), $1$ (diamond), $10$ (square), $100$ (circle). The dashed curves are
calculations based on appropriate approximate expressions of $Q$ for each case,
Eqs.~(\ref{Qfw}) (weak modulation), (\ref{Qss}) (strong, slow modulation), and
(\ref{eqFML01}) (strong, fast modulation). } \label{pd1}
\end{figure}

We investigate the overall effect of the bath fluctuation on
the photon statistics for the steady-state case as the fluctuation rate $R$ is
varied from slow to fast modulation regime. To characterize the overall
fluctuation behavior of the photon statistics, we define an order parameter
$q$, \be q\equiv {\Gamma\over \pi \Omega^2} \int_{-\infty}^{\infty} Q(\omega_L)
{\rm{d}}\omega_L, \label{qdef} \ee where $Q$ in the steady state is given in
Eqs.~(\ref{QltDen}) and (\ref{QltNum}). Before we discuss the behavior of $q$
it is worthwhile mentioning that the lineshape is normalized to a constant
regardless of $R$ and $\Gamma$, \be \int_{-\infty}^{\infty}\la I(\omega_L)\ra
{\rm{d}}\omega_L={\pi \Omega^2\over 2}, \label{Inorm} \ee which can be easily
verified from Eq.~(\ref{eqApzz3mmm}). In contrast, $q$ exhibits nontrivial
behavior reminiscent of a phase transition. In Figs.~\ref{pd0} and \ref{pd1} we
show $q$ versus $R/\Gamma$. The figures clearly demonstrate how the photon
statistics of SMS in the presence of the spectral diffusion becomes Poissonian
as $R \to \infty$ or $\nu\to 0$ (i.e. $q\to 0$ when $R\to \infty$ or $\nu \to
0$).

We first discuss the {\it strong modulation regime},
$\nu\gg \Gamma$, shown in Fig.~\ref{pd0}
with $\nu/\Gamma=100$ in this case.
Depending on the fluctuation rate, there are three distinct regimes:

\noindent
(a) In the slow modulation regime, $R\ll\Gamma$, $q$ decreases
as $1/R$. The approximate calculation (dot-dashed line)
based on the slow modulation approximation,
Eq.~(\ref{Qss}) shows good agreement with the exact calculation.

\noindent (b) When $R$ is such that $\Gamma \ll R \ll \nu$ [case 5], the
intermediate regime is achieved, and $q$ starts to show a plateau behavior. The
plateau behavior is found whenever $Q=2\la I (\omega_L)\ra /\Gamma$, which
yields $q=1$ as can be easily seen from Eqs.~(\ref{qdef}) and (\ref{Inorm}). As
$R$ is increased further such that $\nu \ll R$, the fast modulation regime is
reached, and $q$ still shows a plateau until $R\simeq \nu^2/\Gamma$. In this
regime the Lorentzian behavior of $Q$ is observed in Eq.~(\ref{Qfs}) for
$\Gamma_f\gg \Gamma$.

\noindent
(c) When the bath fluctuation becomes extremely fast
such that $R\gg \nu^2/\Gamma$, the splitting behavior of $Q$ is observed, as discussed
in Eq.~(\ref{Qfs}) for $\Gamma_f\ll \Gamma$, and then $q\propto 1/R$
similar to the slow modulation regime.
The approximate value of $q$
based on fast modulation approximation, Eq.~(\ref{eqFML01}) (dotted line)
shows good agreement with the exact calculation
found using Eqs.~(\ref{QltDen}) and (\ref{QltNum}).
Finally, when $R\to \infty$, $Q=0$.
As mentioned this is expected since the molecule cannot interact with a very fast
bath hence the photon statistics becomes Poissonian.

Now we discuss the effect of magnitude of frequency
fluctuation, $\nu$,
on $q$ in Fig.~\ref{pd1}.
We have calculated $q$ as a function of $1/R$ as $\nu$ is varied
from $\nu/\Gamma=100$ (circles) to $\nu/\Gamma=0.01$ (downward triangles).
We see in Fig.~\ref{pd1} that the $(q,R/\Gamma)$ diagram exhibits
a behavior similar to a phase transition as $\nu$ is varied.
In the strong modulation regime ($\nu/\Gamma \gg 1$),
three distinct regimes appear in the $(q,R/\Gamma)$ diagram
($R\leq \Gamma$, $\Gamma \leq R \leq \nu^2/\Gamma$, and $R \geq \nu^2/\Gamma$),
while in the weak modulation
regime ($\nu/\Gamma \ll 1$) $q$ always decreases as $1/R$.
When three parameters, $\nu$, $\Gamma$, and $R$,
have similar values, $\nu\sim \Gamma \sim R$,
there appears a ``critical point''
in the ``phase diagram''.

Table \ref{table2-2} summarizes various expressions for the lineshape and $Q$
found in the limiting cases of the two state jump model investigated in this
work. For simplicity, we have set $\eta=1$. We see that although the
fluctuation model itself is a simple one, rich behaviors are found. We believe
that these behaviors are generic (although we do not have a mathematical
proof). In the weak modulation cases, both $\la I\ra$ and $Q$ can be described
by a single expression, irrespective of the fluctuation rate $R$. However, in
the strong modulation cases, $\la I\ra$ and $Q$ change their qualitative
features as $R$ changes.


\begin{table}
\begin{center}
\noindent
\begin{tabular}{|c | c| c| c|}
\hline
             & {slow} & {intermediate} & {fast}  \\
\hline & $R \ll \nu \ll \Gamma $ : case 2 &  $\nu \ll R \ll \Gamma $ : case 6
&  $\nu \ll \Gamma \ll R$ : case 4 \\
\cline{2-4} \raisebox{-7ex}[0pt] {weak} & \multicolumn{3}{|c|}
{\begin{tabular}{ccc}
 \\
 & $\displaystyle \la I\ra \sim{\Omega^2\Gamma \over
4\left[\omega_L^2+{\Gamma^2\over 4}\right]} $ &
\\
& &
\end{tabular}}\\
\cline{2-4}
 &
\multicolumn{3}{|c|}
{\begin{tabular}{ccc}
\\
& $\displaystyle Q
\sim {\Gamma_f \over \la I\ra} \left({\rm d}\la I\ra\over {\rm
d}\omega_L\right)^2 $ &
\\
& &
\end{tabular}}
\\
\hline
& $R\ll \Gamma \ll \nu$ : case 1 & $\Gamma \ll R \ll \nu$ : case 5
& $\Gamma \ll \nu \ll R$ : case 3 \\
\cline{2-4} \raisebox{-7ex}[0pt] {strong} &
\begin{tabular}{c}
\\
 $  \displaystyle
 \la I\ra \sim{I_++I_-\over 2} $
 \\ \\
\end{tabular} &
\begin{tabular}{c}
\\
$ \displaystyle \la I\ra $
$\displaystyle\sim {\Omega^2 \nu^2 R
\over(\omega_L^2-\nu^2)^2+4R^2\omega_L^2} $
\\ \\
\end{tabular}
& \begin{tabular}{c}
\\
$ \displaystyle \la I\ra \sim {\Omega^2 (\Gamma+\Gamma_f)\over 4
\left[\omega_L^2+{ (\Gamma+\Gamma_f)^2\over 4}\right]}$
\\ \\
\end{tabular}
\\
\cline{2-4}
&
\begin{tabular}{c}
$ \displaystyle Q
\sim{\la I\ra \over R}$ \\
\end{tabular}
&
\begin{tabular}{c}
$ \displaystyle
 Q\sim
{2 \la I\ra \over \Gamma} $
\end{tabular}
&
\begin{tabular}{c}
\\
$ \displaystyle Q $
$\displaystyle \sim \left\{
\begin{tabular}{ll}
\large $
{2\la I\ra \over \Gamma}$ & $\hspace{-0.0cm}\Gamma_f\gg \Gamma$ 
\vspace{.2cm}
\\
\large
$ {\Gamma_f \over \la I\ra} \left({\rm d}\la I\ra\over {\rm
d}\omega_L\right)^2$ & $\hspace{-0.0cm}\Gamma_f\ll \Gamma$
\end{tabular}\right.\hspace{-0.3cm}
$
\\
\\
\end{tabular} \\
\hline
\end{tabular}


\caption{
Parameter regimes investigated in this work are classified with
expressions for the lineshape, $\la I\ra$, and $Q$ for each case. $I_{\pm}$
have been defined in Eq.~(\ref{eqSML04}).
\label{table2-2}}
\end{center}
\end{table}

\section{Connection to Experiments}
\label{SecExp} Single molecule spectroscopy has begun to reveal the microscopic
nature of low temperature glasses\cite{fleury-jlum-93,boiron-cp-99,geva-jpcb-97,brown-jcp-98,barkai-prl-00,barkai-jcp-00}. An
important question is whether the standard tunneling model of low temperature
glass developed by Anderson, Halperin, Varma\cite{anderson-philos-71} and
Phillips\cite{phillips-jltp-72} is valid or not. As far as macroscopic
measurements of acoustic, thermal, and optical properties are concerned, this
model has proved to be compatible with experimental
results\cite{esquinazi-tunn-98}. However, on a more microscopic level we do not
have much experimental or theoretical proof (or disproof) that the model is
valid. At the heart of the standard tunneling model is the concept of a two
level system(TLS). At very low temperatures, the complicated multidimensional
potential energy surface of the glass system is presumed to reduce to a
multitude of non--interacting double well potentials whose two local minima
correspond to reorientations of clusters of atoms or
molecules\cite{heuer-prl-93}. Hence the complicated behavior of glasses is
reduced to a simple picture of many non--identical and non--interacting TLSs.
For a different perspective on the nature of low temperature excitation in
glasses, see Ref.~\cite{lubchenko-prl-01}.

Geva and Skinner\cite{geva-jpcb-97} have provided a theoretical interpretation
of the static lineshape properties in a glass (i.e.~$\la W\ra$). The theory
relied on the standard tunneling model and the Kubo--Anderson approach as means
to quantify the lineshape behavior (i.e.~the time dependent fluctuations of $W$
are neglected). In Ref.~\cite{barkai-prl-00}, the distribution of static
lineshapes in a glass was found analytically and the relation of this problem
to L\'evy statistics was demonstrated.

Orrit and coworkers\cite{zumbusch-prl-93,fleury-jlum-93} have measured spectral
trails as well as the lineshapes and the fluorescent intensity correlation
function $g^{(2)}(t)$. In Ref.~\cite{boiron-cp-99} spectral trails of 70
molecules were investigated and 22 exhibited behaviors that seemed incompatible
with the standard tunneling model. While the number of molecules investigated
is not sufficient to determine whether the standard tunneling model is valid,
the experiments are approaching a direct verification of this model.

Our theory in the slow modulation limit describes SMS experiments in glasses.
The TLSs in the glass flip between their up and down states with a rate $R$
determined by the coupling of TLS to phonons, the energy asymmetry, and the
tunneling matrix element of the TLS\cite{anderson-philos-71,phillips-jltp-72}.
When a TLS makes a transition from the up to the down state, or vice versa, a
frequency shift $\nu\propto{1/r^3}$ occurs in the absorption frequency of a SM,
where $r$ is the distance between the SM and the TLS. The ${1/r^3}$ dependence
is due to an elastic dipole interaction between the SM and the TLS. In a low
temperature glass, the density of TLSs is very low, hence one finds in
experiment that the SM is coupled strongly to only a few TLSs. In some cases,
when one TLS is in the vicinity of the SM, it is a reasonable approximation to
neglect all the background TLSs. In this case, our theory describes SMS for
chromophores in glasses with a single TLS strongly coupled to SM. Extension of
our work to coupling of SM to many TLSs is important, and can be done in a
straightforward manner provided that the TLSs are not interacting with each
other.

Fleury {\it et al.}\cite{fleury-jlum-93} measured $g^{(2)}(t)$ for a single
terrylene molecule coupled to single TLS in polyethylene matrix. They showed
that their experimental results are well described by
\be
g^{(2)}(\tau)=1+{R_{+}R_{-} (I_{+} - I_{-})^2\over
(R_{+}I_{+}+R_{-}I_{-})^2}e^{-(R_{+}+R_{-})\tau}, \label{expg2} \ee
where $R_{+}$ and $R_{-}$ are the upward and downward transition rates,
respectively. These two rates, $R_+$ and $R_-$ are due to the asymmetry of the
TLS. This result is compatible with our result for $Q$ in the slow modulation
limit. When Eq.~(\ref{expg2}) is used in Eq.~(\ref{Qg2intro}) with
$R_{+}=R_{-}=R$ for the symmetric transition case considered in this work, we
exactly reproduce the result of $Q(T)$ for the slow modulation given in
Eq.~(\ref{eqSML09}). [The asymmetric rate case $R_{+}\ne R_{-}$ is also readily
formulated for $Q$ in the slow modulation limit, and again leads to a result
compatible with Eq.~(\ref{expg2}).] Hence, at least in this limit, our results
are in an agreement with the experiment.

As far as we are aware, however, measurements of photon counting statistics for
SMS in fast modulation regimes have not been made yet. The theory presented
here suggests that even in the fast modulation regime, the deviation from
Poisson statistics due to a spectral diffusion process might be observed under
suitable experimental situations, for example, when the contributions of other
mechanisms such as the quantum mechanical anti-bunching process and the
blinking process due to the triplet state are known {\it a priori}. In this
case, it can give more information on the distribution of the fluctuation
rates, the strength of the chromophore-environment interaction, and the bath
dynamics than the lineshape measurement alone.

\section{Further Discussions}
\label{SecDisc} All along in this work we have specified the conditions under
which the present model is valid. Here we will emphasize the validity and the
physical limitations of our model. We will also discuss other possible
approaches to the problem at hand.

In the present work, we have used classical photon counting statistics in the
weak laser field limit. In the case of strong laser intensity, quantum
mechanical effects on the photon counting statistics are expected to be
important. From theories developed to describe two level atoms interacting with
a photon field {\em in the absence of environmental fluctuations}, it is known
that, for strong field cases, deviations from classical Poissonian statistics
can become significant\cite{mandel-optical-95,plenio-rmp-98}. One of the well
known quantum mechanical effects on the counting statistics is the photon
anti-bunching
effect\cite{fleury-prl-00,lounis-nature-00,basche-prl-92,michler-nature-00,short-prl-83,schubert-prl-92}.
In this case, a sub-Poissonian behavior is obtained, $-1<Q_{\rm {qm}}<0$, where
the subscript ``qm'' stands for quantum mechanical contribution. It is clear
that when the spectral diffusion process is significant (i.~e.~cases 1 and 2)
any quantum mechanical correction to $Q$ is negligibly small. However, in the
fast modulation case where we typically found a small value of $Q$, for
example, $Q\sim 10^{-4}$, quantum mechanical corrections due to anti-bunching
phenomenon might be important unless experiments under extremely weak fields
can be performed such that $|Q_{{\rm qm}}| < Q_{{\rm sd}}$, where the subscript
``sd'' stands for spectral diffusion. The interplay between truly quantum
mechanical effects and the fast dynamics of the bath is left for future work.

In this context it is worthwhile to recall the quantum jump approach developed
in the quantum optics community. In this approach, an emission of a photon
corresponds to a quantum jump from the excited state to the ground state. For a
molecule with two levels, this means that right after each emission event,
$\rho_{ee}=0$ (i.~e.~the system is in the ground state). Within the classical
approach this type of wavefunction collapse never occurs. Instead, the emission
event is described with the probability of emission per unit time being
$\Gamma\rho_{ee}(t)$, where $\rho_{ee}(t)$ is described by stochastic Bloch
equation. At least in principle, the quantum jump approach, also known as the
Monte Carlo wavefunction approach\cite{dum-pra-92,plenio-rmp-98,dalibard-prl-92,mollmer-josab-93,makarov-jcp-99-1,makarov-jcp-01},
can be adapted to calculate the photon statistics of a SM in the presence of
spectral diffusion.

Another important source of fluctuation in SMS is due to the triplet state
dynamics. Indeed, one of our basic assumptions was the description of the
electronic transition of the  molecule in terms of a two state model. Blinking
behavior is found in many SMS experiments\cite{bernard-jcp-93,basche-nature-95,brouwer-prl-98,vandenbout-sci-97,yip-jpca-98}.
Due to the existence of metastable triplet states (usually long lived) the
molecule switches from the bright to the dark states (i.~e.~when molecule is
shelved in the triplet state, no fluorescence is recorded). Kim and
Knight\cite{kim-pra-87} pointed out that $Q$ can become very large in the case
of the metastable three level system {\em in the absence of spectral
diffusion}. This is especially the case when the lifetime of the metastable
state is long. Molski {\it et al.}\cite{molski-cpl-00a,molski-cpl-00b} have
considered the effect of the triplet state blinking on the photon counting
statistics of SMS.

Therefore at least three sources of fluctuations can contribute to the measured
value of $Q$ in SMS ; (i) $Q_{\rm {qm}}$, well investigated in quantum optics
community, (ii) $Q_{\rm {triplet}}$, which can be described using the approach
of Ref.~\cite{kim-pra-87}, and now we have calculated the third contribution to
$Q$, (iii) $Q_{\rm {sd}}$. Our approach is designed to describe a situation for
which the spectral diffusion process is dominant over the others.

It is interesting to see if one can experimentally distinguish $Q_{\rm {sd}}$
and $Q_{\rm{triplet}}$ for a SM in a condensed environment. One may think of
the following {\em gedanken experiment}: consider a case where the SM jumps
between the bright to the dark state, and assume that we can identify the dark
state when the SM is in the metastable triplet state. Further, let us assume
that dark state is long lived compared to the time between emission events in
the bright state. Then, at least in principle, one may filter out the effect of
the dark triplet state on $Q$ especially when the timescale of the spectral
diffusion process is short compared with the dark period by measuring the
photon statistics during the bright period.

\section{Concluding Remarks}
\label{SecCon} In this paper we have developed a stochastic theory of single
molecule fluorescence spectroscopy. Fluctuations described by $Q$ are evaluated
in terms of a three--time correlation function $C_3\left( \tau_1, \tau_2 ,
\tau_3 \right)$ related to the response function in nonlinear spectroscopy.
This function depends on the characteristics of the spectral diffusion process.
Important time ordering properties of the three--time correlation function were
investigated here in detail. Since the fluctuations (i.e., $Q$) depend on the
three--time correlation function, necessarily they contain more information
than the lineshape which depends on the one--time correlation function
$C_1\left( \tau_1 \right)$ via the Wiener--Khintchine theorem.

We have evaluated the three--time correlation function and $Q$ for a stochastic
model for a bath with an arbitrary timescale. The exact results for $Q$ permit
a better understanding of the non--Poissonian photon statistics of the single
molecule induced by spectral diffusion. Depending on the bath timescale,
different time orderings contribute to the lineshape fluctuations and in the
fast modulation regime all time orderings contribute. The theory predicts that
$Q$ is small in the fast modulation regime, increasing as the timescale of the
bath (i.e. $1/R$) is increased. We have found nontrivial behavior of $Q$ as
the bath fluctuation becomes fast. Results obtained in this work are applicable
to the experiment in the slow to the intermediate modulation regime (provided
that detection efficiency is high), and our results in the more challenging
fast modulation regime give the theoretical limitations on the measurement
accuracy needed to detect $Q>0$.

The model system considered in this work is simple enough to allow an exact
solution, but still complicated enough to exhibit nontrivial behavior.
Extensions of the present work is certainly possible in several important
aspects. It is worthwhile to consider photon counting statistics for a more
complicated chromophore-bath model, for example, the case of many TLSs coupled
to the chromophore, to see to what extent the results obtained in this work
would remain generic. Also the effects of a triplet bottleneck state on the
photon counting statistics can be investigated as a generalization of the
theory presented here. Another direction for the extension of the present
theory is to formulate the theory of SMS starting from the microscopic model of
the bath dynamics (e.~g.~the harmonic oscillator bath model). Effects of the
interplay between the bath fluctuation and the quantum mechanical photon
statistics on SMS is also left for future work.

The standard assumption of Markovian processes (e.g. the Poissonian
Kubo--Anderson processes considered here) fails to explain the statistical
properties of emission for certain single ``molecular'' systems such as quantum
dots\cite{kuno-jcp-00,neuhauser-prl-00,shimizu-prb-01}. Instead of the usual
Poissonian processes, a power-law process has been found in those systems. For
such highly non--Markovian dynamics stationarity is never reached and hence our
approach as well as the Wiener--Khintchine theorem does not apply. This problem
has been investigated in Ref.~\cite{jung-qd-02}.


\section[Appendix A: Calculation of Lineshape]{Appendix A: Calculation of Lineshape}
\label{AppMar}
In this appendix, we calculate the lineshape for our working
example. We set $J=1$ and $\omega_0=0$ and as mentioned the stochastic
frequency modulation follows $\Delta \omega (t) = \nu h(t)$, where $h(t)$
describes a two state telegraph process with $h(t)=+1$ (up) or $h(t) = -1$
(down). Transitions from state up to down and down to up are described by the
rate $R$.

We use the marginal averaging approach\cite{shore-josab-84,burshtein-jetp-66,shore-coherent-90} to calculate the
average lineshape. Briefly, the method gives a general prescription for the
calculation of averages $\langle y \rangle$ ($y$ is a vector) described by
stochastic equation $\dot{y} = M(t) y$, where $M(t)$ is a matrix whose elements
are fluctuating according to a Poissonian process (see
Refs.~\cite{shore-josab-84,shore-coherent-90} for details). We define the
marginal averages, $\langle v(t) \rangle_x$ , $\langle u(t) \rangle_x$ and
$\langle w(t) \rangle_x$  where $x=+$ or $x=-$ denotes the state of the two
state process at time $t$. For the stochastic Bloch equation, the evolution
equation for the marginal averages are
\be
\left(
\begin{array}{c}
\langle \dot{u} \rangle_{+} \\
\langle \dot{v} \rangle_{+} \\
\langle \dot{w} \rangle_{+} \\
\langle \dot{u} \rangle_{-} \\
\langle \dot{v} \rangle_{-} \\
\langle \dot{w} \rangle_{-}
\end{array}
\right) &=&
\left(
\begin{array}{c c c c c c}
-R - {\Gamma \over 2} &  \delta_{L}^{+}   & 0 \ &  R \  & 0 \ & 0 \\
-  \delta_{L}^{+} \ &  - R - {\Gamma\over 2} \  & - \Omega  \  & 0 \  &  R \  & 0 \\
0 \  & \Omega \ & - \Gamma - R  \ &   0 \  &  0 \  & R \\
R \ &  0 \  & 0 \  & - R  - {\Gamma \over 2} \  & \delta_{L}^{-}  \  &  0 \\
0 \  & R \  & 0 \ & -\delta_{L}^{-}\  & - R -{ \Gamma \over 2} \  &  -\Omega  \\
0 \  & 0 \  & R \ &  0 \  & \Omega \ &  - R - \Gamma
\end{array}
\right)
\no \\
&\times&
\left(
\begin{array}{c}
\langle u \rangle_{+} \\
\langle v \rangle_{+} \\
\langle w \rangle_{+} \\
\langle u \rangle_{-} \\
\langle v \rangle_{-} \\
\langle w \rangle_{-}
\end{array}
\right)
+ \left(
\begin{array}{c}
0 \\
0 \\
-\Gamma/2 \\
0 \\
0 \\
-\Gamma/2
\end{array}
\right)
,
\label{eqA2} \ee
where $\delta_{L}^{\pm}= \delta_{L} \mp  \nu$. The steady state solution is
found by using a symbolic program such as {\it Mathematica}\cite{math},
\begin{equation}
\langle I\left( \omega_L \right) \rangle = {1 \over 2}  \Omega\left( \langle
v_{st} \rangle_- + \langle  v_{st} \rangle_+ \right) = {1 \over 2}  \Gamma
\left(  \langle w_{st}\rangle_{+} + \langle w_{st} \rangle_- + 1 \right),
\label{eqA3}
\end{equation}
where $\left(  \langle w_{st}\rangle_{+}+\langle w_{st} \rangle_{-}+1\right)/2
$ represents the steady state occupation of the excited level. We find
\begin{equation}
\langle I \left(\omega_L \right) \rangle ={A \over B},   \label{Iex}
\end{equation}
\begin{eqnarray}
A & = & \Gamma \Omega^2 \left[(\Gamma+2R)
\left(\Gamma(4\omega_L^2+(\Gamma+4R)^2)
+4\nu^2(\Gamma+4R)\right)+2\Omega^2\Gamma(\Gamma+4R)\right],
\no \\
B&=&\Gamma (\Gamma+2R) \left( ( 4(\omega_L^2-\nu^2)-\Gamma(\Gamma+4R))^2
+16\omega_L^2(\Gamma+2R) \right) + 4\Omega^2\left(4\omega_L^2\Gamma(\Gamma+3R)
\right.
\no \\
&&\left. + (\Gamma^2+4\Gamma R+4\nu^2)(\Gamma^2+3\Gamma R+4 R^2)\right)
+4\Omega^4\Gamma (\Gamma+4R).  \no
\end{eqnarray}

\noindent {\it \bf Remark 1} When $R \to 0$, it is easy to show that $\langle
I\left( \omega_L \right) \rangle$ is a sum of two Lorentzians centered at $\pm
\nu$, \be \la I(\omega_L) \ra = {\Omega\over 2} (v_{+} + v_{-}), \label{eqA4}
\ee where $v_{\pm}$ are steady state solution of Bloch equation for two level
atom (see Ref.~\cite{cohentann-atom-93})
\begin{equation}
v_{\pm} = { \Omega \over 2}{ \Gamma/2 \over (\delta_{L}^{\pm})^2 + \Gamma^2/4 +
\Omega^2 / 2}. \label{eqA5}
\end{equation}

\noindent {\it\bf Remark 2} If $\Omega \to 0$, the solution can also be found
based on the Wiener--Khintchine formula, using the weights
$\hat{P}_{ij}^{-1}(s)$ defined in Appendix A.
\begin{equation}
\langle I\left( \omega_L \right) \rangle = {\Omega^2 \over 4} \mbox{Re}\left[
\sum_{i=\pm,j=\pm} \hat{P}_{ij}^{-1}\left( i \omega_L + \Gamma/2\right)\right]
\label{eqWK1}
\end{equation}
which gives \be &&\langle I\left( \omega_L \right) \rangle = {\Omega^2 (4\Gamma
\omega_L^2+(\Gamma+4 R)(\Gamma^2+4\Gamma R+4\nu^2)) \over
(4(\omega_L^2-\nu^2)-\Gamma(\Gamma+4R) )^2+16\omega_L^2(\Gamma+2R)^2}. \ee
\label{eqWK2} Now if $\Gamma \to 0$ we get the well known result of Kubo
\begin{equation}
 \langle I\left( \omega_L \right) \rangle = \frac{{\Omega}^2 {\nu}^2\,R}
{\left( \omega_L^2 - \nu^2 \right)^2 + 4 R^2 \omega_L^2 }, \label{IG0}
\end{equation}
then in the slow modulation limit, $\nu\gg R$ the line $\langle
I\left(\omega_L\right) \rangle$ exhibits splitting (i.e., two peaks at $\pm
\nu$) while for the fast modulation regime $\nu\ll R$ the line is a Lorentzian
centered at $\omega_L=0$ and  motional narrowing is observed.

\noindent {\it\bf Remark 3}  If $\nu\to 0$ the solution reduces to the well
known Bloch equation solution of a stable two level atom, which is independent
of $R$.

\noindent {\it\bf Remark 4} We have assumed that occupation of state $+$ and
state $-$ are equal. More general case, limited to weak laser intensity regime,
was considered in Ref.~\cite{reilly1-jcp-94}.

\section{Appendix B: Perturbation Expansion}
\label{AppPert}
In this appendix, we find expressions for the photon current
using the perturbation expansion. We also use the Lorentz oscillator model to derive
similar results based on a classical picture.

We use the stochastic Bloch equation\cite{shore-josab-84,colmenares-theochem-97} to investigate $\rho_{ee}(t)$ in
the limit of weak external laser field when we expect $\rho_{ee} \simeq 0$,
$\rho_{gg}\simeq 1$ for times $\Gamma T\gg 1$. We rewrite
Eqs.~(\ref{eq:udot})-(\ref{eq:wdot})
\be { {\rm d} \tilde{\rho}_{ee} \over {\rm d} t } &=& - \Gamma
\tilde{\rho}_{ee} + { i \Omega \over 2} \left( \tilde{\rho}_{eg} -
\tilde{\rho}_{ge} \right),
\label{eqA00} \\
{ {\rm d}\tilde{\rho}_{ge} \over {\rm d} t} &=& - \left[ i \delta_L (t) +
\Gamma/2 \right] \tilde{\rho}_{ge} - { i \Omega \over 2} \left(
\tilde{\rho}_{ee} - \tilde{\rho}_{gg} \right), \label{eqA01} \ee
where $\tilde{\rho}_{eg}=\rho_{eg}e^{ i \omega_L t} $,
$\tilde{\rho}_{ge}=\rho_{ge}e^{- i \omega_L t} $, $\tilde{\rho}_{ee}=\rho_{ee}$
and $\tilde{\rho}_{gg}=\rho_{gg}$. Using Eqs.~(\ref{eqA00}),(\ref{eqA01}), the
normalization condition $\tilde{\rho}_{ee}+\tilde{\rho}_{gg}=1$, and
$\tilde{\rho}_{ge}=\mbox{C.C.}[\tilde{\rho}_{eg}]$, the four matrix elements of
the density matrix can be determined in principle when the initial conditions
and the stochastic trajectory $\Delta\omega(t)$  are specified. We use the
perturbation expansion \be \tilde{\rho}_{ee}(t) = \tilde{\rho}^{(0)}_{ee}(t) +
\Omega \tilde{\rho}^{(1)}_{ee}(t) + \Omega^2  \tilde{\rho}^{(2)}_{ee}(t) + \cdots, \label{eqA00-1}\\
\tilde{\rho}_{ge}(t) = \tilde{\rho}^{(0)}_{ge}(t) + \Omega
\tilde{\rho}^{(1)}_{ge}(t) + \Omega^2  \tilde{\rho}^{(2)}_{ge}(t) + \cdots ,
\label{eqA02} \ee and initially
$\tilde{\rho}^{(i)}_{ee}(t=0)=\tilde{\rho}^{(i)}_{eg}(t=0)=0$ for
$i=1,2,3,\cdots$. We insert Eqs.~(\ref{eqA00-1}),(\ref{eqA02}) into
Eqs.~(\ref{eqA00}),(\ref{eqA01}), and first consider only the zeroth order
terms in $\Omega$. We find
 $\tilde{\rho}^{(0)}_{ee}(t) = \tilde{\rho}_{ee}(0)\exp(-\Gamma t )$,
this is expected since when the laser field is absent, population in the
excited state is decreasing due to spontaneous emission. The off-diagonal term
is $\tilde{\rho}_{ge}^{(0)}(t) = \tilde{\rho}_{ge}^{(0)} (0) \exp[ - \Gamma t
/2 - i \int_0^t{\rm d} t' \delta_L (t')] $, this term is described by the
dynamics of a Kubo--Anderson classical oscillator, $\dot{x} = [- i \delta_L(t)
-\Gamma/2] x$\cite{kubo-statphys2-91}.

\subsection*{First Order Terms}
 The first order term is described by the equation
\begin{equation}
{ {\rm d} \tilde{\rho}_{ee}^{(1)} \over {\rm d} t }= - \Gamma
\tilde{\rho}_{ee}^{(1)} + {i \over 2}
 \left( \tilde{\rho}_{eg}^{(0)} - \tilde{\rho}_{ge}^{(0)}\right).
\label{eqAzzz}
\end{equation}
This equations yields the solution
\begin{equation}
\tilde{\rho}_{ee}^{(1)}(t) =  e^{ - \Gamma t} \int_0^t e^{\Gamma t'}  {\rm d}
t' \mbox{Im} \left[ \tilde{\rho}_{ge}^{(0)}(t')\right]. \label{eqA03}
\end{equation}
One can show that  $\rho_{ee}^{(1)}(t)$ is unimportant for times $\Gamma t \gg
1$, like all the other terms which depend on the initial condition.

For the off-diagonal term we find
\begin{equation}
{ {\rm d} \tilde{\rho}_{ge}^{(1)} \over {\rm d} t} = - \left[ i \delta_L(t)+
\Gamma/2\right]\tilde{\rho}_{ge}^{(1)} -{ i \over 2} \left( 2
\tilde{\rho}_{ee}^{(0)} - 1\right). \label{eqAqzz}
\end{equation}
The transient term $\tilde{\rho}_{ee}^{(0)}=\rho_{ee}(0)\exp( - \Gamma t)$ is
unimportant, and Eq.~(\ref{eqAqzz}) yields
\begin{equation}
\tilde{\rho}_{ge}^{(1)} \left( t \right) = { i \over 2} \int_0^t {\rm d} t_1
\exp\left[ - i \int_{t_1}^t {\rm d} t' \delta_L (t')  - \Gamma (t - t_1)/2
\right]. \label{eqA03a}
\end{equation}
Using $v(t) = \Omega \mbox{Im}\left[ \tilde{\rho}_{ge}^{(1)}(t)\right]$ we find
Eq.~(\ref{eqA03bmmm}).

According to the discussion in the text the number of photons absorbed in time
interval $(0,T)$ is determined by time integration of the photon current
$ W \simeq \Omega \int_0^T {\rm d} t v(t)$.
Using Eq.~(\ref{eqA03bmmm}) and definition of $\delta_L(t)$, we obtain
Eq.~(\ref{eqApzz1}). It is convenient to rewrite Eq.~(\ref{eqApzz1}) also in
the following form
\be W
=
{\Omega^2 \over 4} \int_0^T {\rm d} t_2 \int_0^T {\rm d} t_1
e^{ - i \omega_L (t_2 - t_1) - \Gamma|t_2 - t_1|/2 + i \int_{t_1}^{t_2} \Delta
\omega(t') {\rm d} t'}. \label{eqApzz1a} \ee

We calculate the average number of counts $\langle W \rangle$ using
Eq.~(\ref{eqApzz1}). The integration variables are changed to $\tau=t_2 - t_1$
and $t_1$,  and for such a transformation the Jacobian is unity. Integration of
$t_1$ is carried out from $0$ to $T-\tau$, resulting in
\begin{equation}
{\langle W \rangle} = { \Omega^2 T \over 2 } \mbox{Re}\left[ \int_0^T {\rm d}
\tau \left( 1 - {\tau \over T} \right) e^{- i\omega_L \tau - \Gamma \tau/2}
C_1^{-1}(\tau) \right] \label{eqApzz2}
\end{equation}
and $C_1^{l}(\tau) = \langle e^{-i\, l \int_0^{\tau}  \Delta \omega (t') {\rm
d} t ' } \rangle$ is the one--time correlation function. In the limit of $T \to
\infty$ we find Eq.~(\ref{eqApzz3mmm}).

 Using Eq.~(\ref{eqApzz1a}) the fluctuations are determined by
\be &&\langle W^2 \rangle = {\Omega^4 \over 16 }
\int_0^T\int_0^T\int_0^T\int_0^T {\rm d} t_1 {\rm d} t_2 {\rm d} t_3 {\rm d}
t_4
e^{ - i \omega_L ( t_2 - t_1 + t_4 - t_3) - \Gamma(|t_1 -t_2| + |t_3 - t_4|)/2} \no \\
&&\mbox{\makebox[0.5in]{ }}\times\left \langle \exp \left[ i
\int_{t_1}^{t_2}{\rm d} t' \Delta \omega (t')
  + i \int_{t_3}^{t_4}{\rm d} t' \Delta \omega (t') \right] \right \rangle,
\label{eqApzz4AP} \ee
and changing integration variables, $t_3 \to t_4$ and $t_4\to t_3$, yields
Eq.~(\ref{eqApzz4}). We replaced $t_3$ and $t_4$ to get a pulse shape similar
to that in the three--time photon echo experiment (when $t_1<t_2<t_3<t_4$).
Notice that the derivation did not assume a specific type of random process
$\Delta \omega(t)$ and our results are not limited to the two state telegraph
process we analyze in the text.

\subsection*{Second Order Terms}
According to Eq.~(\ref{eqInt1}) and (\ref{eqid1}) $\Gamma \int_0^T \rho_{ee}(t)
{\rm d} t = \Omega \int_0^T v(t) {\rm d} t \gg 1$. We now show that this
equation is valid within perturbation theory. For this aim we must consider
second order perturbation theory. The  equation for the second order term
\begin{equation}
{{\rm d} \tilde{\rho}^{(2)} _{ee}(t) \over {\rm d} t }
 = - \Gamma \tilde{\rho}_{ee}^{(2)} + \mbox{Im}\left[ \tilde{\rho}^{(1)}_{ge}(t)\right]
\label{eqA03c}
\end{equation}
yields
\be \tilde{\rho}^{(2)}_{ee} (t ) ={ 1 \over 2} e^{ - \Gamma t } \mbox{Re}
\left[ \int_0^t {\rm d} t_2 \int_0^{t_2} {\rm d} t_1
e^{ \Gamma( t_1 + t_2) /2 - i \omega_L(t_2 - t_1) + i \int_{t_1}^{t_2} \Delta
\omega (t') {\rm d} t'}\right] \label{eqA04} \ee
%
where we have neglected terms depending on initial condition. Since
$\tilde{\rho}_{ee}(t) \simeq \Omega^2 \tilde{\rho}_{ee}^{(2)}(t)$ we see that
the response is quadratic with respect to the Rabi frequency as we expect from
symmetry (population in excited state does not depend on sign of ${\bf E}_0)$.
Using Eq.~(\ref{eqA04}) we find
\be \rho_{ee}(t)={1\over 4}  \Omega^2{e^{ - \Gamma t }} \int_0^t {\rm d} t_2
\int_0^t {\rm d} t_1
e^{\Gamma(t_1 + t_2)/2 - i \omega_L(t_2 - t_1) + i \int_{t_1}^{t_2} \Delta
\omega(t') {\rm d} t'}.
\ee
Now, we consider the standard ensemble measurement and average
 $\rho^{(2)}_{ee}(t)$
with respect to history of the process $\Delta \omega (t')$.
 Assuming stationarity and a changing variables,
we find in the limit of $T \to \infty$
\begin{equation}
\langle \rho^{(2)}_{ee} \left( \infty \right ) \rangle = { \Omega ^2 \over 2
\Gamma} \mbox{Re} \left[ \int_0^{\infty} {\rm d} \tau   C_1^{-1}(\tau) e^{ -
\Gamma \tau/2  - i \omega_L \tau } \right], \label{eqA06}
\end{equation}
which is the lineshape. From Eq.~(\ref{eqApzz3mmm}) we see that $\Gamma \langle
\rho^{(2)}_{ee} (\infty )\rangle = \lim_{T \to \infty} \langle W  / T
\rangle=\langle I(\omega_L) \rangle$. Thus the theory of the averaged lineshape
can be based on either $v$ (first order perturbation theory) or on $\rho_{ee}$
(second order perturbation theory).

Instead of the averages, let us consider the stochastic variables
$\Gamma\Omega^2 \int_0^T \tilde{\rho}_{ee}^{(2)} (t) {\rm d} t$ and $\Omega^2
\int_0^T v^{(1)} (t) {\rm d} t$, where $v^{(1)}=\mbox{Im} \left[
\tilde{\rho}_{eg}^{(1)}(t) \right]$. Using Eq.~(\ref{eqA04}),  and the Laplace
$T \to s$ transform
\begin{equation}
\Omega^2 \int_0^T {\rm d} T e^{-s T} \left[ \Gamma \int_0^T {\rm d} t
\tilde{\rho}_{ee}^{(2)} (t)\right] ={\Omega^2 \over 2 s} {\Gamma \over \Gamma +
s} \hat{f}_{\rm sto}(s), \label{eqLTpee}
\end{equation}
where \be &&\hat{f}_{\rm sto}\left( s \right) = \int_0^\infty {\rm d} t e^{- s
t}
{\rm Re}\left[ \int_0^t {\rm d} t_{1} e^{ - \Gamma( t - t_1)/2 - i \omega_L ( t
- t_1) + i \int_{t_1}^t \Delta \omega (t') {\rm d} t'} \right] \label{eqLTpee1}
\ee
is a functional of the stochastic function $\Delta \omega(t)$. Using
Eq.~(\ref{eqApzz1}) we find
\be \int_0^\infty {\rm d} T  e^{ - s T} \left[ \Omega^2 \int_0^T {\rm d}t
v^{(1)} ( t )  \right] = {\Omega^2  \over 2 s} \hat{f}_{\rm sto}(s).
\label{eqLTpee2} \ee
Comparing Eq.~(\ref{eqLTpee}) and Eq.~(\ref{eqLTpee2}) we see that  a theory
based on $\rho_{ee}$ or on $v$ are not entirely identical. However, for long
times $\Gamma T \gg 1$, we may use small  $s$ behavior (i.e., $\Gamma/ (s +
\Gamma)\simeq1$) and $\Gamma \int_0^T \rho_{ee}(t) {\rm d} t = \Omega \int_0^T
v(t) {\rm d} t$ as expected.

\subsection*{Classical Lorentz Model}
 The stochastic Bloch equation is a semiphenomenological
equation with some elements of quantum mechanics in it. To understand better
whether our results are quantum mechanical in origin we analyze a classical
model. Lorentz invented the theory of classical, linear interaction of light
with matter. Here we investigate a stochastic Lorentz oscillator model. We
follow Allen and Eberley\cite{allen-resonance-75} who considered the
deterministic model in detail. The classical model is also helpful because its
physical interpretation is clear. We show that for weak laser intensity, the
stochastic Bloch equations are equivalent to classical Lorentz approach.

We consider the equation for harmonic dipole $|e| x(t)$ in the driving field,
${\cal{E}}(t)=E \cos(\omega_L t)$,
\begin{equation}
\ddot{x} +  \Gamma \dot{x} + \omega_0^2(t) x = { |e| \over  m  } { \cal{E} }(t)
\label{eqB01}
\end{equation}
where all symbols have their usual meanings and $\omega_0(t) = \omega_0 + \nu
h(t)$ is a stochastic time--dependent frequency and $| \nu h(t) | \ll
\omega_0$. All along this section we use symbols which appear also in the Bloch
formalism since their meanings in the Bloch and in the Lorentz models are
identical, as we show below.

We decompose $x$ into two parts.
\begin{equation}
x(t) = x_0 \left[ u \cos\left( \omega_L t \right) - v \sin\left( \omega_L t
\right) \right], \label{eqB02}
\end{equation}
$x_0$ is a time--independent constant, while $u$ and $v$ vary slowly in time.
The work done by the laser force $F= |e| E \cos( \omega_L t )$ on the particle
is ${\rm d} {\cal{W}} = - F{\rm d} x$ hence $ {{\rm d} {\cal {W}} \over {\rm d}
t} = - F{\dot{x}}$. Using Eq.~(\ref{eqB02}) we find
\begin{equation}
{{\rm d} \overline{\cal{W}} \over {\rm d} t}={1\over 2} |e| E x_0 \omega_L v
\label{eqB03}
\end{equation}
and the overbar denotes average over rapid laser oscillations (e.g., we assume
that the noise term $h(t)$ evolves slowly if compared with the laser period $ 2
\pi / \omega_L$). Since $h(t)$ is stochastic so is $v(t)$ hence the power ${
{\rm d} \overline{\cal{W}} \over {\rm d} t}$ is also a stochastic function.

As in Ref.~\cite{allen-resonance-75}, we assume that $u$ and $v$ vary slowly in
time such that
\begin{equation}
|\dot{u}| \ll \omega_L |v| , \ \ |\ddot{u}| \ll \omega_{L}^2 |u| , \ \
|\dot{v}| \ll \omega_L |u|,  \ \ |\ddot{v}| \ll \omega_{L}^2 |v| ,
\label{eqB04}
\end{equation}
then insert Eq.~(\ref{eqB02}) into Eq.~(\ref{eqB01}) and find two equations for
the envelopes $u$ and $v$,
\be
\dot{u} &=&  \left[\omega_L -  \omega_0(t)\right] v - {\Gamma u\over 2}, \label{eqB05} \\
\dot{v} &=& -\left[\omega_L - \omega_0(t)\right] u - {\Gamma v\over 2} +
{{|e|E\over 2 m \omega_L x_0}}, \label{eqB06} \ee
where the relation $[\omega_0^2(t)  - \omega_L^2]/ ( 2 \omega_L) \simeq
\omega_0(t) - \omega_L$ was used. Comparing Eqs.~(\ref{eqB03}), (\ref{eqB05}),
(\ref{eqB06}) with Eqs.~(\ref{eq05}), (\ref{eq:udot})-(\ref{eq:wdot}), we will
now show that in the weak laser intensity limit the Bloch equation describes
the dynamics described by the Lorentz model. To see this clearly, note that
when $\Omega\to 0$, $\rho_{ee}\simeq 0$, and hence $w\simeq -1/2$. Therefore if
we replace $-\Omega w$ in the Bloch equation,
Eqs.~(\ref{eq:udot})-(\ref{eq:wdot}) with $\Omega/2$, the Bloch equations for
$u$ and $v$ become uncoupled from that for $w$. Using this approximation we
find \be
\dot{u} &=&  \delta_L(t) v - {\Gamma u\over 2}, \label{eqB06-1} \\
\dot{v} &=& -\delta_L(t) u - {\Gamma v\over 2} + {\Omega\over 2}. \label{eqB07}
\ee

It is clearly seen that the Bloch equation in the weak intensity limit
[Eq.~(\ref{eqB06-1}) and (\ref{eqB07})] has the same structure as the Lorentz
equation [Eq.~(\ref{eqB05}) and (\ref{eqB06})]. Note that two parameters, $x_0$
and $m$ only appear in the Lorentz model while two other parameters,
$\Omega=-{\bf d}_{eg}\cdot {\bf E}/\hbar$ and $\hbar$ in the Bloch equation. To
make the equivalence between these two approaches complete, the following
relations can be deduced by comparing Eqs.~(\ref{eqB05}), (\ref{eqB06}) with
Eqs.~(\ref{eqB06-1}), (\ref{eqB07}), and Eq.~(\ref{eqB03}) with
Eq.~(\ref{eq05}),
\be
\Omega\leftrightarrow {|e|E\over m\omega_L x_0},  \, \, \, \, \, {\hbar \Omega}
\leftrightarrow {|e|x_0 E\over 2},  \no \ee or \be d_{eg}\leftrightarrow
-{1\over 2}|e| x_{0}, \, \, \, \, \, \hbar\omega_L \leftrightarrow {1\over 2}
m\omega_L^2 x_0^2. \no
\ee
To conclude, when the laser intensity is not strong
the stochastic phenomenological Bloch equation describes classical behavior.

\section{Appendix A: Exact Calculations of $\langle W\rangle$ and $\langle W^2\rangle$}
\label{AppWex}
In this appendix we use straightforward complex analysis and find
\begin{equation}
 \langle W^2 \rangle =
{\Omega^4 \over 16} {\cal L}^{-1} \left\{ \sum_{i=1}^5 \hat{\xi}_i(s) +\rm{C.C.}\right\},
\label{AppC1}
\end{equation}
where
\be
\hat{\xi}_1(s)&=&
{1\over s^2} \sum_{i,j,k,l}
\hat{P}^{-1}_{ij}\left(s+s_{+}\right)
\hat{P}^0_{jk}\left(s\right)
\hat{P}_{kl}^{1}\left(s+s_{-}\right), \\
\hat{\xi}_2(s)&=&
{1\over s^2} \sum_{i,j,k,l}
\hat{P}^{-1}_{ij}\left(s+s_{+}\right)
\hat{P}^0_{jk}\left(s +\Gamma\right)
\hat{P}_{kl}^{1}\left(s+s_{-}\right), \\
\hat{\xi}_3 (s)&=&
{1\over s^2} \sum_{i,j,k,l}
\hat{P}^{1}_{ij}\left(s+s_{-}\right)
\hat{P}^0_{jk}\left(s\right)
\hat{P}_{kl}^{1}\left(s+s_{-}\right),  \\
\hat{\xi}_4 (s)&=&
{1\over s^2} \sum_{i,j,k,l}
\hat{P}^{1}_{ij}\left(s+s_{-}\right)
\hat{P}^0_{jk}\left(s +\Gamma\right)
\hat{P}_{kl}^{1}\left(s+s_{-}\right),  \\
\hat{\xi}_5 (s) &=&
{2\over s^2} \sum_{i,j,k,l}
\hat{P}^{1}_{ij}\left(s+s_{-}\right)
\hat{P}^2_{jk}\left(s +2 s_{-}\right)
\hat{P}_{kl}^{1}\left(s+s_{-}\right), \label{eqAE02}
\ee
and $s_{\pm} = \Gamma/2 \pm i \omega_L$.
The Laplace transforms of the weights $P_{ij}^{a}(\tau)$ are
\be
\hat{P}_{++}^0(s)&=& \hat{P}_{--}^0(s)  = {R + s \over s \left( s + 2 R\right)},   \label{eqLTW1}\\
\hat{P}_{+-}^0(s)&=&  \hat{P}_{-+}^0(s) = {R \over s \left( s + 2 R\right)},
\\
\hat{P}_{++}^{1}(s)&=&  \hat{P}_{--}^{-1}(s) = {R + s - i \nu \over \left(s - s_1\right) \left( s - s_2 \right)},
 \\
\hat{P}_{+-}^{1}(s) &=& \hat{P}_{-+}^{-1}(s) =
 \hat{P}_{-+}^{1}(s) = \hat{P}_{+-}^{-1}(s) =
 {R  \over \left(s - s_1\right) \left( s - s_2 \right)},  \\
\hat{P}_{--}^{1}(s) &=&  \hat{P}_{++}^{-1}(s) = {R + s + i \nu \over \left(s - s_1\right) \left( s - s_2 \right)}, \no \\
\hat{P}_{ij}^{2}(s) &=&  \hat{P}_{ij}^{1}(s)|_{\nu \to 2 \nu},
\label{eqLTW}
\ee
where
$$ s_{1,2} = - R \pm \sqrt{R^2 - \nu^2}.$$
The inverse Laplace transforms of $\hat{\xi}_{i}(s)$ in Eq.~(\ref{eqAE02})
are calculated using standard methods of complex analysis to yield $\xi_{i}(T)$:
\be
\xi_1(T) = \sum_{m=1}^5
\left.\left[s^3 \hat{\xi}_1(s)(s-z_m)\right]\right|_{s=z_m} f_1(z_m,T),
\label{eqaa03}
\ee
where the simple poles $z_i$ are given by
\be
z_1 &=& - s_{+}  + s_1 = -{\Gamma \over 2} - i \omega_L - R + \sqrt{R^2 - \nu^2}, \no
\\
z_2 &=& - s_{+} +  s_2 = -{\Gamma \over 2} - i \omega_L - R - \sqrt{R^2 - \nu^2}, \no
\\
z_3 &=& - s_{-} +  s_1 = -{\Gamma \over 2} + i \omega_L - R + \sqrt{R^2 - \nu^2}, \no
\\
z_4 &=& - s_{-} +  s_2 = -{\Gamma \over 2} + i \omega_L - R - \sqrt{R^2 - \nu^2}, \no
\\
z_5&=&-2R, \no
\label{eqaa04}
\ee
and $f_1(z,T)$ is
\begin{equation}
f_1(z,T)=-{T^2\over 2 z} - {T \over z^2} - {1 - e^{zT} \over z^3}.
\label{eqaa05}
\end{equation}
Notice that if $\omega_L=0$, $z_1=z_3$ and $z_2=z_4$, then the poles become
second order. Also, we can neglect exponential decays $\exp(z_i T)$ for $i=1-4$
since $\Gamma T \gg 1$, and the term $\exp(z_5 T)$ is important when $R T \le
1$.
\be
\xi_2(T) = \sum_{m=1}^6 \left.\left[s^2 \hat{\xi}_2(s)(s-z_m)\right]\right |_{s=z_m}f_2(z_m,T),
\label{eqaa06}
\ee
where $z_1=-s_{+} + s_1$, $z_2=-s_{+}+s_2$,
$z_3 = -s_{-}+s_1$, $z_4=-s_{-}+s_2$,
 $z_5=-\Gamma$, $z_6=-2 R - \Gamma$
and
\be
f_2(z,T)=-{T\over  z} - {1 - e^{zT} \over z^2}.
\label{eqaa07}
\ee
The expression $s^2 \hat{\xi}_2(s)(s-z_m)\left. \right |_{s=z_m}$ in Eq.~(\ref{eqaa06}) is the residue of
$s^2 \hat{\xi}_2(s)$ when $s=z_m$.
\be
\xi_3(T)
&
=
&\left.\left[s^3(s - z_3) \hat{\xi}_3(s)\right]\right |_{s=z_3}f_1(z_3,T)
\no \\
&
+
&
\sum_{m=1}^2 \left.\left[s^3(s - z_m)^2 \hat{\xi}_3(s)\right]\right |_{s=z_m}f_3(z_m,T)
\no \\
&+&
\sum_{m=1}^2 \left.\left[{{\rm d}\over{\rm d}s} s^3(s - z_m)^2 \hat{\xi}_3(s)\right]
\right |_{s=z_m}f_1(z_m,T),
\label{eqaa07a}
\ee
where
$z_1=-s_{-}+s_1$, $z_2=-s_{-}+ s_2$, $z_3=-2 R$
and
\begin{equation}
f_3(z,T) = {T^2 \over 2 z^2} + {2 T \over z^3} + {T e^{zT} \over z^3} + {3 (1-e^{zT}) \over z^4}.
\label{eqaa08}
\end{equation}
\be
\xi_4(T)
&
=
&
\sum_{m=1}^2 \left.\left[s^2 (s - z_m)^2 \hat{\xi}_4(s)\right]\right |_{s = z_m} f_4(z_m,T) \no \\
&
+
&
\sum_{m=1}^2 \left. \left[{{\rm d} \over {\rm d}s} s^2 (s - z_m)^2  \hat{\xi}_4(s)\right]
\right |_{s=z_m} f_2(z_m,T) \no \\
&+&\sum_{m=3}^4 \left. \left[s^2 (s - z_m) \hat{\xi}_4(s)\right]\right |_{s=z_m} f_2(z_m,T),
\label{eqaa09}
\ee
where $z_1=-s_{-}+s_1$, $z_2= - s_{-} + s_2$, $z_3= - \Gamma$,
$z_4= - 2 R - \Gamma$ and
\begin{equation}
f_4(z,T) = {T \over  z^2} + {T e^{zT} \over z^2} + {2 (1-e^{zT}) \over z^3}.
\label{eqaa10}
\end{equation}
Finally,
\be
%
\xi_5(T)
&
=
&
\sum_{m=1}^2 \left.\left[ s^2 (s - z_m)^2  \hat{\xi}_5(s)\right]\right |_{s=z_m}  f_4(z,T) \no \\
&+&\sum_{m=1}^2 \left.\left[{{\rm d} \over {\rm d}s} s^2 (s - z_m)^2 \hat{\xi}_5(s)\right]
\right |_{s=z_m} f_2(z_m,T) \no \\
&+&\sum_{m=3}^4 \left.\left[s^2 (s - z_m) \hat{\xi}_5(s)\right]\right |_{s = z_m} f_2(z_m,T)
\label{eqaa11}
\ee
where
$z_1=-s_{-}+s_1$, $z_2=-s_{-}+s_2$, $z_3=- 2 s_{-}+\tilde{s}_1$ and
$z_4=-2 s_{-}+\tilde{s}_2$ and $\tilde{s}_{1,2} = -R \pm \sqrt{R^2 - 4 \nu^2}$.

From Eq.~(\ref{eqApzz2}) the average
counting number can be written as
\be
\langle W \rangle = {\Omega^2 \over 4}{\cal{L}}^{-1} \left\{{1\over s^2}\left[
\h{C}_1^{-1}(s+s_+)+\rm{C.C.}\right]\right\},
\ee
which leads to
Eq.~(\ref{eqmain1}) with Eq.~(\ref{C1}).
%
It is also easy to show that
\be \la W\ra  = {\Omega^2 \over 4} {\cal{L}}^{-1} \left\{ {1 \over s_1 - s_2}
\left[ - s_2 f_2\left( s_1 - s_{+},T\right) + s_1 f_2\left( s_2 -
s_{+},T\right) \right] + \rm{C.C.} \right\}, \label{eqmain1T} \ee
where $f_2(z,T)$ was defined in Eq.~(\ref{eqaa07}).
In the limit of large $T$ we have
\begin{equation}
 \langle W  \rangle \simeq {\Omega^2 \over 8} T \left[
\sum_{i,j} \hat{P}^{-}_{ij}\left(s_{+}\right) + \rm{C.C.} \right].
\label{eqmainT22}
\end{equation}

\section{Appendix B: $Q$ in the Long Time Limit}
\label{AppQlt} The exact expression for the  $Q$ parameter in the long time
limit is given by $Q=\mbox{Numerator}[Q]/\mbox{Denominator}[Q]$, where
\be
&&\mbox{Denominator}[Q]=
  \Gamma R \left[ \Gamma^3 + 8 \Gamma^2 R + 16 {\nu}^2 R +
     4 \Gamma \left( {\nu}^2 + 4 R^2 + \omega_L^2 \right)  \right]
 \nonumber \\
   &&\mbox{\makebox[1.15 in]{}} \times \left[ \Gamma^4 + 4 \Gamma^3 R + 16 \Gamma R \left( {\nu}^2 + \omega_L^2 \right)  +
     4 \Gamma^2 \left( 2 {\nu}^2 + R^2 + 2 \omega_L^2 \right)  \right.
 \nonumber \\
   &&\mbox{\makebox[1.15 in]{}} \left.   +  16 \left( {\nu}^4 - 2 {\nu}^2 \omega_L^2 + R^2 \omega_L^2 + \omega_L^4 \right)
  \right]
 \nonumber \\
  &&\mbox{\makebox[1.15 in]{}} \times \left[ \Gamma^4 + 8 \Gamma^3 R + 32 \Gamma R \left( {\nu}^2 + \omega_L^2 \right)  +
   8 \Gamma^2 \left( {\nu}^2 + 2 R^2 + \omega_L^2 \right) \right.
 \nonumber \\
    &&\mbox{\makebox[1.15 in]{}} \left.  + 16 \left( {\nu}^4 - 2 {\nu}^2 \omega_L^2 + 4 R^2 \omega_L^2 + \omega_L^4 \right)        \right]^2
\label{QltDen} \ee
and
\be
&&\mbox{Numerator[Q]}=
   64 {\nu}^2  {\Omega}^2
   \left[ \Gamma^{11} \omega_L^2 + 28 \Gamma^{10} R \omega_L^2 \right.
\no \\
     &&\mbox{\makebox[1 in]{ }}\left. + 4 \Gamma^9 \left( 85 R^2 \omega_L^2 + 4 \omega_L^4 +
        {\nu}^2 \left( R^2 + 4 \omega_L^2 \right)  \right) \right.
\nonumber \\
   &&\mbox{\makebox[1 in]{ }}   \left. + 16 \Gamma^8 R \left( 146 R^2 \omega_L^2 + 21 \omega_L^4 +
        {\nu}^2 \left( 4 R^2 + 21 \omega_L^2 \right)  \right)  \right.
\nonumber \\
 &&\mbox{\makebox[1 in]{ }}    \left. + 512 \Gamma^2 {\nu}^2 R^3 \left( 11 {\nu}^6 + 92 R^4 \omega_L^2 +
        96 R^2 \omega_L^4 + 3 \omega_L^6 \right. \right.
\nonumber \\
 &&\mbox{\makebox[1 in]{ }}  \left. \left. +  {\nu}^4 \left( 18 R^2 - 19 \omega_L^2 \right)  +
        5 {\nu}^2 \left( 18 R^2 \omega_L^2 + \omega_L^4 \right)  \right)  \right.
\nonumber \\
  && \mbox{\makebox[1 in]{ }}  \left. + 16 \Gamma^7 \left( 620 R^4 \omega_L^2 + 183 R^2 \omega_L^4 + 6 \omega_L^6  \right. \right.
\no \\
   && \mbox{\makebox[1 in]{ }}  \left.  \left. + {\nu}^4 \left( 4 R^2 + 6 \omega_L^2 \right) +
       {\nu}^2 \left( 25 R^4 + 182 R^2 \omega_L^2 + 4 \omega_L^4 \right)  \right) \right.
\nonumber \\
   &&\mbox{\makebox[1 in]{ }}  \left. + 32 \Gamma^6 R \left( 832 R^4 \omega_L^2 + 424 R^2 \omega_L^4 + 42 \omega_L^6  \right. \right.
\no \\
   && \mbox{\makebox[1 in]{ }}  \left. \left. +  {\nu}^4 \left( 25 R^2 + 42 \omega_L^2 \right)  + {\nu}^2
         \left( 38 R^4 + 437 R^2 \omega_L^2 + 28 \omega_L^4 \right)
 \right)  \right.
\nonumber \\
   &&\mbox{\makebox[1 in]{ }}  \left. +128 \Gamma^4 R \left( {\nu}^6 \left( 27 R^2 + 14 \omega_L^2 \right)  +
        2 \left( 10 R^2 + 7 \omega_L^2 \right)  \right. \right.
\no \\
   &&\mbox{\makebox[1 in]{ }}   \left. \left.  \left. \times ( 4 R^2 \omega_L + \omega_L^3 \right)^2 +
2 {\nu}^4 \left( 28 R^4 + 65 R^2 \omega_L^2 - 7 \omega_L^4 \right)  \right.
\right.
 \no \\
 &&\mbox{\makebox[1 in]{ }} \left. \left. + {\nu}^2 \left( 8 R^6 + 625 R^4 \omega_L^2 + 207 R^2 \omega_L^4 - 14 \omega_L^6
           \right)  \right)  \right.
\nonumber \\
 && \mbox{\makebox[1 in]{ }} \left.    + 1024 \Gamma {\nu}^2 R^2
      \left( {\nu}^8 + 8 R^4 \omega_L^4 + 3 R^2 \omega_L^6 + \omega_L^8 \right. \right.
\no \\
 && \mbox{\makebox[1 in]{ }}\left. \left.  + {\nu}^6 \left( 7 R^2 - 4 \omega_L^2 \right)  +
{\nu}^4 \left( -11 R^2 \omega_L^2 + 6 \omega_L^4 \right)   \right. \right.
\no \\
&&\mbox{\makebox[1 in]{ }} \left. \left. \left. +{\nu}^2 \left( 40 R^4 \omega_L^2 + R^2 \omega_L^4 - 4
\omega_L^6 \right)  \right) \right. \right.
\nonumber \\
  &&\mbox{\makebox[1 in]{ }} \left. + 64 \Gamma^5 \left( 688 R^6 \omega_L^2 + 552 R^4 \omega_L^4 + 111 R^2 \omega_L^6 + 4 \omega_L^8  \right.\right.
\no \\
   &&\mbox{\makebox[1 in]{ }} \left. \left.   +{\nu}^6 \left( 6 R^2 + 4 \omega_L^2 \right)   +
           {\nu}^4 \left( 57 R^4 + 105 R^2 \omega_L^2 - 4 \omega_L^4 \right)  \right. \right.
\no \\
  &&\mbox{\makebox[1 in]{ }} \left. \left.
     +   {\nu}^2 \left( 28 R^6 + 653 R^4 \omega_L^2 + 102 R^2 \omega_L^4 - 4 \omega_L^6
           \right)  \right) \right.
\nonumber \\
 &&\mbox{\makebox[1 in]{ }} \left.  + 2048 {\nu}^2 R^3 \left( {\nu}^8 - 4 {\nu}^6 \omega_L^2 + 4 R^4 \omega_L^4 + 5 R^2 \omega_L^6 + \omega_L^8 \right.\right.
\no\\
 && \mbox{\makebox[1 in]{ }}\left.\left. +{\nu}^4 \left( 5 R^2 \omega_L^2 + 6 \omega_L^4 \right)  -
2 {\nu}^2 \left( 5 R^2 \omega_L^4 + 2 \omega_L^6 \right)  \right)  \right.
\no \\
    && \mbox{\makebox[1 in]{ }} \left.  + 256 \Gamma^3 \left( {\nu}^8 \left( 4 R^2 + \omega_L^2 \right)  +
\omega_L^2 \left( R^2 + \omega_L^2 \right)  \left( 4 R^2 + \omega_L^2 \right)^3
\right. \right.
\no \\
    && \mbox{\makebox[1 in]{ }} \left. \left. +  {\nu}^6 \left( 39 R^4 + 4 R^2 \omega_L^2 - 4 \omega_L^4 \right)  \right. \right.
\nonumber \\
  &&  \mbox{\makebox[1 in]{ }} \left. \left. + {\nu}^4 \left( 20 R^6 + 114 R^4 \omega_L^2 - 7 R^2 \omega_L^4 + 6 \omega_L^6 \right) \right. \right.
\nonumber \\
 && \mbox{\makebox[1 in]{ }} \left.\left. + {\nu}^2 \left( 356 R^6 \omega_L^2 + 215 R^4 \omega_L^4 - 14 R^2 \omega_L^6 -  4 \omega_L^8 \right)
\right)  \right].
\label{QltNum}
\end{eqnarray}
To obtain these results we find the small $s$ expansion of the Laplace
transforms $\langle \hat{W}^2 \rangle$ and $\langle \hat{W} \rangle$. These
give in a standard way the long time behavior of the averages $\langle W^2
\rangle$ and $\langle W \rangle$ with which $Q$ is found. These equations were
derived using {\it Mathematica}\cite{math}, without which the calculation is
cumbersome. Note that $Q$ is an even function of $\omega_L$ and $\nu$, as
expected from symmetry. For some special cases, discussed in the text, the
exact results are much simplified.

\noindent {\it \bf Remark 1} In the limit of $\Gamma\rightarrow 0$, which
corresponds to the limit of Kubo's lineshape theory, we get \be Q={2\nu^2
\Omega^2 R \over \Gamma [(\omega_L^2-\nu^2)^2+4R^2\omega_L^2]}, \label{QG0} \ee
\noindent {\it \bf Remark 2} For $\omega_L=0$, we get \be Q=
  \frac{256 {\Omega^2} {\nu}^4 R \left( \Gamma + 2 R \right) }
   {\Gamma \left( \Gamma + 4 R \right)  \left( \Gamma^2 + 4 {\nu}^2 + 2 \Gamma R \right)
     {\left( \Gamma^2 + 4 {\nu}^2 + 4 \Gamma R \right) }^2}.
\label{Qltw0}
\ee

\section{Appendix C: $Q$ in the Slow Modulation Regime}
\label{AppQs} In the two--state random walk model where the ``velocity'' of the
particle is either $I_+$ or $I_-$, we can conveniently calculate the first and
second moments of the ``coordinate'' $W$ by introducing the generating function
$G(k,T)$, \be G(k,T)=\left\la \exp\left (-i k \int_0^T I(t) {\rm
d}t\right)\right\ra=\left \la e^{-ikW}\right \ra, \label{AppE1} \ee where $k$
is an auxiliary variable. For the stochastic process where the ``velocity'' of
the particle alternates between between $I_+$ and $I_-$ with transition rate
$R$, we can easily evaluate ${G}(k,T)$ using the Laplace $s \to T$
transformation, \be
G(k,T)
={\cal{L}}^{-1}\left\{ s + ik (I_+ + I_- )/2 +2R \over (s + ik I_+ +R) (s + ik I_- +R) -R^2\right\}.
\ee
Now from Eq.~(\ref{AppE1}) we have $\la W\ra$ and $\la W^2\ra $
\be
\la W\ra&=&\left \la \int_0^{T} I(t) \rm{d} t \right\ra = -\left.{\partial G(k,T)
\over \partial ik}\right |_{k=0},  \\
\la W^2\ra &=&\left \la \left [\int_0^{T} I(t) \rm{d} t\right]^2 \right\ra
= \left.{\partial^2 G(k,T) \over \partial (ik)^2}\right |_{k=0},
\ee
By taking derivatives of $G(k,T)$ we find
\be
-\left.{\partial G(k,T) \over \partial ik}\right |_{k=0}
&=&{\cal{L}}^{-1}\left\{ {I_+ + I_-\over 2 s^2}\right\},  \\
\left.{\partial^2 G(k,T)\over \partial (ik)^2}\right |_{k=0}
&=&{\cal{L}}^{-1}\left\{{(I_+^2 + I_-^2)\over s^2(s+2R)}
+{(I_+ + I_-)^2 R\over s^3(s+2R)}\right\},
\ee
which yield after the inverse Laplace transform,
\be
\la W\ra &=&\left({I_+ + I_- \over 2}\right) T,    \\
\la W^2\ra &=& {(I_+ + I_-)^2\over 4} T^2 + {(I_+ - I_-)^2 \over 4}\left({T\over R}+{e^{-2RT}-1\over 2R^2}\right).
\label{AppE5}
\ee
$Q$ is given by Eq.~(\ref{eqSML09})
in the slow modulation regime.

\section{Appendix D: $Q$ in the Fast Modulation Regime}
\label{AppQf} Based on the approximation introduced in the text we can
calculate $Q$ in the fast modulation regime. Once the factorization of the
three--time correlation functions is made in Eq.~(\ref{factor}),
$\hat{\xi}_i(s)$, the functions determining $\la W^2\ra$ [Eq.~(\ref{eqAE02})]
can be written as
\be \h\xi_{1}(s)&=&{2\over s^2}
\h{C}_{1}^{-1}(s+s_+)\h{C}_{1}^{0}(s)\h{C}_{1}^{1}(s+s_{-}),  \\
\h\xi_{2}(s)&=&{2\over s^2}
\h{C}_{1}^{-1}(s+s_+)\h{C}_{1}^{0}(s+\Gamma)\h{C}_{1}^{1}(s+s_-),  \\
\h\xi_{3}(s)&=&{2\over s^2}
\h{C}_{1}^{1}(s+s_-)\h{C}_{1}^{0}(s)\h{C}_{1}^{1}(s+s_-),  \\
\h\xi_{4}(s)&=&{2\over s^2}\h{C}_{1}^{1}(s+s_-)\h{C}_{1}^{0}(s+\Gamma)
\h{C}_{1}^{1}(s+s_-),  \\
\h\xi_{5}(s)&=&{4\over s^2}
            \h{C}_{1}^{1}(s+s_-)\h{C}_{1}^{2}(s+2s_-)\h{C}_{1}^{1}(s+s_-), \label{AQf1}
\ee where $s_{\pm}=\Gamma/2\pm i\omega_L$ as defined in the Appendix A. Then $\la W\ra$ and $\la W^2\ra$ are calculated from Eqs.
(\ref{eqmain}) and (\ref{eqmain1}), \be \la W \ra &=&{\Omega^2 \over 4}
{\cal{L}}^{-1}\left\{ {1\over s^2} {F}_1(s)\right\}, \label{AQf2} \\
\la W^2 \ra
&=&{\Omega^4\over 8}{\cal{L}}^{-1} \left\{{1\over s^3}{F}_2(s)\right\}, \label{AQf3}
\ee
where
\be
F_1(s)&=&c_1(s)+\rm{C.C.},   \\
F_2(s)&=&\left(1+{s\over s+\Gamma}\right) F_1(s)^2 + 2 s (c_1(s)^2 c_2(s)+ \rm{C.C.}),  \\
c_1(s)&\equiv&\h{C}_1^{1}(s+s_-), \\
c_2(s)&\equiv&\h{C}_1^{2}(s+2s_-). \label{AQf4}
\ee
After the cumulant approximation is made for $C_1^{l}(t)$ and the long time
limit is taken in Eq.~(\ref{cumul}) $c_1(z)$ and $c_2(z)$ are simply given by
\be
c_1(s)&=&{1\over s+s_{-}+\Gamma_f/2},   \\
c_2(s)&=&{1\over s+2s_{-}+2\Gamma_f}.  \label{AQf5}
\ee

Since only the long time limit is relevant for the calculation of
$Q$
in the fast modulation regime we make expansions of $\la W\ra$ and $\la W^2 \ra$
around $s=0$, and find in the long time limit
\be
Q={\Omega^2\over 2}{F'_2(0)-F_1(0)F'_1(0)\over F_1(0)}. \label{AQf6}
\ee
Note that Eq.~(\ref{AQf6}) is valid once the factorization approximation
is made irrespective of the second order cumulant approximation.
After performing a lengthy but straightforward algebra
using Eqs.~(\ref{AQf4})-(\ref{AQf6}), we obtain the result of
$Q$ in the fast modulation regime given in Eq.~(\ref{eqFML01}).
%

\newcommand{\noopsort}[1]{} \newcommand{\printfirst}[2]{#1}
  \newcommand{\singleletter}[1]{#1} \newcommand{\switchargs}[2]{#2#1}

\end{document}